\documentclass[3p,times,twocolumn]{elsarticle}
\usepackage{ecrc}
\usepackage{multicol}
\usepackage{graphicx}
\usepackage{booktabs}
\usepackage{slashed}
\usepackage{amssymb,bm,mathrsfs,bbm,amscd}
\usepackage[tbtags]{amsmath}
\usepackage{lastpage}
\usepackage{multirow}
\usepackage{textcomp}
\usepackage{epstopdf}
\usepackage{float}
\usepackage{footnote}
\usepackage{hyperref}
\usepackage{indentfirst}
\usepackage{color}
\usepackage{soul}
\usepackage{url}
\volume{}
\firstpage{1}
\journalname{ journalname }
\runauth{}
\jid{ASTROPART PHYS}
\jnltitlelogo{jnltitlelogo}
\usepackage{amssymb}
\usepackage[figuresright]{rotating}
\begin{document}
\begin{frontmatter}
\setlength{\abovecaptionskip}{4pt plus1pt minus1pt}
\setlength{\belowcaptionskip}{4pt plus1pt minus1pt}
\setlength{\abovedisplayskip}{6pt plus1pt minus1pt}
\setlength{\belowdisplayskip}{6pt plus1pt minus1pt}
\addtolength{\thinmuskip}{-1mu}
\addtolength{\medmuskip}{-2mu}
\addtolength{\thickmuskip}{-2mu}
\setlength{\belowrulesep}{0pt}
\setlength{\aboverulesep}{0pt}
\setlength{\arraycolsep}{2pt}
\setlength{\parindent}{2em}
\dochead{}
\title{Study of longitudinal development of air showers in the knee energy range 
}
\author[1]{Feng Zhang}
\address[1]{School of Physical Science and Technology, Southwest Jiaotong University, Chengdu 610031, Sichuan, China}
\address[2]{Department of Physics, Bamidele Olumilua University of Education, Science and Technology, Ikere-Ekiti, Nigeria.}
\author[1]{Hu Liu\corref{mycorrespondingauthor}}
\cortext[mycorrespondingauthor]{Corresponding author.}
\ead{huliu@swjtu.edu.cn}
\author[1]{Fengrong Zhu\corref{mycorrespondingauthor}}
\ead{zhufr@swjtu.edu.cn}
\author[1,2]{Jacob Oloketuyi}

\begin{abstract}
Ground-based cosmic ray experiments detect cosmic ray mainly by measuring the longitudinal and lateral distribution of secondary particles produced in the extensive air shower (EAS). The EAS of cosmic ray in the knee energy region is simulated via CORSIKA software. Several simulation samples with different energy, composition and zenith angles were carried out to understand the longitudinal development of electron, muon and Cherenkov light in EAS. All the results presented were obtained assuming an observation plane at an altitude of 4400 m a.s.l. The differences of longitudinal development between electron and Cherenkov light were studied, and the reconstruction uncertainty of shower maximum for electron from Cherenkov light was estimated to be 10-15$g/cm^{2}$ for nuclei above 1 PeV. The performances of energy measurement and the composition discrimination ability based on longitudinal development were studied and compared with that from lateral distribution. It was found that number of electron per depth at its shower maximum has the smallest shower-to-shower fluctuations, but the shower-to-shower fluctuations of electron density measured at observation level was very close to it when the appropriate zenith angle was employed. The shower-to-shower fluctuations of shower maximum for electron is 50-55 $g/cm^2$ for proton, and 20-25 $g/cm^2$ for iron, but the composition discrimination ability between nuclei from muon density measured at observation level is much better than the shower maximum variable from longitudinal development. The hadronic model dependencies of the longitudinal development and lateral distribution were also discussed.
\end{abstract}

\end{frontmatter}

\section{Introduction}
\label{introduction}
Cosmic ray are energetic particles from space. The most important features of cosmic ray spectrum are the so-called ``knee'' around 3 × $10^{15}$ eV and the ``ankle'' around 5 × $10^{18}$ eV~\cite{BLUMER2009293}. The origin of the ``knee'' of cosmic ray energy spectra is still under debate, however, it is widely believed to be a strong constraint for acceleration and propagation models~\cite{Erlykin_1997, Berezhko_2007}. One method to understand the origin of the ``knee'' of cosmic ray spectrum is directly measuring the spectrum and isotropy of cosmic ray, especially the energy spectra and the ``knee'' of individual species~\cite{Erlykin_1997, Berezhko_2007, def_A}. \\\indent
For particles around the ``knee'' energy, currently they are only detected by ground-based experiments through the extensive air shower (denoted as EAS hereafter) of the primary particles. There are several secondary particles produced in EAS, most of which are gamma rays, electrons/positrons, muons and neutrons/anti-neutrons. Meanwhile, Cherenkov light and fluorescence light are produced during the propagation of charged secondary particles (mostly electrons and positrons) in the atmosphere.  The density of secondary particles at the observation level versus the perpendicular distance to the shower axis (namely, the lateral distribution) has been well studied in the literature~\cite{EASliu,LateralDis,LaterDis2,LaterDis3}. This was used to reconstruct the core position, energy and the type information of primary particles. But, more so, it is an important way for detecting the cosmic ray by EAS. The number of secondary particles versus the slant depth of atmosphere traversed (namely the longitudinal development) also contains energy and type information of the primary particles, which is another important way for detecting cosmic ray by EAS. \\\indent 
The longitudinal development of EAS can be studied through Cherenkov light, fluorescence light and muon, because they interact very rarely with the atmosphere during propagation, while the other secondary components interact with the atmosphere very frequently before reaching the observation level. The longitudinal developments of Cherenkov light and fluorescence light were normally measured by telescope, which converged the Cherenkov/fluorescence light on the focal plane to have an image of the EAS. Since Cherenkov light and fluorescence light were produced mostly by electrons along the trajectory, the longitudinal developments of Cherenkov light and fluorescence light were regarded as a measurement of the longitudinal development of electrons. The longitudinal development of muon can be reconstructed from the arrival time of muon at the observation level as documented in the literature~\cite{Longimuon,Longimuon2,Longimuon3} \\\indent
Basically, ground-based experiments detect the primary particles by measuring the longitudinal development and lateral distribution of EAS. For example, HESS, MAGIC and VERITAS experiments detect gamma rays by measuring the longitudinal development of Cherenkov light, the reconstructed angle resolution within $0.1^\circ$, and the energy resolution around 15\%~\cite{HESS,MAGIC,VERITAS}. WFCTA of LHAASO experiment detect cosmic ray nuclei also by measuring the image of Cherenkov light, the reconstructed energy resolution for proton around 10—15\% in the ``knee'' energy range, the length, width and angular distance variable (the angular distance between the center of the image and the direction of the primary particles) from the image, were used to identify the proton from other nuclei~\cite{paperof_WFCTA_YZY}. Pierre Auger experiment, Telescope Array experiment and HiRes experiment constructed several telescopes to measure the longitudinal development of fluorescence light from cosmic ray nuclei around and above the ``ankle'' energy, measuring the all-particle spectrum and composition of cosmic ray ~\cite{Auger,TA,HiRes}. The experiments that measure the lateral distribution of the secondary particles from EAS, e.g., KASCADE experiment, HAWC experiment and many others were also reported~\cite{KASCADEproton,HAWC,ARGO,TIBET}.  \indent
In the last decades, there had been more hybrid experiments that measured both the longitudinal development and lateral distribution, e.g. the LHAASO experiment, the Pierre Auger experiment, the Telescope Array experiment and others, measuring the EAS comprehensively~\cite{LHAASO_Design,Auger,TA}. However, some details about them remain unclear. It is therefore important to understand the advantages and disadvantages of the two detection methods, and combine them to have a better performance. \indent
In this paper, the differences of longitudinal developments between electron and Cherenkov light were investigated. The energy and composition measurements based on longitudinal development of EAS were compared with that based on lateral distribution. The uncertainties of energy and composition measurements based on longitudinal development and lateral distribution were both composed of two parts, with the first from the shower-to-shower fluctuations, and the other from the measurement uncertainties due to the detector. The measurement uncertainties due to detector were not studied in this paper, since it is different for different detectors, and can be reduced with better designed detectors. So the performances derived in this paper can be regarded as an upper limit for the variables studied, and they can still provide important information for the selection of detector type, detector design and data analysis for physical measurements.\indent
The paper is organized as below, the simulation procedures and parameters are introduced in section \ref{simulation}. The properties of longitudinal development, the differences between electron and Cherenkov light are presented in section \ref{long_dev}. The performances of energy measurement and composition discrimination from longitudinal development of secondary particles, and their comparison with those from lateral distribution are studied in section \ref{compa_study}. The differences between $EPOS-LHC$ and $QGSJet-II-04$ models are presented in section \ref{diff_hadronic}, and finally section \ref{summary} presents the summary. 

\section{Simulation}
\label{simulation}
The cascade processes of primary particles in the atmosphere were simulated with CORSIKA software (v77410)~\cite{Corsika}. For high-energy hadronic interaction processes, the $EPOS-LHC$ and $QGSJet-II-04$ models are employed, the main results are based on $EPOS-LHC$ model. The $Fluka$ and $EGS4$ models were employed for low-energy hadronic interaction processes and electromagnetic interaction processes, respectively. There are five mass groups, namely proton, helium, CNO group (the mass number, noted as A, is 14), MgAlSi group (A is 27) and iron. All of which were simulated with fixed energy at $log_{10}(E/GeV)$=5.1, 5.3, 5.5, 5.7, 5.9, 6.1, 6.5, 6.9, 7.3, 7.7, 8.1, 8.5 and 8.9, respectively, corresponding to the energy range from below the ``knee'' to approaching the ``ankle''. The altitude of detection plane was set to  4400 m a.s.l., and the horizontal and vertical components of the earth magnetic field at the observation site were set to be 34.618 $\mu T$ and 36.13 $\mu T$, respectively. Because Cherenkov light has different properties of longitudinal development for different zenith angles (denoted as $\theta$ below), to reduce the amount of simulation, three zenith angles ($\theta$=$0^\circ$, $28^\circ$ and $45^\circ$) were simulated, which corresponds to a maximum atmosphere depth of $\sim850$ $g/cm^{2}$ with the choosing altitude. The azimuth angle of primary particles is uniformly distributed in the range of $[0^{\circ},360^{\circ}]$. \indent
The secondary components produced in EAS include Cherenkov light, electron, gamma rays, muon, neutron and so on~\cite{EASliu}. The threshold energies for hadrons, muons, electrons and photons were set to 0.1, 0.1,  0.001 and 0.001 GeV, respectively. The longitudinal development of secondary particles was recorded in steps of 10 $g/cm^{2}$. Cherenkov light was produced and propagated to the observation level. To reduce the size of Cherenkov data in Cherenkov light simulation, only the energy of primary particles less than $10^{7.4}$ GeV were simulated, and only Cherenkov photons inside several regions at observation level were recorded. Each region corresponds to one telescope, and all photon bunches reaching spheres with 3m radius around the center of the region were detected and their position, and arrival direction were stored. \indent
In the production of secondary particles and Cherenkov light, the atmospheric refractive index and its evolution with height are taken into account, and the density of the atmosphere along the altitude is described by the US standard atmosphere model. The wavelength distribution of Cherenkov light is also simulated. \indent
The perpendicular distance between the telescope and shower axis (denoted as $R_{p}$ below) was fixed at $R_{p}$=50, 100, 150, 200, 300 and 400 m, respectively. In reality, Cherenkov light was absorbed inside the ozone region and scattered by atmosphere and aerosol. Telescopes were built to converge the Cherenkov light on the focal plane to have an image of the EAS, however, they were dependent on the atmosphere models, the aerosol models and the parameters of the telescope. This study focused on the physical properties of longitudinal development in CORSIKA data therefore, no telescope simulation was performed. \indent
\section{Longitudinal development}
\label{long_dev}
During the cascade process of the EAS, the number of secondary particles first increased as the EAS developed. It was then decreased due to energy loss. Moreover, two important parameters can be extracted from the longitudinal development. For electron and muon, one is the slant atmosphere depth at which the number of secondary particles per atmospheric depth bin reached maximum (denoted as $X^{max}_{e}$ for electron and $X^{max}_{\mu}$ for muon), another one is the number of secondary particles per atmospheric depth bin at $X^{max}_{e}$/$X^{max}_{\mu}$ (denoted as $N^{max}_{e}$ for electron and $N^{max}_{\mu}$ for muon). Meanwhile, for Cherenkov light, one is the slant atmosphere depth at which the number of Cherenkov photons per atmospheric depth bin observed by a telescope located $R_{p}$ away from the shower axis reached maximum (denoted as $X^{max}_{Cer}$), another one is the number of Cherenkov light at $X^{max}_{Cer}$ (denoted as $N^{max}_{Cer}$. As can be seen below, $X^{max}_{Cer}$ is not an intrinsic properties of air shower, it is different for telescopes located at different distance to the shower axis, so a $R_{p}$ value is attached when $X^{max}_{Cer}$ is giving. In general, $X^{max}_{e}$, $X^{max}_{\mu}$ and $X^{max}_{Cer}$ can be used to distinguish between different composition. $N^{max}_{e}$, $N^{max}_{\mu}$ and $N^{max}_{Cer}$ are  good energy estimators. In this section, the longitudinal developments of electrons, muons and Cherenkov light will be discussed. \indent
\subsection{Electron and muon}
\label{longitudinal_eu}
\begin{figure*}[htbp]
\begin{minipage}[t]{1.\linewidth} 
\includegraphics[width=0.5\linewidth]{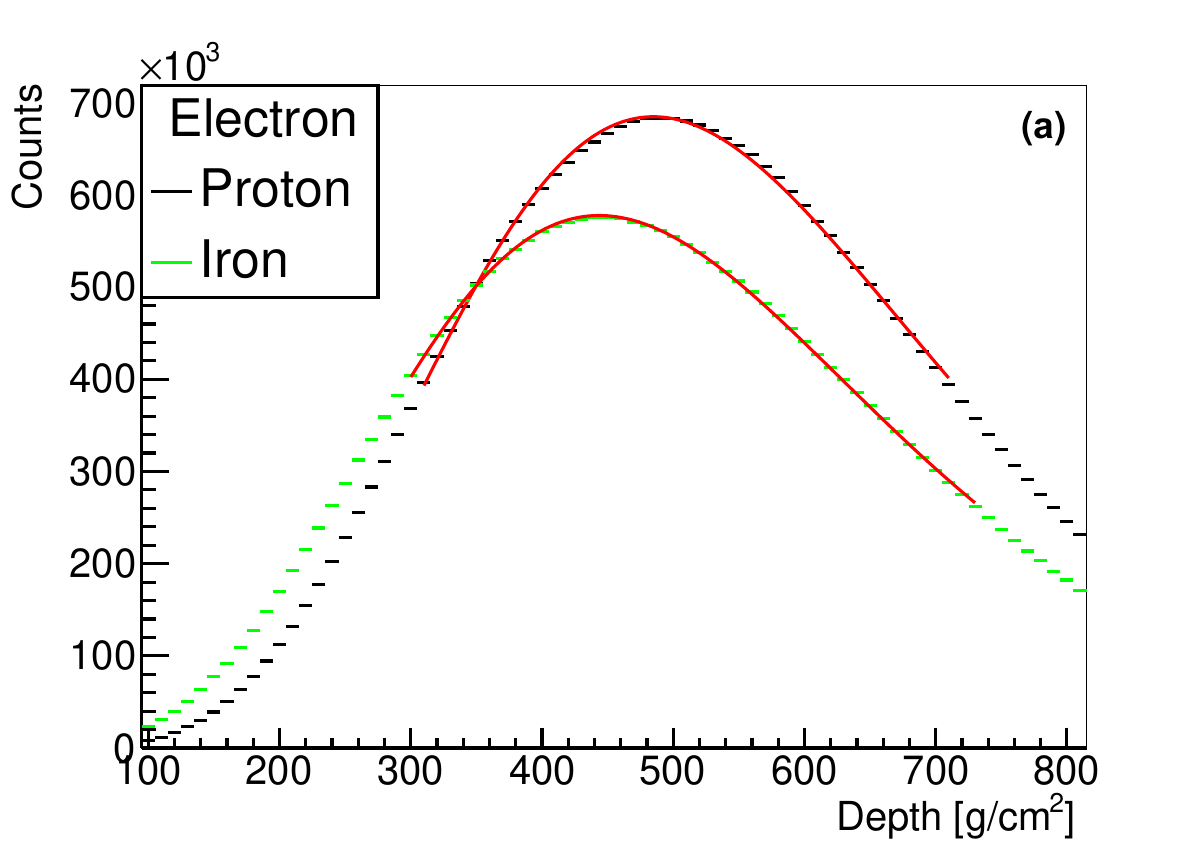} 
\hspace{0cm}
\includegraphics[width=0.5\linewidth]{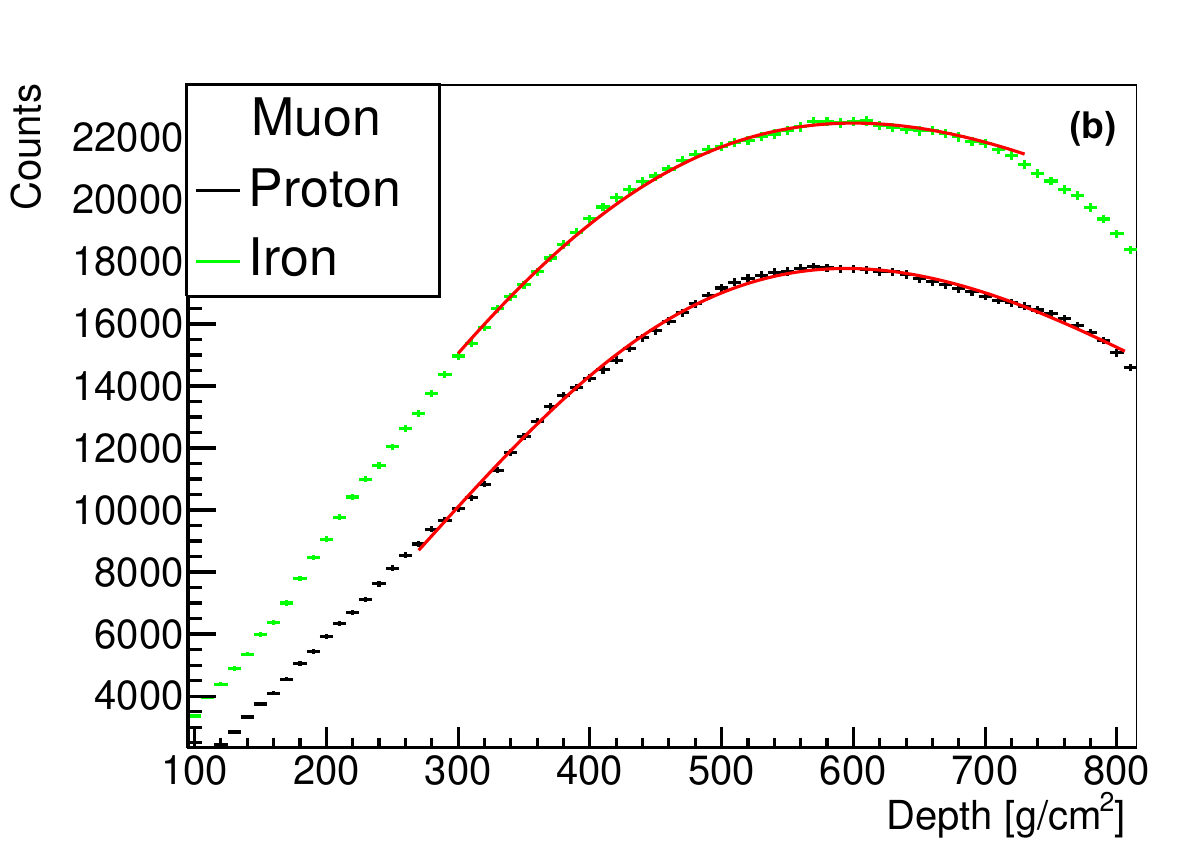}
\hspace{0cm}
\includegraphics[width=0.5\linewidth]{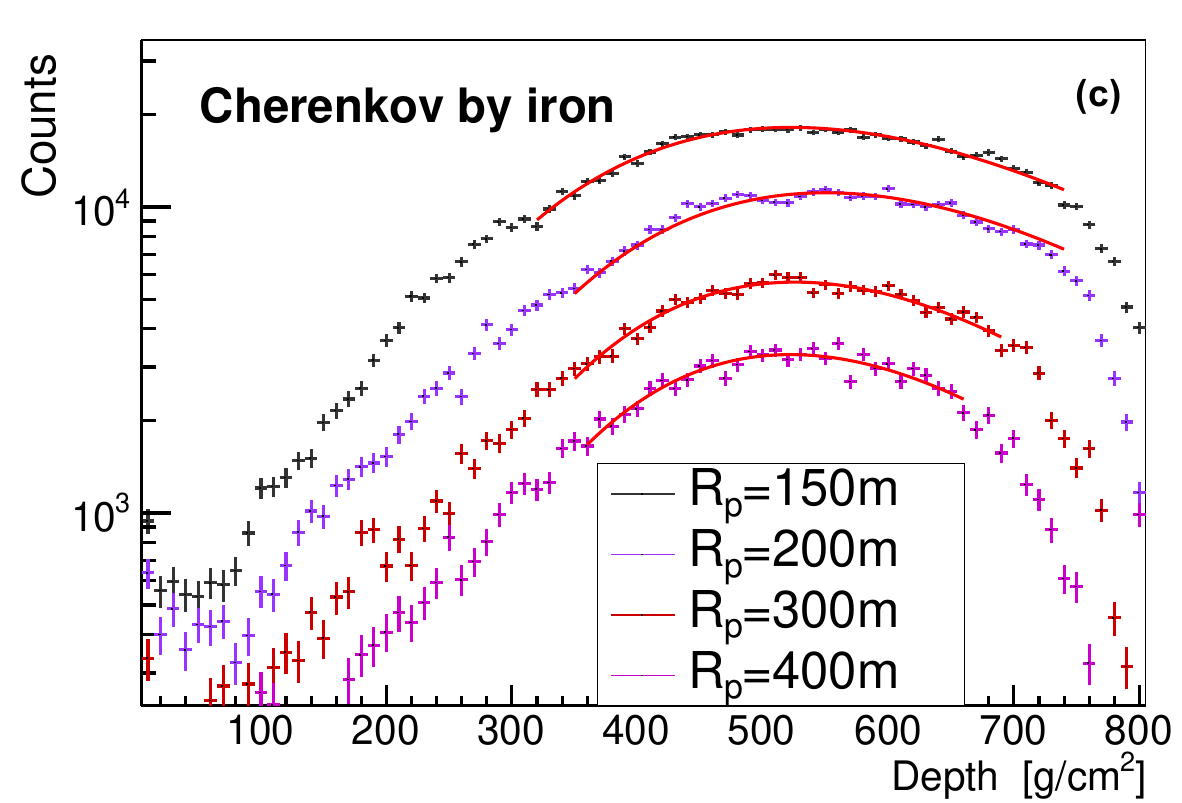}
\hspace{0cm}
\includegraphics[width=0.5\linewidth]{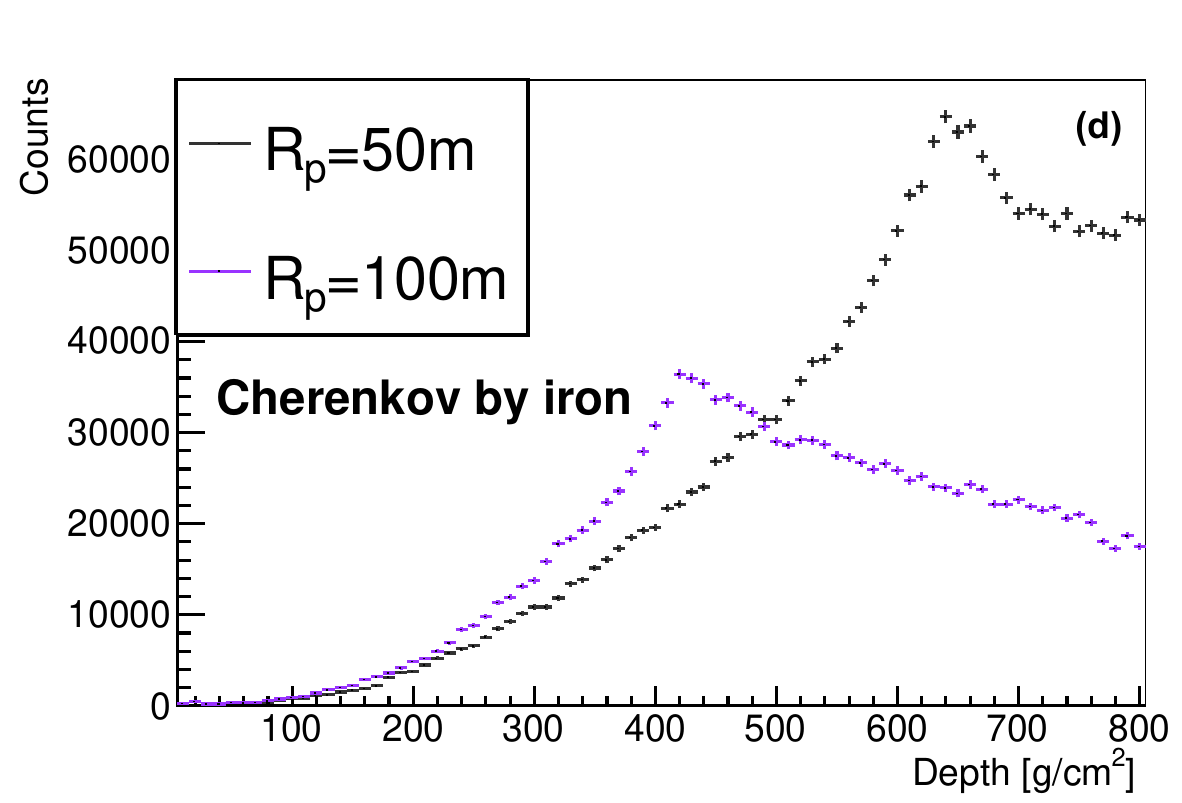}
\caption{Upper panel: The longitudinal developments of electron (a)}and muon (b) in the shower induced by proton and iron, respectively. Histograms with different colors indicate different primary compositions. Lower panel: The longitudinal developments of Cherenkov light with different perpendicular distances from the shower axis (or different $R_{p}$ values) in an iron-induced shower. Colored lines indicate different $R_{p}$ values, $R_{p}$=150, 200, 300 and 400 m in the lower left plot (c), while $R_{p}$=50, 100 m in the lower right plot (d). The number of electron, muon and Chrenkov light with $R_{p}=>$150m was fitted by formula \ref{eq1} as shown in red lines, while the Cherenkov light with $R_{p}<=$100m was not fitted since they are dominated by the Cherenkov pool.
\label{longi_all}
\end{minipage}
\end{figure*}

The longitudinal developments of electron and muon for proton and iron nuclei showers with energy $log_{10}(E/GeV)$=6.1 were extracted directly from CORSIKA data, and are shown in the upper panel of Fig. \ref{longi_all}. The x-axis is the traversed atmospheric depth from entering the atmosphere to the position of electron/muon, and the y-axis is the number of electron/muon per atmospheric depth bin. \indent
The longitudinal development can be parameterized by several functions including, the Gaisser-Hillas function, Greisen function and Gaussian-in-age function~\cite{ABUZAYYAD20011,Aab_2019,PHD_ZBK}. The fitting results of the three functions are close to each other~\cite{ABUZAYYAD20011,PHD_ZBK}. To reduce the number of parameters, the Gaussian-in-age function (formula \ref{eq1}) was employed to parameterize the longitudinal development of electron, muon and Cherenkov light; where X is the slant depth of atmosphere traversed, $X_{max}$ is the slant depth of atmosphere when EAS reached shower maximum, $N_{max}$ is the number of particles per slant depth bin at shower maximum, $\sigma$ indicates the width of longitudinal development curve, and s is the shower age, which is defined as $s$=$3X/(X+2X_{max})$. The red lines in the upper panel of Fig. \ref{longi_all} are the fitting results of the longitudinal developments of electron and muon. For proton-induced shower, the $\chi^{2}/dof$ are around 1.9 and 0.95 for electron and muon, respectively, and they are similar for iron-induced shower.\indent

\begin{equation}
N(X) = N_{max}e^{-\frac{(s-1)^{2}}{2\sigma^2}}\ \ \ (s=\frac{3X}{X+2X_{max}})  \tag{1}
\label{eq1}
\end{equation}

\subsection{Cherenkov light}
\label{longitudinal_ch}
\begin{figure*}[htbp]
\begin{minipage}[t]{1.\linewidth}
\centerline{\includegraphics[width=0.8\linewidth]{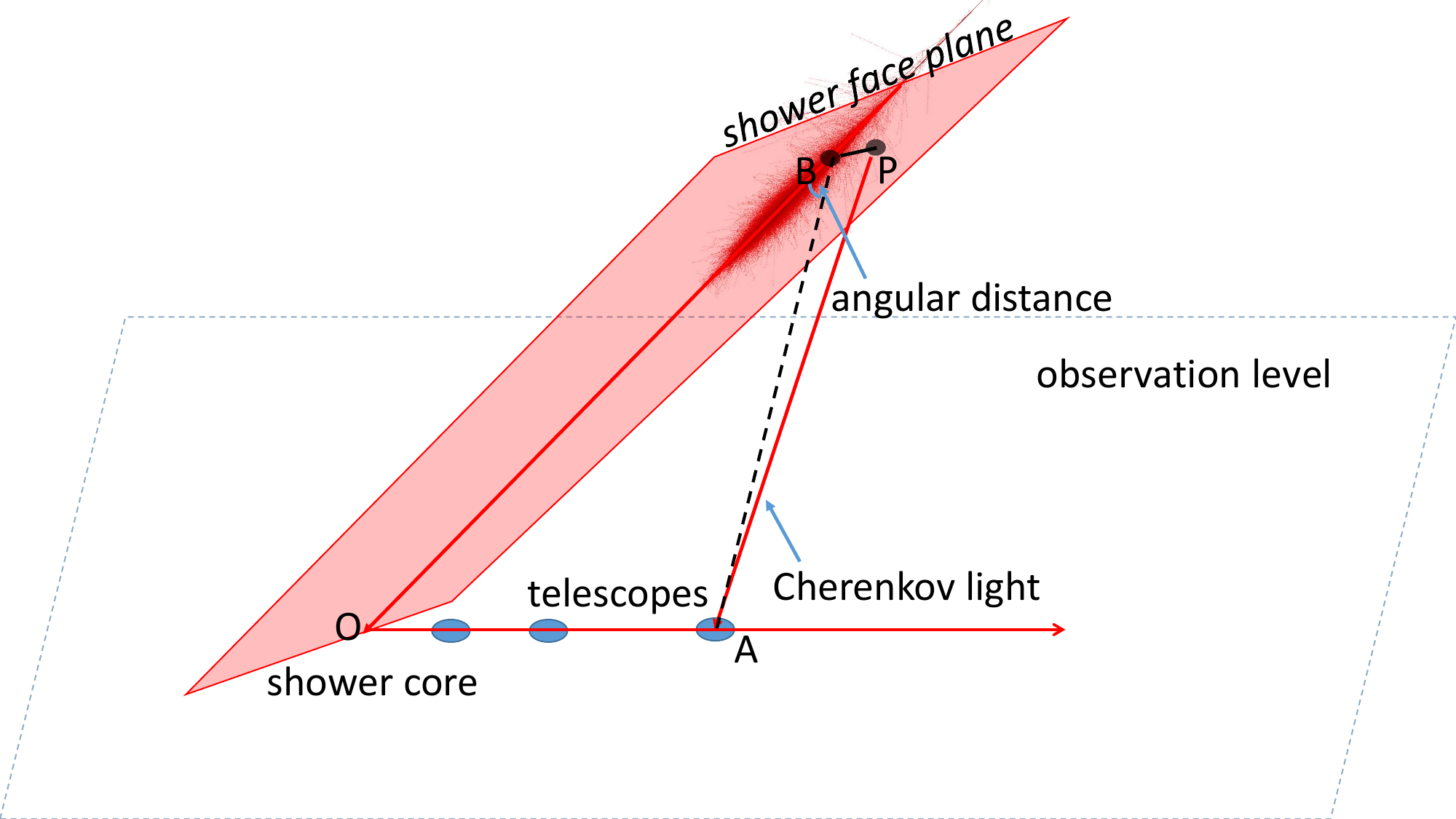}}
\caption{ The reconstruction of emitting position of the Cherenkov light. The plane constructed by the shower axis and vector $\overrightarrow{OA}$ (O is the shower core position, A is the position of the} telescope) is defined as the shower detection plane (SDP plane). The shower face plane (red plane) was constructed with the shower axis and the perpendicular vector of the SDP plane. These two planes are perpendicular to each other. P is the intersection point by projecting the arrival direction of each Cherenkov light back into the shower face plane, which is regarded as the emitting position of the Cherenkov light. The black dashed line BA is the projection of $\overrightarrow{PA}$ at the shower detection plane. 
\label{shower}
\end{minipage} 
\end{figure*}
\vspace{0pt}

For Cherenkov light, only the one detected by the telescope was recorded. The longitudinal developments of Cherenkov light for telescopes at different $R_p$ values for iron-induced shower were shown in the lower panel of Fig. \ref{longi_all}, where the x-axis is the traversed slant depth from entering the atmosphere to the reconstructed emitting position of the Cherenkov light, and the y-axis is the number of Cherenkov light per slant atmospheric depth bin detected by the telescope. The reconstructed emitting position of Cherenkov light was calculated according to the method shown in Fig. \ref{shower}, where the shower detection plane (denoted as SDP plane hereafter) was constructed by the shower axis and vector $\overrightarrow{OA}$ (O is the shower core position, A is the position of the telescope). The shower face plane (red shaded area in Fig. \ref{shower}) was constructed by the shower axis and the perpendicular vector of the SDP plane. These two planes are perpendicular to each other. By projecting the arrival direction of each Cherenkov light back into the shower face plane, the intersection point (marked as P in Fig. \ref{shower}) is the reconstructed emitting position ofthe Cherenkov light~\cite{measure_Xamx_che}. The longitudinal development of Cherenkov light can also be well parameterized with formula \ref{eq1}. The red lines in the lower left panel are the fitting results of the longitudinal developments of Cherenkov light with $R_{p}>100m$ for iron-induced shower, the fitting results for proton-induced shower are also similar.\indent
The longitudinal developments of Cherenkov light for $R_{p}<=100m$ are shown in the lower right panel of Fig. \ref{longi_all}. According to the figure, the longitudinal developments of Cherenkov light for $R_{p}<=100m$ is different from the ones for larger $R_{p}$ values, with an evident narrow peak in the longitudinal developments. In those region, the amount of Cherenkov light reaching the telescopes is highly dominated by the Cherenkov pool, which is centered at the shower core position with a radius of around 125m. The shower maximum of Cherenkov light observed by telescope ($X_{Cer}^{max}$) in this region is highly correlated with $R_{p}$ values, while for Cherenkov light outside the Cherenkov pool region (lower left panel of Fig. \ref{longi_all}), the broader distribution corresponding to the Cherenkov light generated by scattered electron, and the shower maximum of Cherenkov light are less dependent on $R_{p}$ values \cite{measure_Xamx_che}. Therefore, this study focused only on the Cherenkov light that can be observed by the telescopes with $R_{p}>$100 m. \indent

\subsection{Electron versus Cherenkov light}
\begin{figure*}[htbp]
\begin{minipage}[t]{1.\linewidth}
\includegraphics[width=0.5\linewidth]{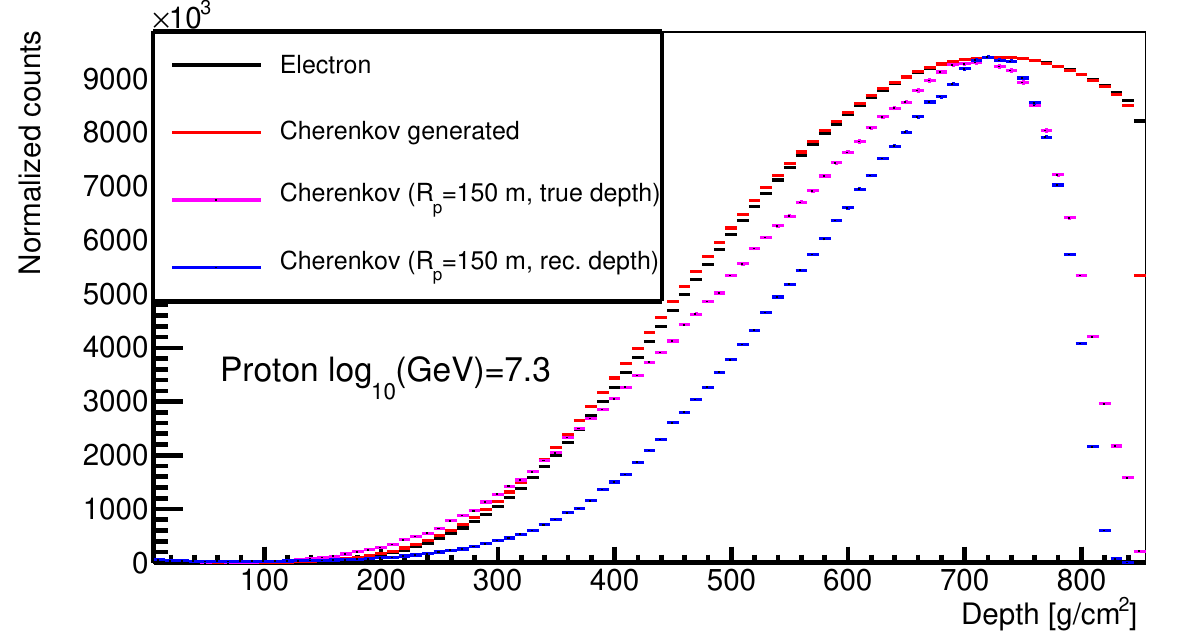}
\hspace{0cm}
\includegraphics[width=0.5\linewidth]{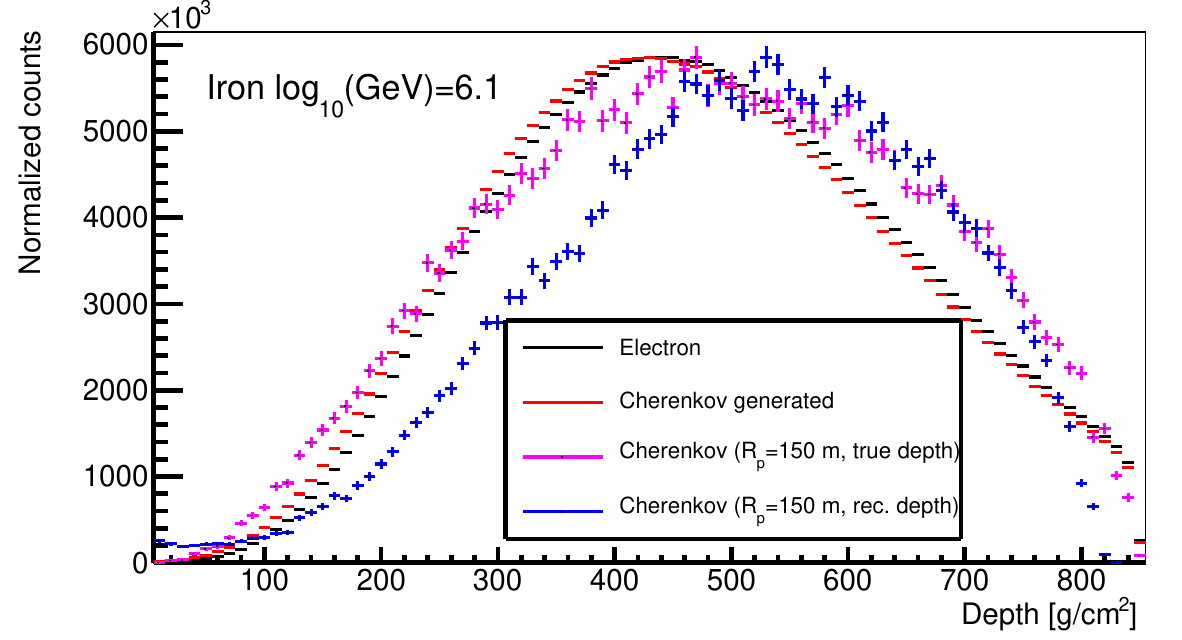}
\caption{The comparison of longitudinal developments between electron (black), all Cherenkov light generated (red, named Cherenkov generated in the legend), Cherenkov light observed by a telescope ($R_{p}$=150 m) with the true slant depth (magenta, named Cherenkov ($R_{p}=150$ m, true depth) in the legend) and Cherenkov light observed by a telescope ($R_{p}$=150 m) with the reconstructed slant depth (blue, named Cherenkov ($R_{p}=150$ m, rec. depth) in the legend). The shower is induced by proton with energy $log_{10}(E/GeV)$=7.3 (left), and iron with energy $log_{10}(E/GeV)$=6.1 (right), the zenith angle of primary particle is 45 degree.}
\label{longdiff_ech}
\end{minipage} 
\end{figure*}

Cherenkov light was produced by the charged secondary particles (mainly electron and positron) of EAS when their speed is larger than the local speed of light in the atmosphere. At low energy, the fluorescent light is very weak, and the longitudinal development of electrons can be studied via the Cherenkov light and radio emitted by electrons. However, Cherenkov light was produced anisotropically, and in principle, the longitudinal development of Cherenkov light observed by a telescope can be different from the longitudinal development of electrons. \indent
The longitudinal development of electron and Cherenkov light for the same event are shown in Fig. \ref{longdiff_ech}, where four longitudinal development lines were constructed and compared. The black histogram is the longitudinal development of electron; the red histogram is the longitudinal development of all Cherenkov light (denoted as Cherenkov generated), which is defined as the entire amount of Cherenkov photons generated at each depth bin versus the slant depth traversed, and the slant depth is the true value recorded in CORSIKA data; the magenta histogram is the longitudinal development of Cherenkov light only for Cherenkov light observed at 150 m away from the shower axis (denoted as Cherenkov ($R_{p}=150$ m, true depth)); the blue histogram is the longitudinal development of Cherenkov light also only for Cherenkov light observed at 150 m away from the shower axis, but the slant depth is the reconstructed depth with the method described in Fig. \ref{shower}. (denoted as Cherenkov ($R_{p}=150$ m, rec. depth)). \indent
The differences between the black and red histograms is caused by the non-uniform Cherenkov light yield along the shower trajectory (due to the evolution of refraction index, atmosphere molecule density and energy distribution of electrons). The difference between red and magenta histograms is due to the propagation effect, because the angular distribution of Cherenkov light generated is not isotropic, and for each slant atmosphere depth, only the Cherenkov light pointing to the direction of telescope can be observed by the telescope, therefore the fraction of Cherenkov light observed by the telescope with respect to the total Cherenkov light generated is different at different slant atmosphere depth, this account for the difference between red and magenta histograms. The difference between magenta and blue histograms is caused by the difference between the reconstructed depth and true depth, this can be understood from the reconstruction method, the slant depth of Cherenkov light emitting position was reconstructed by backtracking the Cherenkov light and finding the cross position with the shower face plane however, the electrons in the shower is not just distributed in the shower face plane, they are distributed in three-dimensional space, this generated the difference between the true slant depth and reconstructed slant depth. As can be seen in Fig. \ref{longdiff_ech}, the difference between longitudinal development of electron and Cherenkov light observed by a telescope is mainly due to the propagation effect and the reconstruction error of slant atmosphere depth.\indent 

\vspace{0pt}

\begin{figure*}[htbp]
\begin{minipage}[t]{1.\linewidth} 
\includegraphics[width=0.5\linewidth]{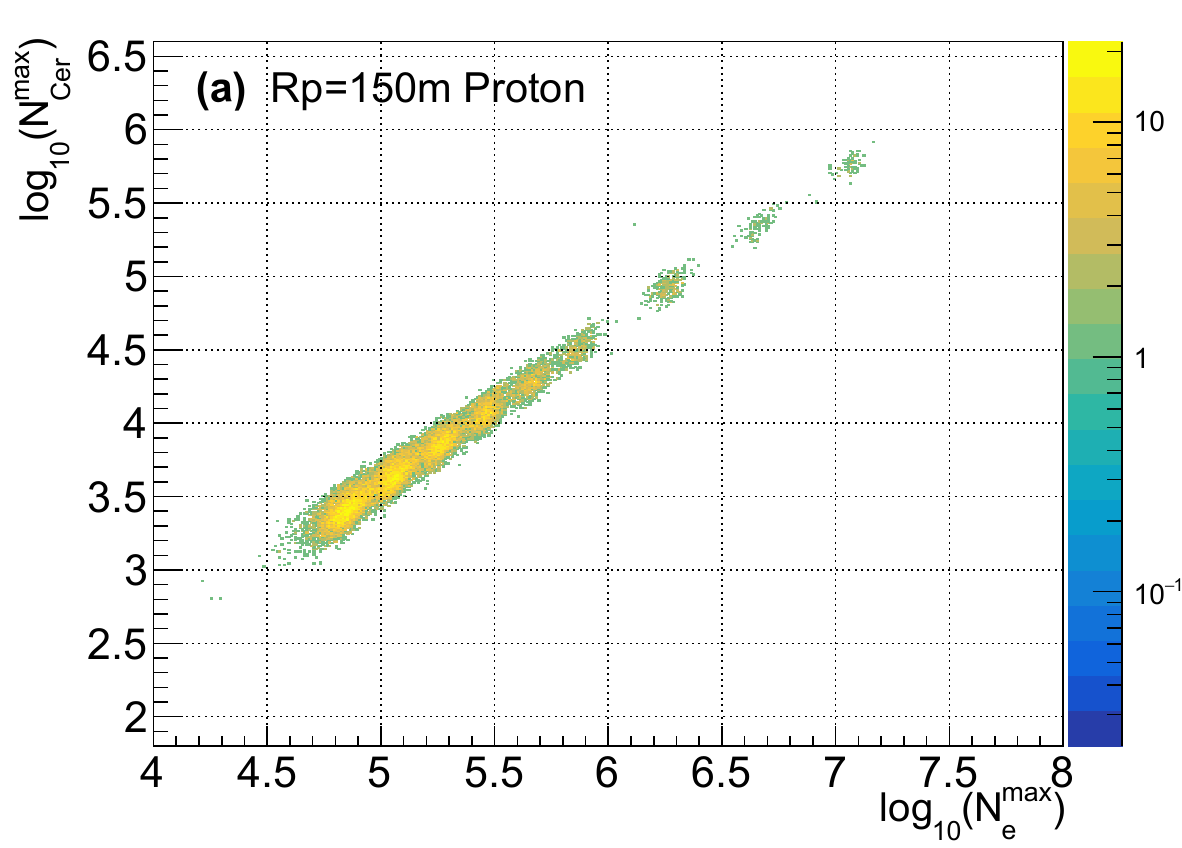}
\hspace{0cm}
\includegraphics[width=0.5\linewidth]{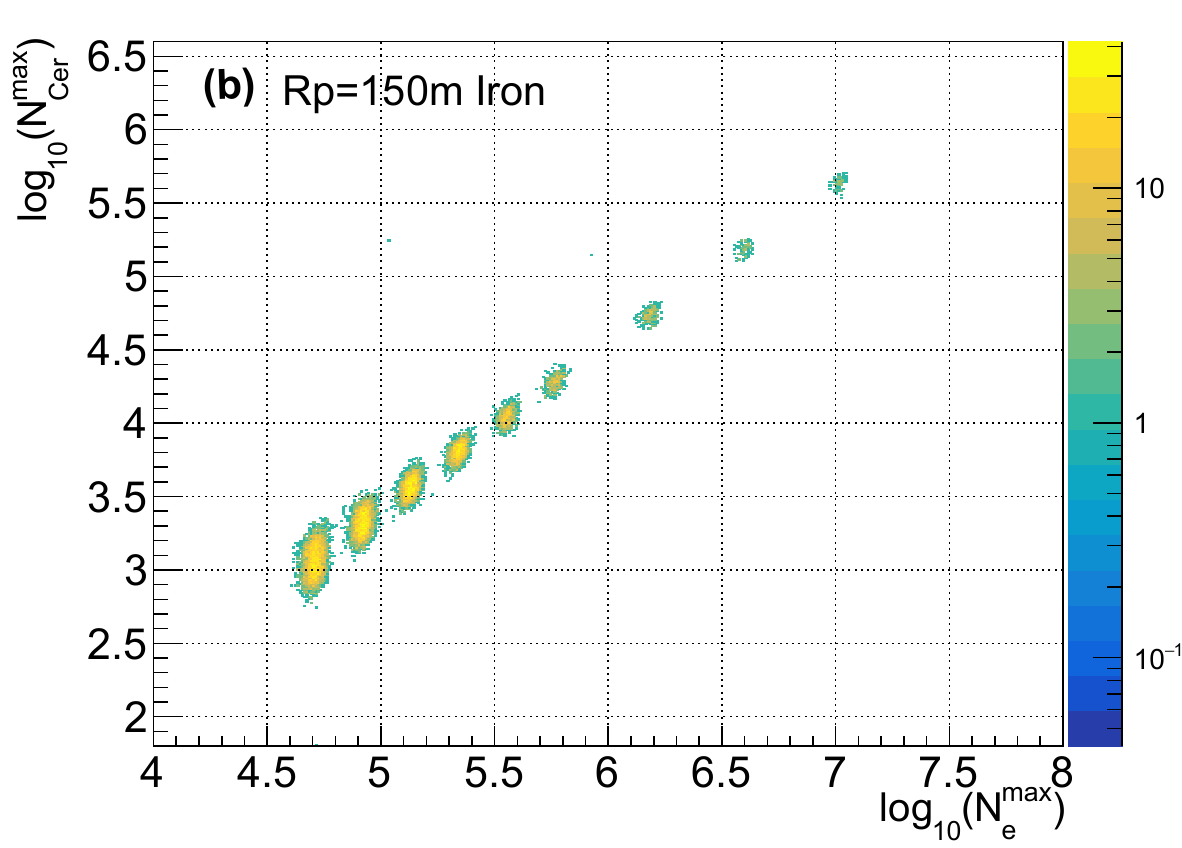}
\hspace{0cm}
\includegraphics[width=0.5\linewidth]{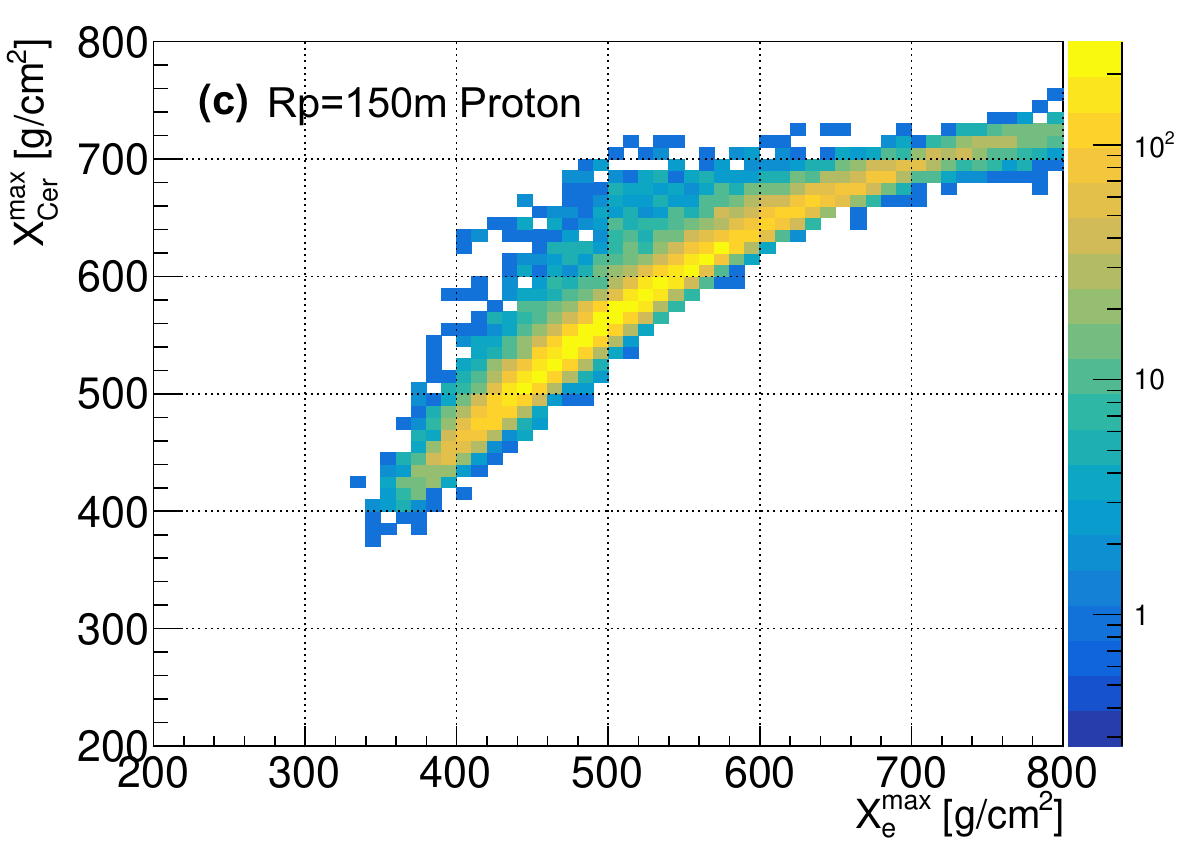}
\hspace{0cm}
\includegraphics[width=0.5\linewidth]{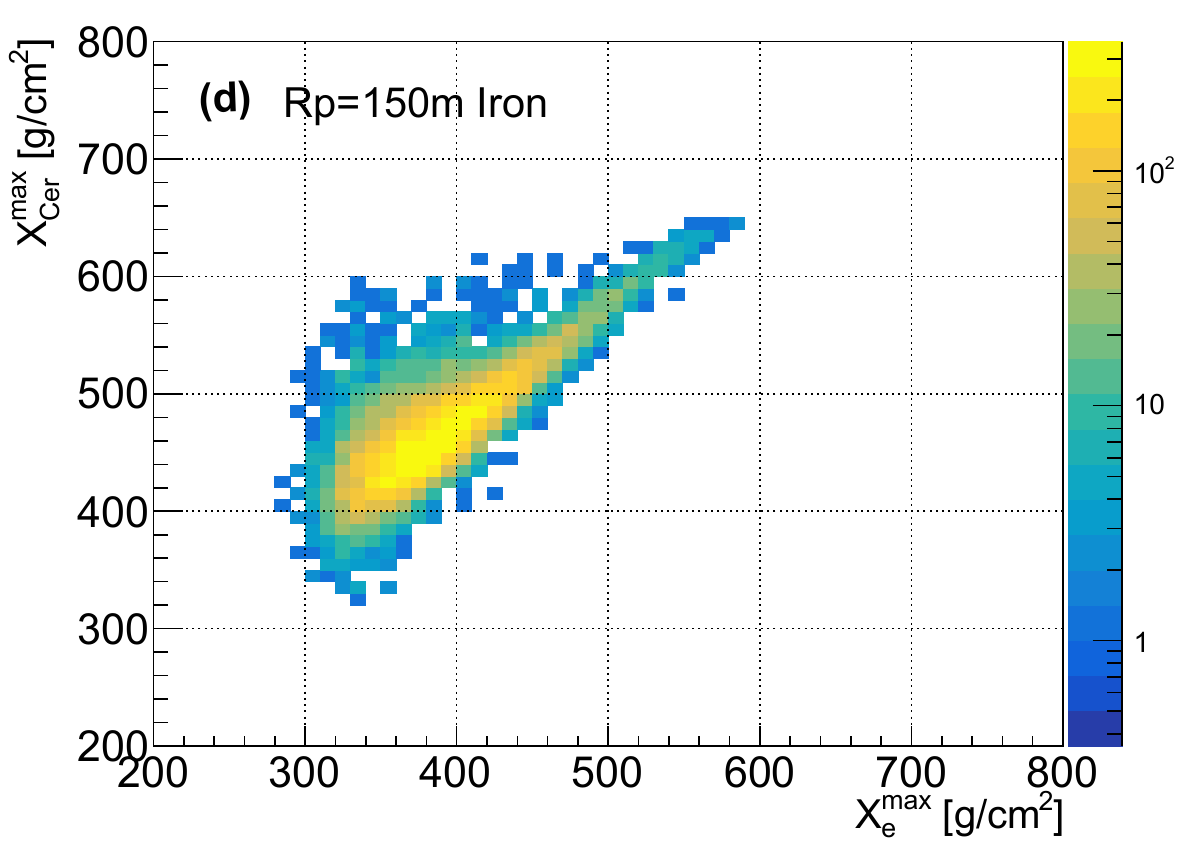}
\caption{The Upper panel (a, b) shows the relationship between $N^{max}_{Cer}$ at $R_{p}$=150m and $N^{max}_{e}$, while the Lower panel (c, d) shows the relationship between $X^{max}_{Cer}$ at $R_{p}$=150m and $X^{max}_{e}$.The shower was induced by proton (a, c) and iron (b, d) and with $R_{p}$=150 m.}
\label{diffN_ch2e}
\end{minipage} 
\end{figure*}
\vspace{0pt}

\begin{figure*}[htbp]
\begin{minipage}[t]{1.\linewidth} 
\includegraphics[width=0.5\linewidth]{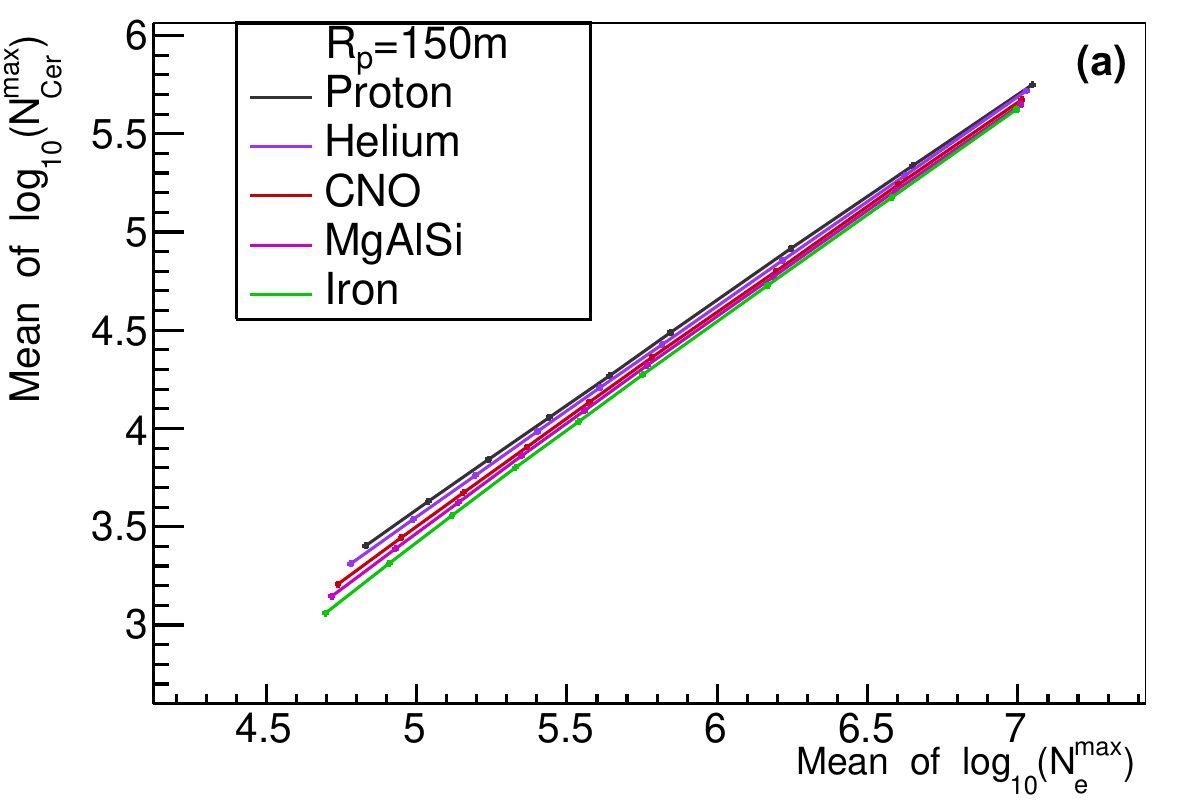}  
\hspace{0cm}
\includegraphics[width=0.5\linewidth]{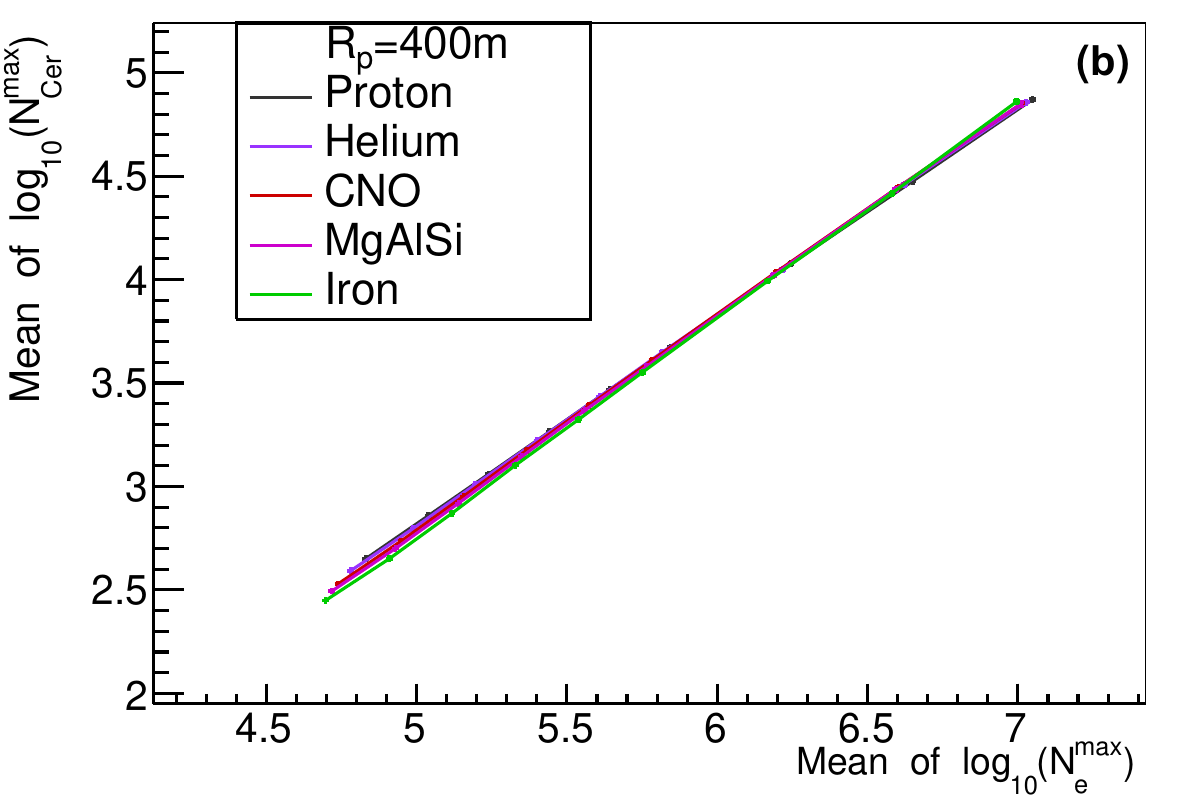}
\hspace{0cm}
\includegraphics[width=0.5\linewidth]{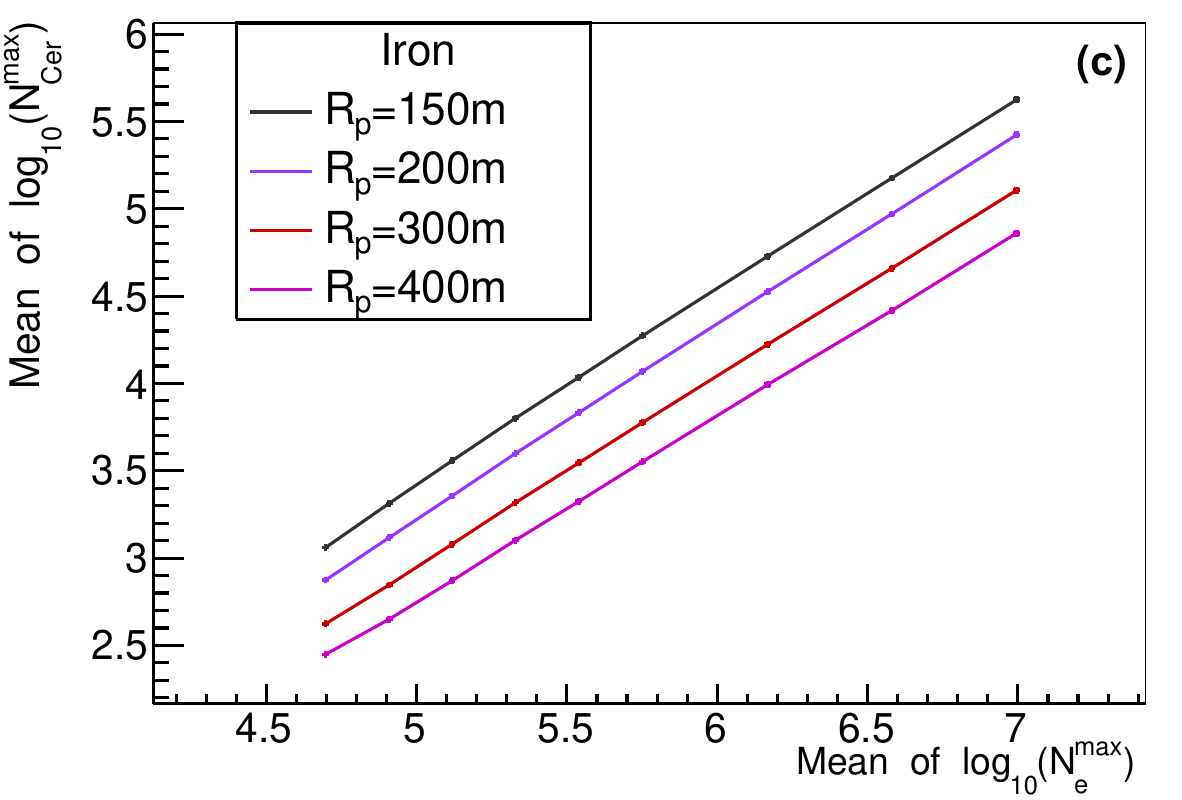}
\hspace{0cm}
\includegraphics[width=0.5\linewidth]{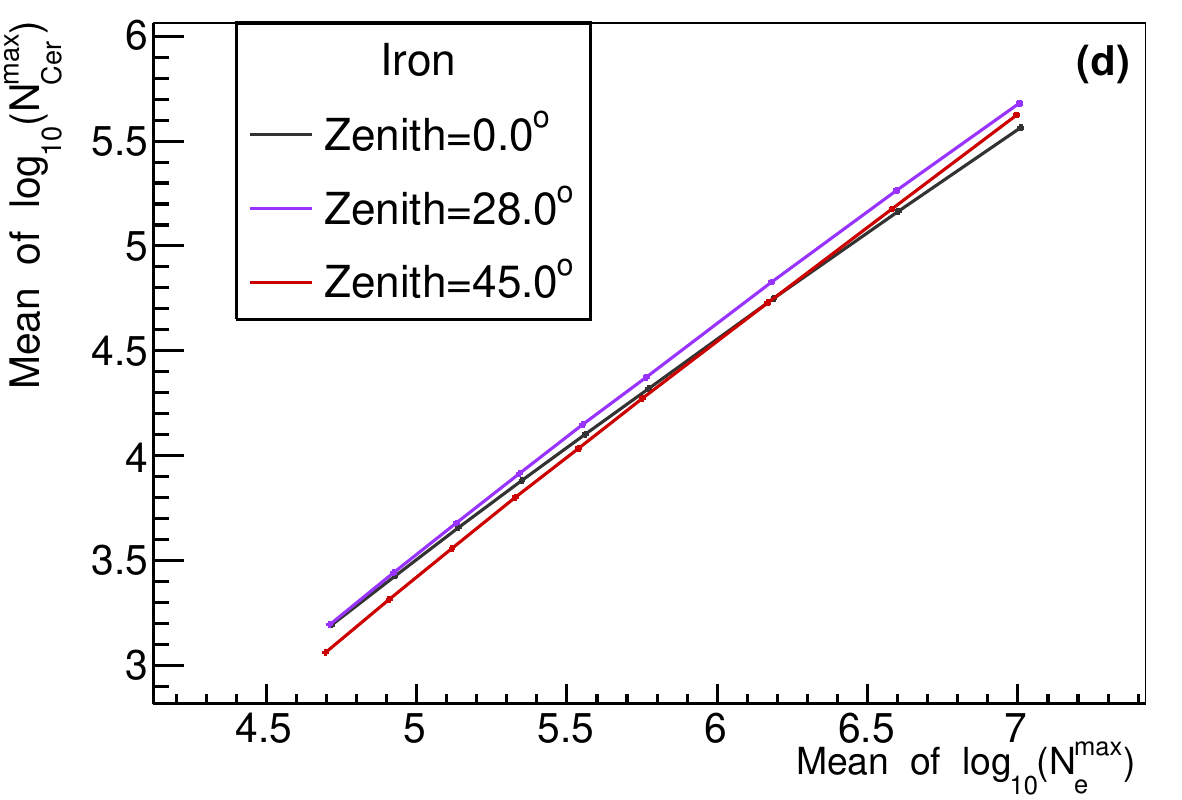}
\caption{The relationship between $N^{max}_{Cer}$ and $N^{max}_{e}$ for different primary particles, different Rp values and differentzenith angles.} The x and y axis are the mean value of a Gaussian function fitted to the $log_{10}(N^{max}_{e})$ distribution and $log_{10}(N^{max}_{Cer})$ distribution by one telescope of the sample. In the upper panel, points with different colors indicate different composition, (a) $R_{p}$=150 and (b) 400 m, Lower left plot (c) shows the relationship for different $R_{p}$ value but the same zenith angle $\theta$=$45^\circ$, lower right plot (d)  shows the relationship for different zenith angles but the same $R_{p}$=150 m.
\label{diffitN_ch2e} 
\end{minipage} 
\end{figure*}

\begin{figure*}[htbp]
\begin{minipage}[t]{1.\linewidth} 
\includegraphics[width=0.5\linewidth]{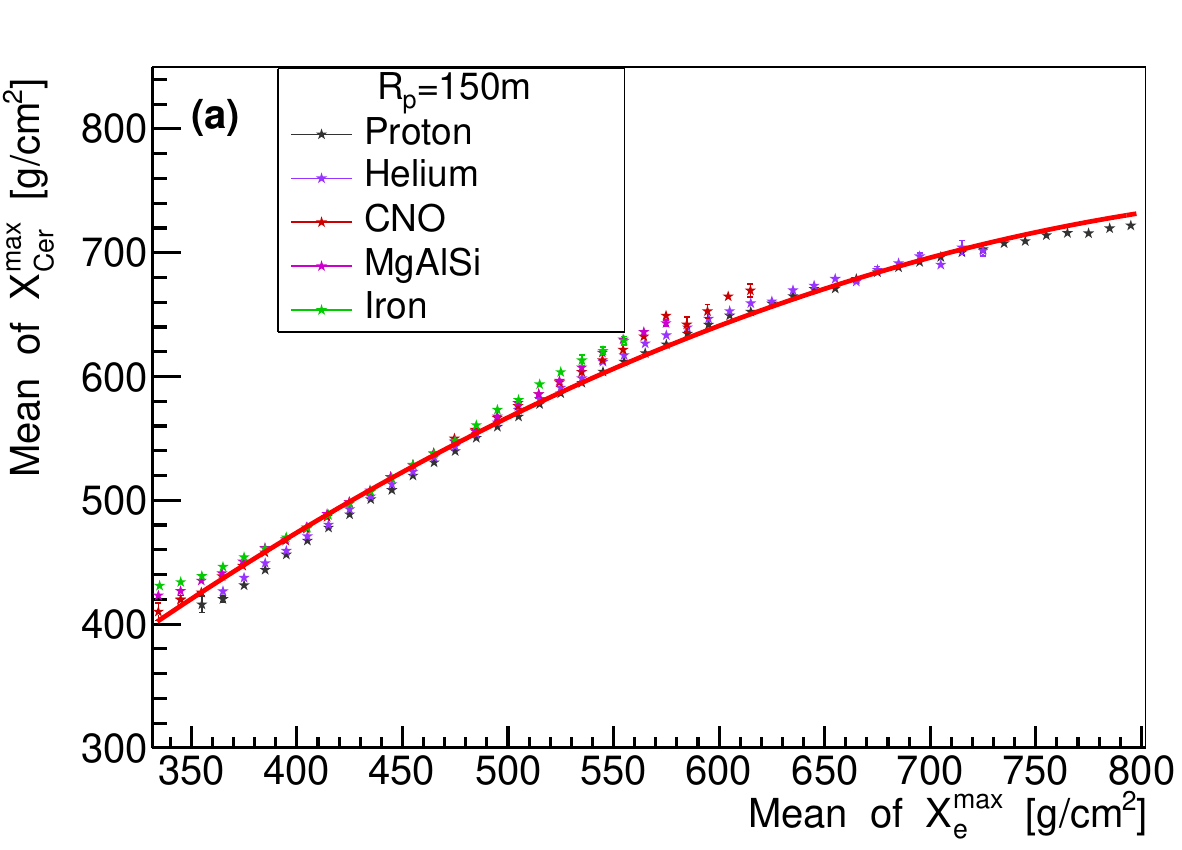}
\hspace{0cm}
\includegraphics[width=0.5\linewidth]{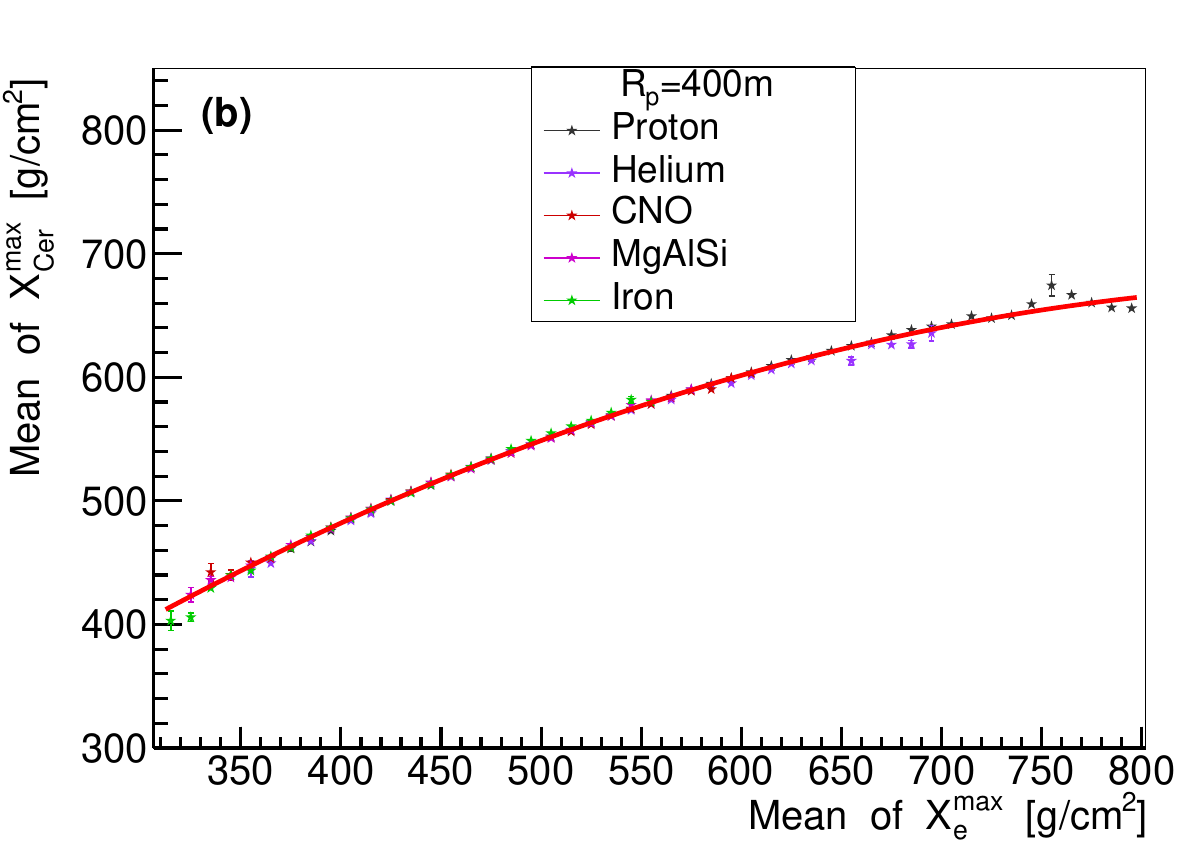}
\hspace{0cm}
\includegraphics[width=0.5\linewidth]{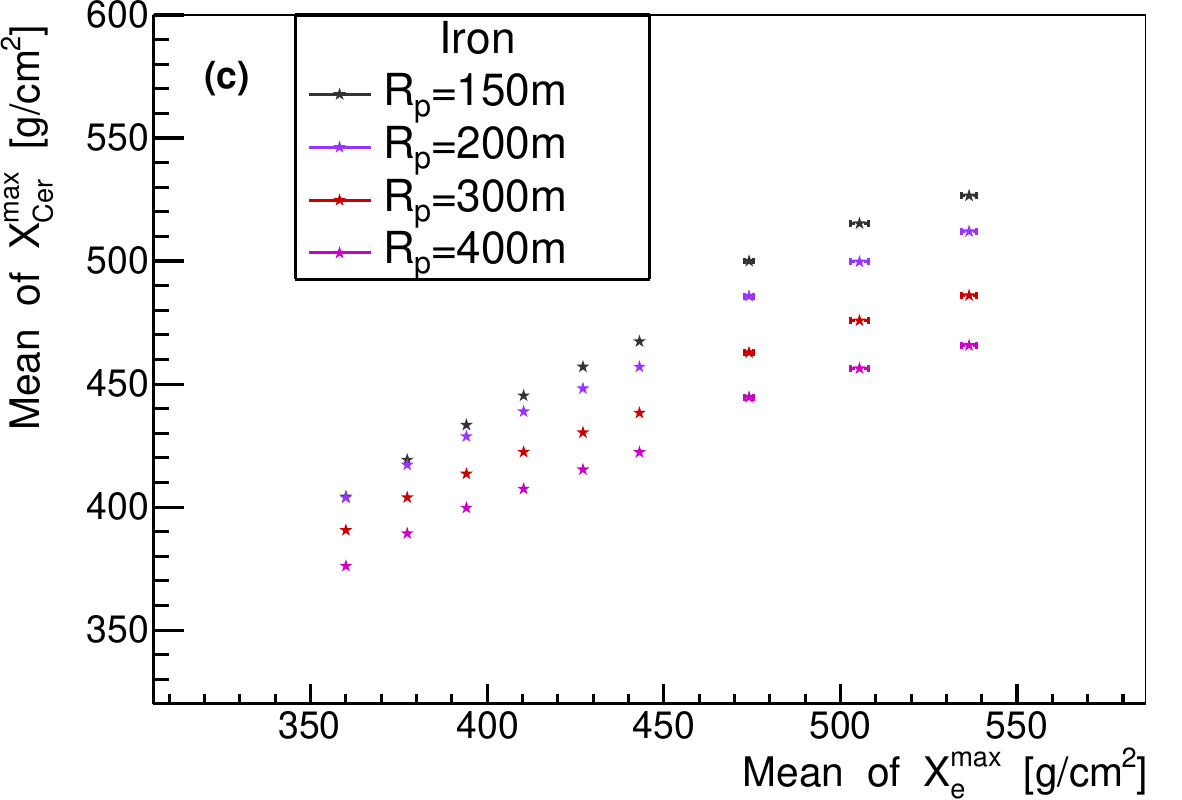}
\hspace{0cm}
\includegraphics[width=0.5\linewidth]{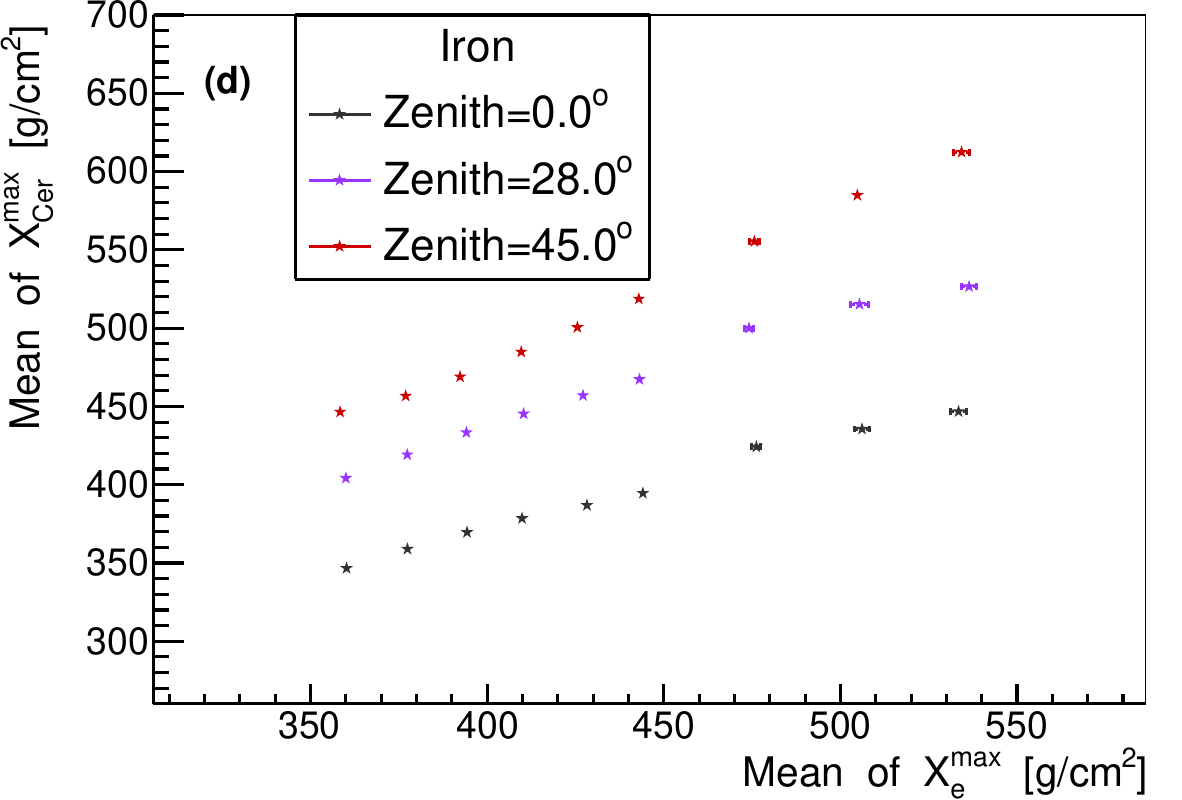}
\caption{The relationship between mean of $X^{max}_{Cer}$ by one telescope and $X^{max}_{e}$ for different primary particles, different $R_{p}$ values and different zenith angles. X and y axis are the mean value of a Gaussian function fitted to the $X^{max}_{e}$ distribution and $X^{max}_{Cer}$ distribution by one telescope of the sample. In the upper panel, points with different colors indicate different composition, $R_{p}$=150 (a) and 400 m (b), while the red line is a parabola function fit to those points. Lower left (c) plot shows the relationship for different $R_{p}$ value but the same zenith angle $\theta$=$28^\circ$, lower right (d) plot shows the relationship for different zenith angles but the same $R_{p}$=150 m.}
\label{diffX_ch2eall}
\end{minipage} 
\end{figure*}

The comparison of $N^{max}_{Cer}$ by one telescope versus $N^{max}_{e}$, and $X^{max}_{Cer}$ by one telescope versus $X^{max}_{e}$ for each event are shown in Fig. \ref{diffN_ch2e}. To study the relationship quantitatively, the mean value of the Gaussian function fitted to $log_{10}(N^{max}_{e})$, $log_{10}(N^{max}_{Cer})$, $X^{max}_{e}$ and $X^{max}_{Cer}$ distribution were evaluated. The relationship of the mean value between $log_{10}(N^{max}_{e})$ distribution and $log_{10}(N^{max}_{Cer})$ distribution for different type of primary particles, different $R_{p}$ values and zenith angles are shown in Fig. \ref{diffitN_ch2e}. As depicted in the figure, the relationship between the mean value of $log_{10}(N^{max}_{e})$ and $log_{10}(N^{max}_{Cer})$ is more dependent on the composition of the primary particles for smaller $R_{p}$ values, and is less dependent on composition for larger $R_{p}$ values. It can then be said that the relationship is heavily dependent on the $R_{p}$ value and slightly dependent on the zenith angle. Therefore, $N^{max}_{Cer}$ by one telescope needs to be corrected for $R_{p}$ value and zenith angle before measuring the energy. As an energy estimator, it is slightly dependent on the composition of the primary particles.\indent
The relationship between the mean value of $X^{max}_{Cer}$ distribution by one telescope and $X^{max}_{e}$ distribution for different type of primary particles, different $R_{p}$ values and different zenith angles are shown in Fig. \ref{diffX_ch2eall}. As can be observed, the relationship is almost composition independent, especially for larger $R_{p}$ values. The relationship can be well fitted with a parabola function (red lines in the upper panel of Fig. \ref{diffX_ch2eall}), which makes $X^{max}_{Cer}$ a good composition independent $X^{max}_{e}$ estimator. However, the relationship is heavily dependent on $R_{p}$ values and zenith angles. The $R_{p}$ dependence may be due to the angular distance which is different for different $R_{p}$ values (see Fig. \ref{shower}), and the Cherenkov light produced anisotropically. The zenith angle dependence may be due to the physical position of the shower maximum, which is different for different zenith angles even though the shower maximum of electron is the same for different zenith angles. Although $X^{max}_{Cer}$ is kind of a complicated composition independent $X^{max}_{e}$ estimator.  $X^{max}_{Cer}$ by one telescope needs to be corrected for $R_{p}$ and zenith angle before measuring the $X^{max}_{e}$. It is also noticed that around 100 TeV, the difference of $X^{max}_{Cer}$ between $R_{p}$=150 m and $R_{p}$ =200m is very small, (lower left plot of  Fig. \ref{diffX_ch2eall}), which is still in agreement with the results by~\cite{measure_Xamx_che}. \\\indent

\vspace{0pt}

\vspace{0pt}

\begin{figure*}[htbp]
\begin{minipage}[t]{1.\linewidth} 
\includegraphics[width=0.34\linewidth]{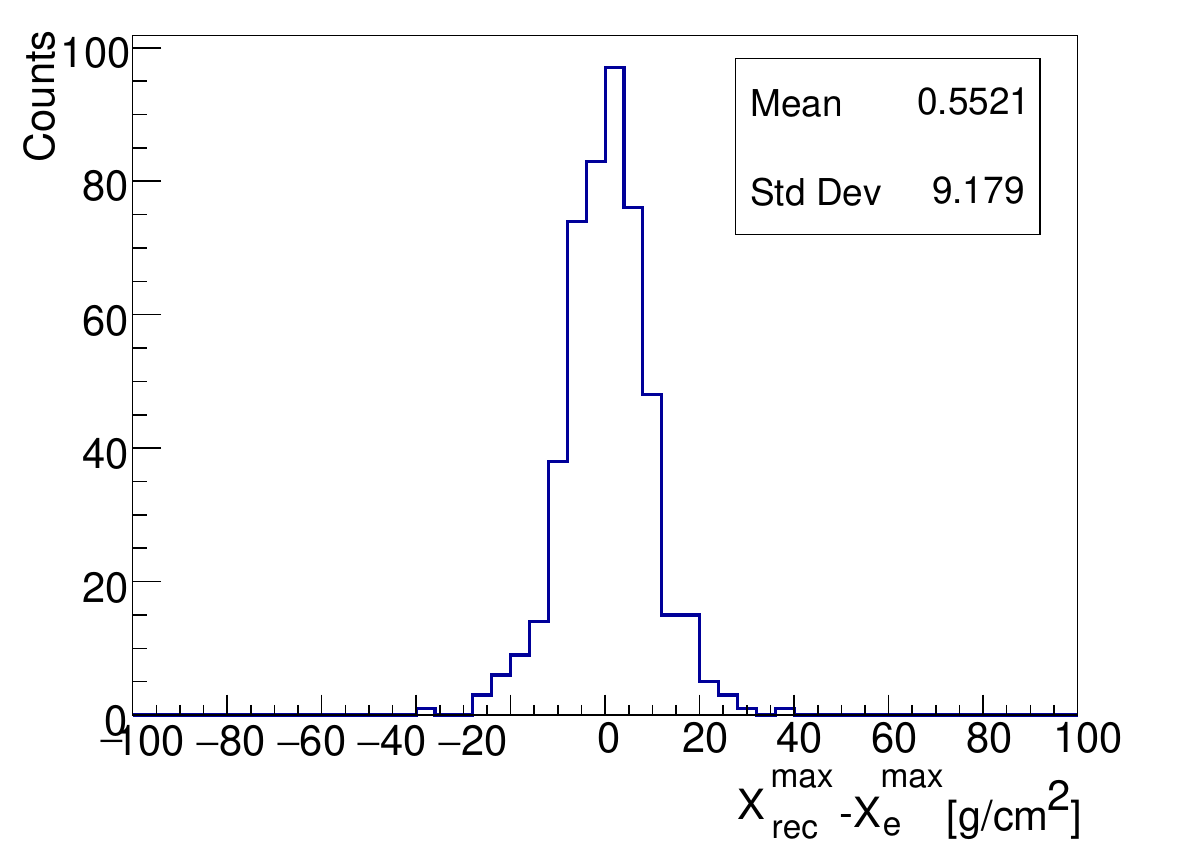}
\hspace{0cm}
\includegraphics[width=0.34\linewidth]{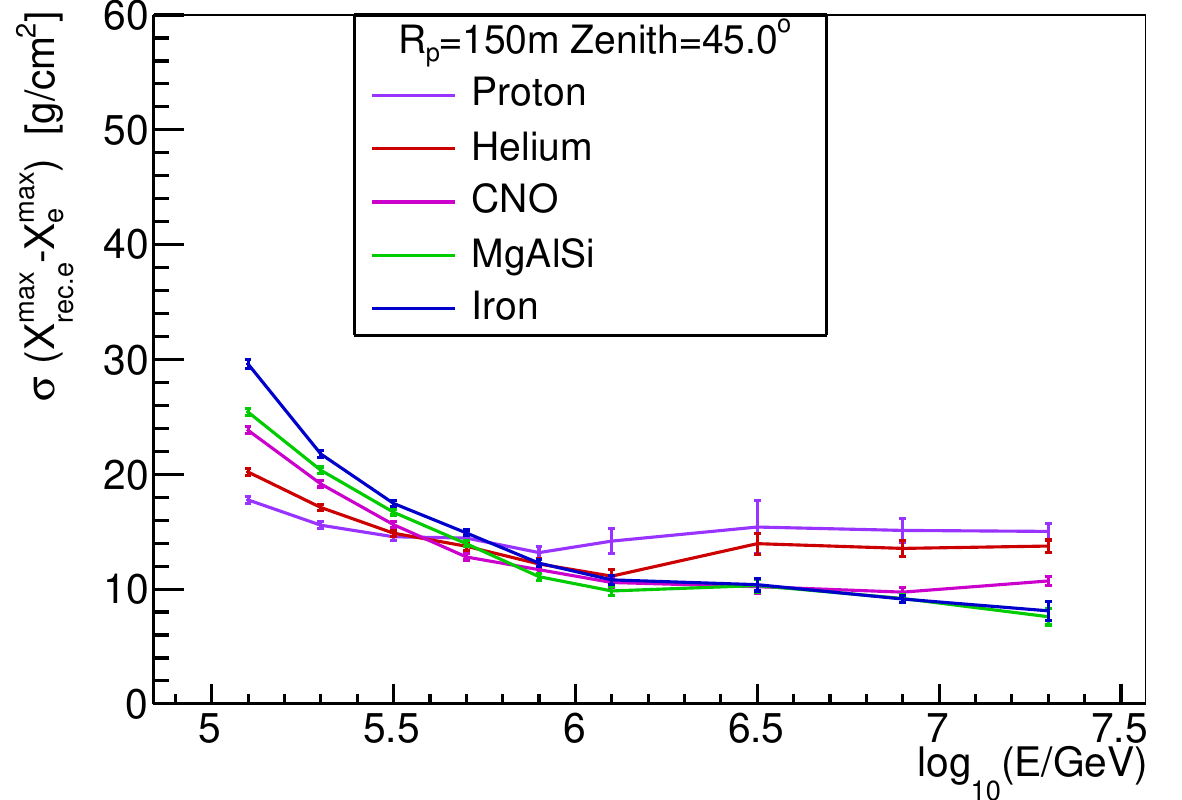}
\hspace{0cm}
\includegraphics[width=0.34\linewidth]{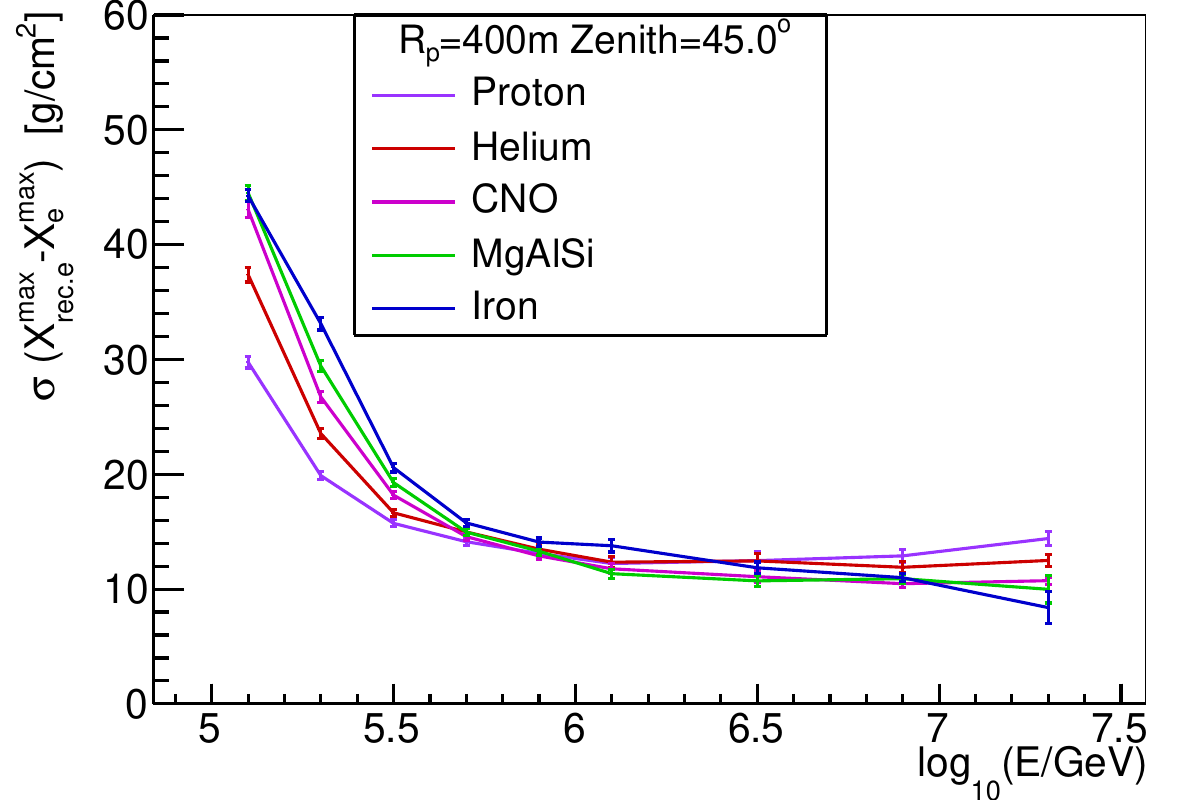}
\caption{Left panel shows the distribution of $X^{max}_{rec. e}$-$X^{max}_{e}$ for the iron inducted shower at energy $log_{10}(E/GeV)$=6.9 with $R_{p}$=150 m, where $X^{max}_{rec. e}$ is the reconstructed $X^{max}_{e}$ by converting $X^{max}_{Cer}$ according to the red line in Fig. \ref{diffX_ch2eall} for each $R_{p}$ and zenith angle bin. Middle and right diagrams show the uncertainty of the reconstruction from $X^{max}_{Cer}$ to $X^{max}_{e}$ for difference particles induced shower with $R_{p}$=150 and 400 m, respectively. Lines with different colors indicate different primary particles. The uncertainty is defined as the sigma value of Gaussian function fitted to $X^{max}_{rec. e}-X^{max}_{e}$ distribution.}
\label{deltaX}
\end{minipage} 
\end{figure*}

Regarding the quality of $X^{max}_{Cer}$ as $X^{max}_{e}$ estimator, the sigma value of the Gaussian function fitted to $X^{max}_{rec. e}$ - $X^{max}_{e}$ distribution was used to estimate the uncertainty of reconstructing $X^{max}_{e}$ with $X^{max}_{Cer}$, where $X^{max}_{rec. e}$ is the reconstructed $X^{max}_{e}$ by converting $X^{max}_{Cer}$ according to the red line in Fig. \ref{diffX_ch2eall} for each $R_{p}$ and zenith angle bin. The typical distribution of $X^{max}_{rec. e}$ - $X^{max}_{e}$, the typical uncertainty for $R_{p}=150$ and $R_{p}=400$ m are shown in Fig. \ref{deltaX}. Which shows an uncertainty of 10-15 $g/cm^{2}$ for the $X^{max}_{e}$ reconstruction for all compositions and $R_{p}$ values when energy is larger than $\sim$ 1PeV. From 100 TeV to 1 PeV, the uncertainty is 10-30 $g/cm^{2}$ for $R_{p}=150$ m, and 10-45 $g/cm^{2}$ for $R_{p}=400$ m.\\\indent

\vspace{0pt}

\subsection{Energy estimator}
\label{relationship_ENmax}
The energy of a primary particles is proportional to the number of secondary particles. The total number or density at a fixed distance from the shower axis of the secondary particles on a fixed observation plane which were widely used to reconstruct the energy of primary particles~\cite{crab_csz,paperof_WFCTA_YZY,Bartoli2015KneeOT}. However, they are composition dependent, which will introduce systematic errors for some measurements, e.g., the all-particle flux of cosmic ray. In principle, secondary particles measured at shower maximum  have better performances. For instance, the number of secondary particles at shower maximum has smaller fluctuations than the measurement at fixed altitude, because the former is less affected by the shower-to-shower fluctuations. Therefore, the number of secondary particles at shower maximum is assumed to be a better energy estimator than the variable measured on a fixed observation plane. \indent
\begin{figure*}[htbp]
\begin{minipage}[t]{1.\linewidth} 
\includegraphics[width=0.34\linewidth]{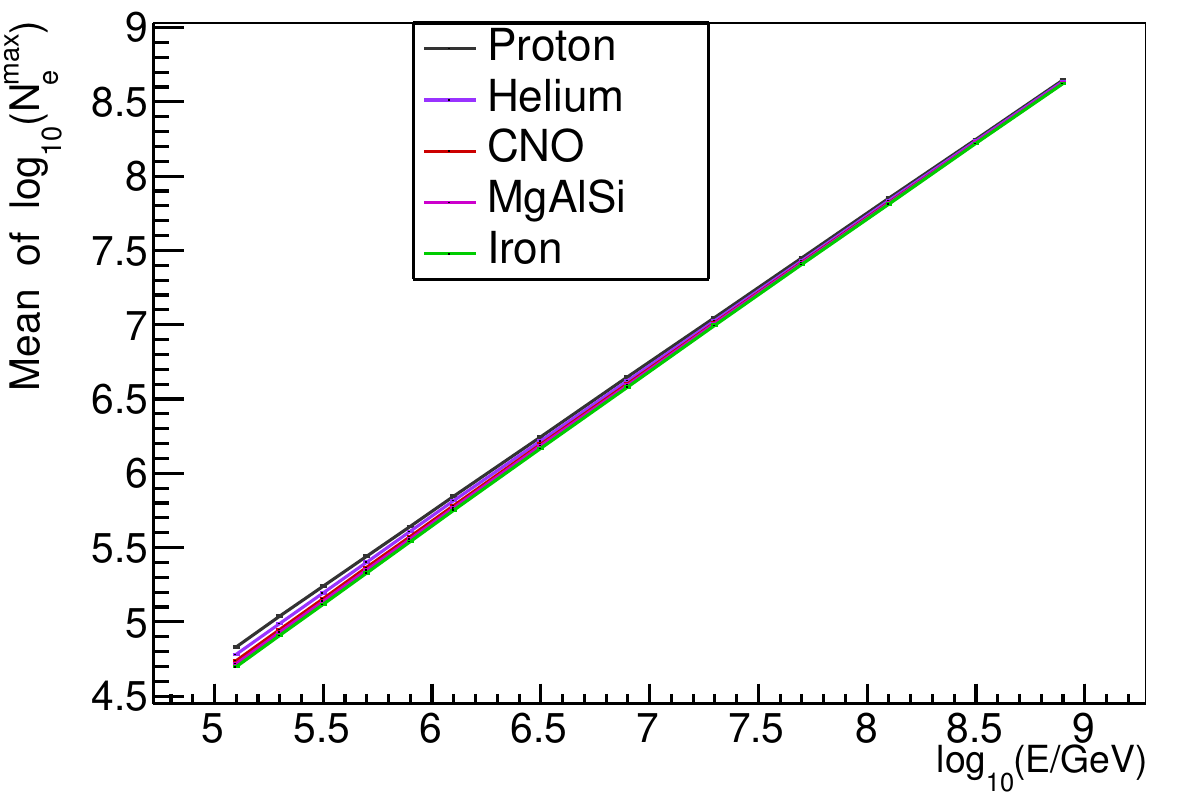}
\hspace{0cm}
\includegraphics[width=0.34\linewidth]{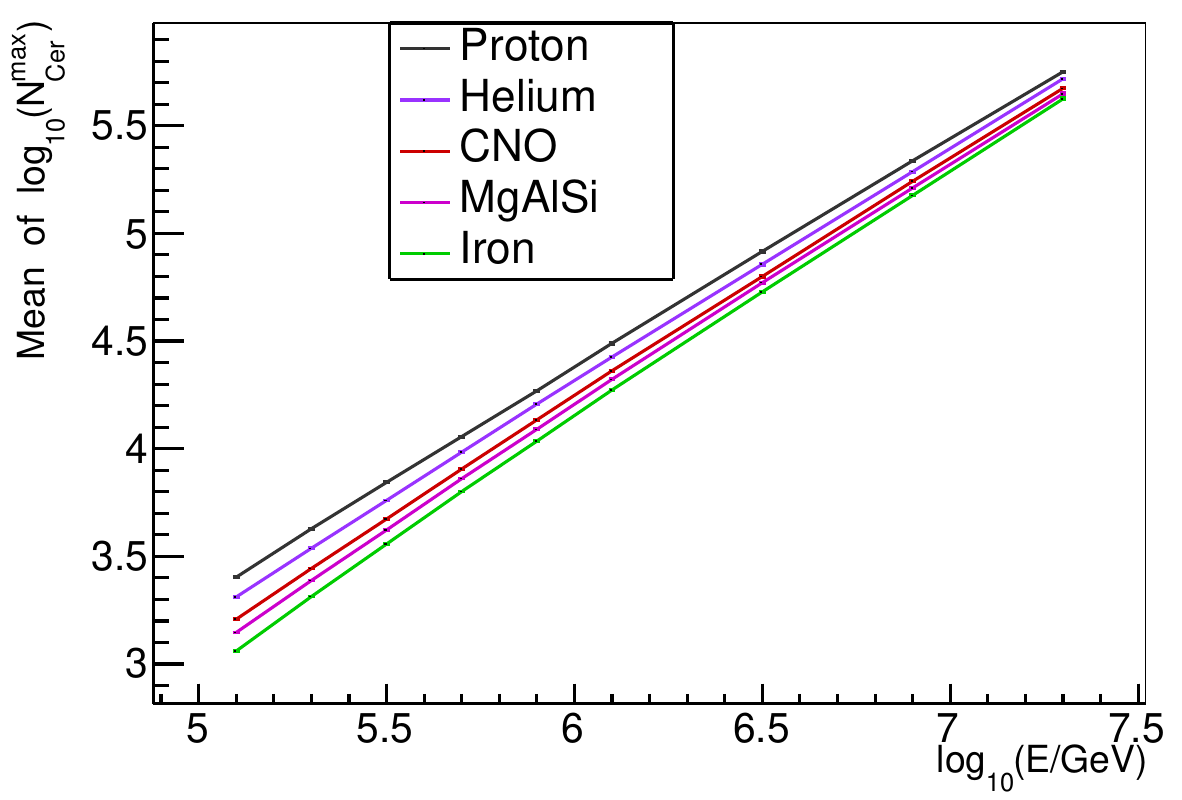}
\hspace{0cm}
\includegraphics[width=0.34\linewidth]{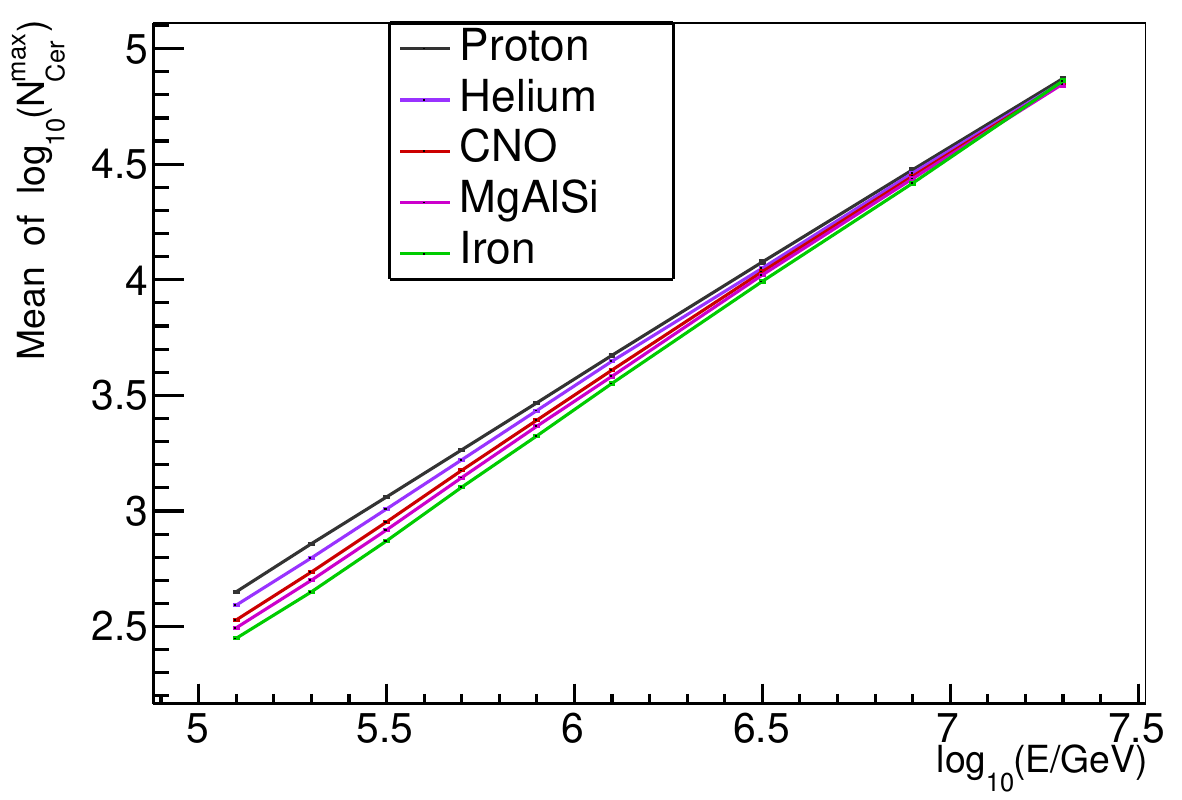}
\caption{The relationship between $N_{max}$ and primary energy for different primary particles. Y-axis is the mean value of a Gaussian function fitted to the $log_{10}(N^{max}_{e})$ distribution (left panel), and $log_{10}(N^{max}_{Cer})$ distribution of the sample (middle and right panel). Lines with different colors indicate different composition. Left panel is the relationship between $log_{10}(N^{max}_e)$ and $log{10}(E/GeV)$, while the middle and right panels are the relationship between $log_{10}(N^{max}_{Cer})$ and $log_{10}(E/GeV)$ with $R_{p}$=150 and 400 m, respectively.}
\label{lnNmax_lgE}
\end{minipage} 
\end{figure*}
\vspace{0pt}

In Fig.\ref{lnNmax_lgE} we show the relationship between the mean value of $log_{10}(N^{max}_{e})$, $log_{10}(N^{max}_{Cer})$ and the energy of primary particles. The x-axis is the true energy of the sample, the y-axis is the mean value of the Gaussian function fitted to the $log_{10}(N^{max}_{e})$ distribution or $log_{10}(N^{max}_{Cer})$ distribution of the samples. Lines with different colors indicate different composition. As illustrated in Fig. \ref{lnNmax_lgE}, the energy reconstructed from $N^{max}_{e}$ is dependent on composition very slightly, especially in higher energy range. However, the energy reconstructed from $N^{max}_{Cer}$ is much more composition dependent. This is because $N^{max}_{Cer}$ is an estimator of $N^{max}_{e}$, but the relationship between $N^{max}_{Cer}$ and $N^{max}_{e}$ is composition dependent, especially for smaller $R_{p}$ values (see Fig. \ref{diffitN_ch2e}). \indent
\begin{figure*}[htbp]
\begin{minipage}[t]{1.\linewidth} 
\centerline{\includegraphics[width=0.6\linewidth]{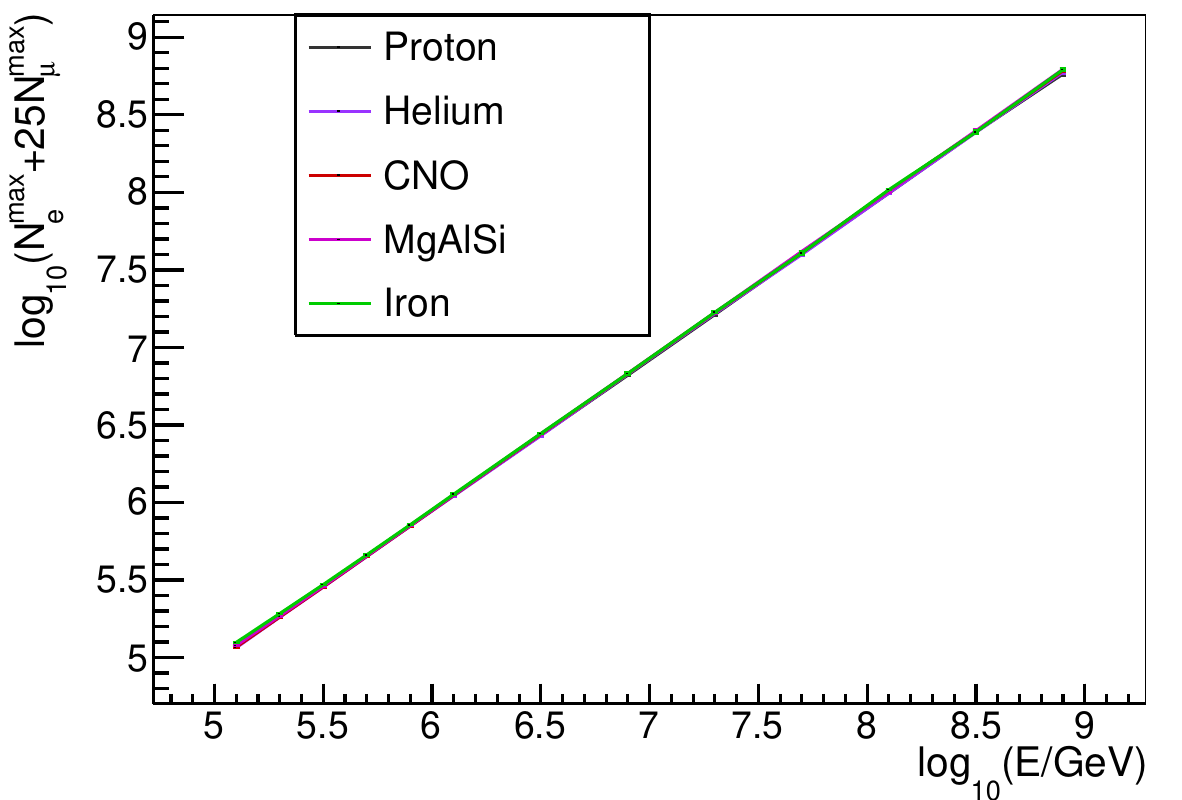}}
\caption{ The relationship between $N^{max}_e+25N^{max}_{\mu}$, and the energy for different primary particles. Y-axis is the mean value of a Gaussian function fitted to the $log_{10}(N^{max}_e+25N^{max}_{\mu})$ distribution.}
\label{log10Ne_25Nu}
\end{minipage} 
\end{figure*}
\vspace{0pt}

To reduce the composition dependencies, $N^{max}_{e}$+25$N^{max}_{\mu}$ is a very good energy estimator \cite{Ne_25Nu}, which is composition independent, because it sums up the energy both from electromagnetic component and hadronic component of EAS. Therefore a new variable based on $N^{max}_{e}$ and $N^{max}_{\mu}$ is constructed as an energy estimator, which is $N^{max}_{e\mu}$=$N^{max}_{e}$+$C\times$ $N^{max}_{\mu}$, where $C$ is a coefficient. The relationship between the mean value of the Gaussian function fitted to $log_{10}(N^{max}_{e\mu})$ distribution and energy of primary particles for $C=25$ is shown in Fig.\ref{log10Ne_25Nu}. Lines with different colors indicate different composition. As it can be seen, the variable $N^{max}_{e\mu}$ is composition independent. The ratio of $N^{max}_{e\mu}$ between different composition is  within $\pm$5\% in the energy region from $log_{10}(E/GeV)=5.1$ to $log_{10}(E/GeV)=8.9$. Therefore $N^{max}_{e}+25N^{max}_{\mu}$ is a very good composition independent energy estimator. Moreover, when the coefficient $C$ changed from 20 to 27, the ratio of $N^{max}_{e\mu}$ between different composition is also similar. \\\indent

\subsection{Composition estimator}
\label{relationship_AXmax}
\begin{figure*}[htbp]
\begin{minipage}[t]{1.\linewidth} 
\includegraphics[width=0.34\linewidth]{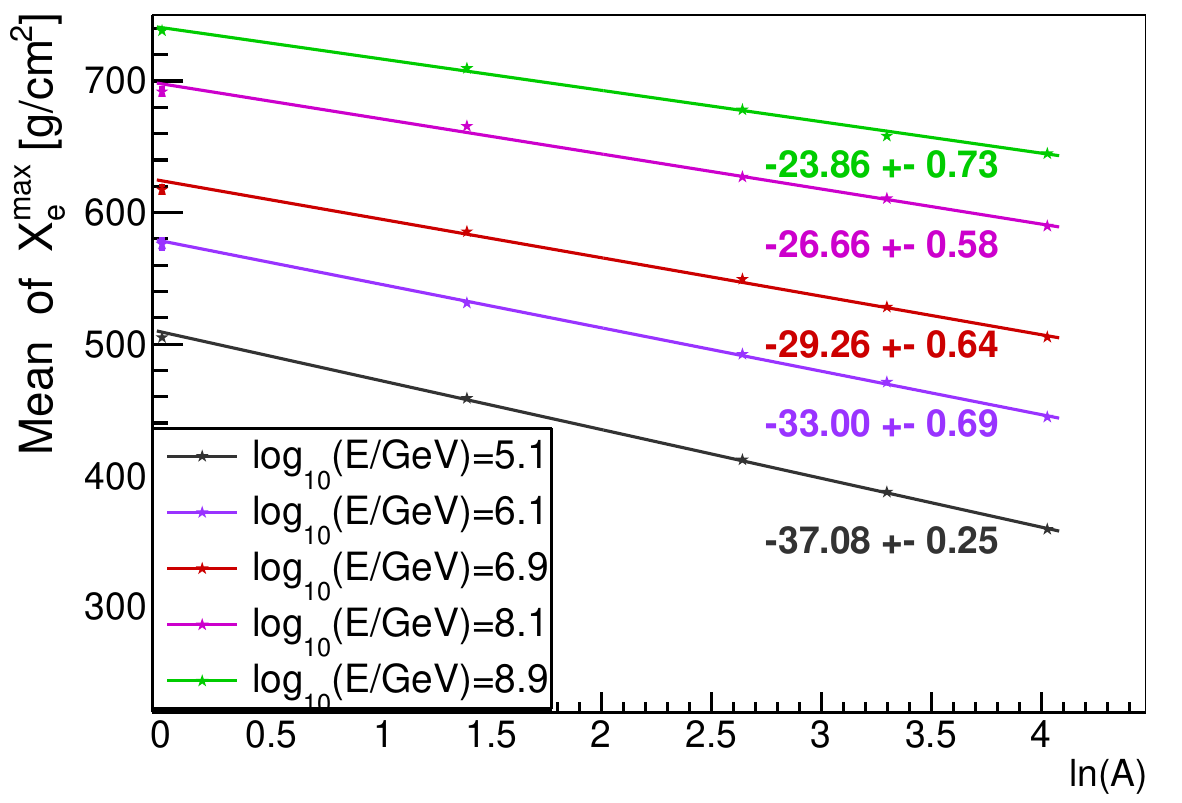}
\hspace{0cm}
\includegraphics[width=0.34\linewidth]{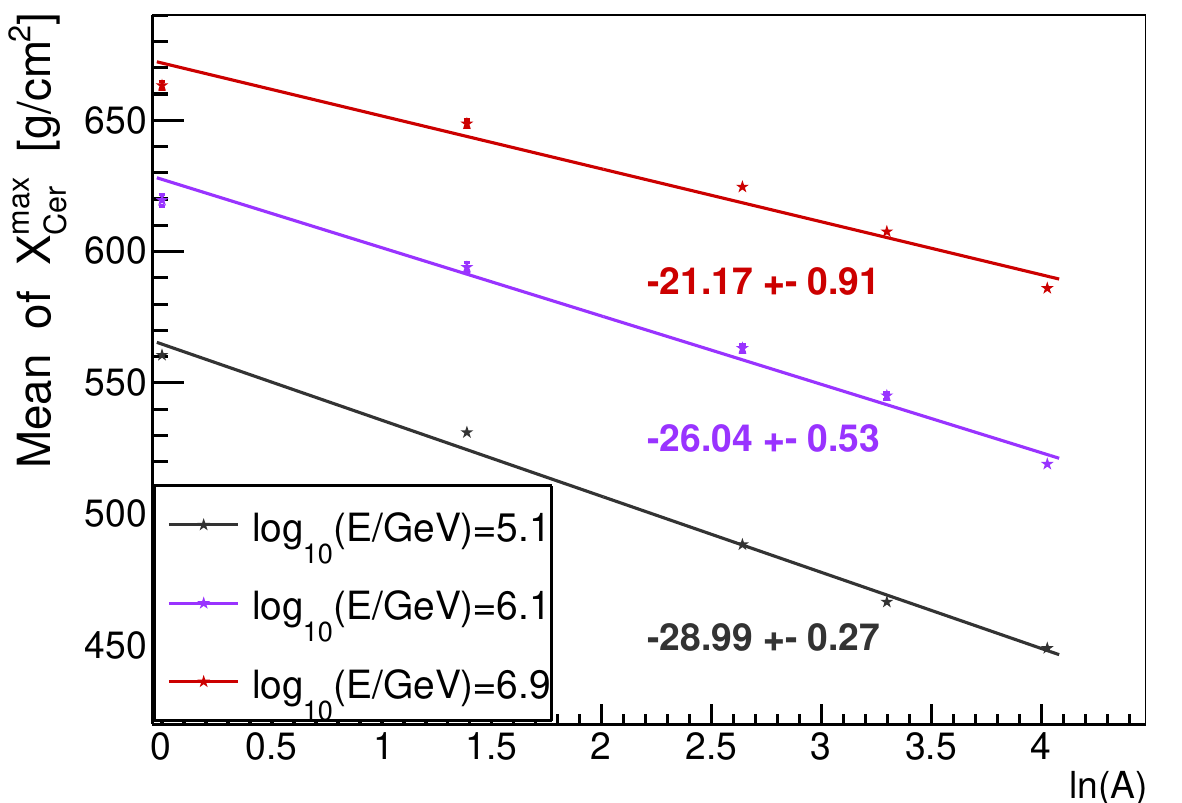}
\hspace{0cm}
\includegraphics[width=0.34\linewidth]{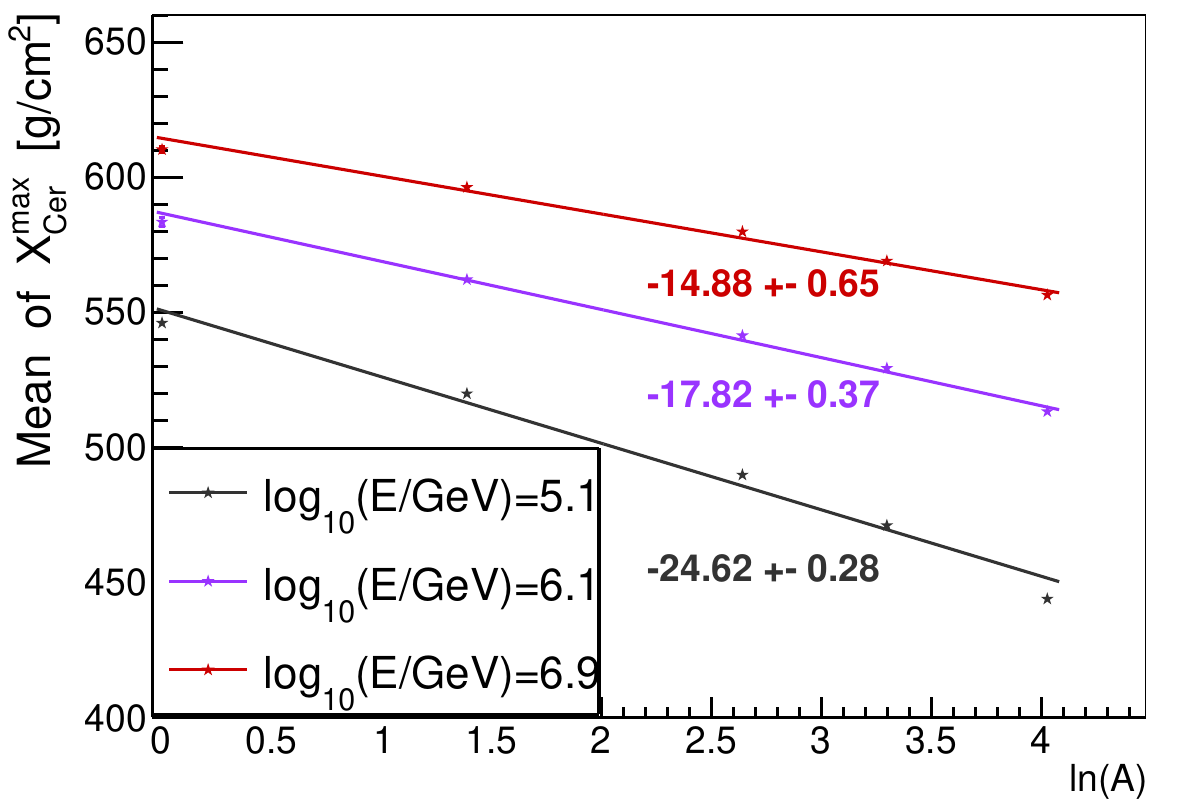}
\caption{Relationship between shower maximum and logarithm mass of primary particles for different energy. Y-axis is the mean value of the $X^{max}_{e}$ and $X^{max}_{Cer}$ distribution. Left panel is $X^{max}_e$ versus lnA, while middle and right panels are $X^{max}_{Cer}$ versus lnA with $R_{p}$=150 and 400 m, respectively. The points with different colors indicate different energy of the primary primary particles. The colored lines are a linear fit to the corresponding points. The values next to the line of each plot indicate the slope of fitted line.}
\label{Xmax_A_diffE}
\end{minipage} 
\end{figure*}

According to ~\cite{Ne_25Nu}, $X^{max}_{e}$ is a good estimator of logarithm mass (denoted as lnA hereafter), or a good composition estimator, due to the fact $X_{max}^{A}$=$X_{max}^{P}$-$\lambda_{r}$lnA, where $X_{max}^{A}$ is the shower maximum for composition with mass number A, $X_{max}^{P}$ is the shower maximum for proton, and $\lambda_{r}$ is the radiation length ($\sim37$ $g/cm^2$). This is widely used in composition identification for different experiments~\cite{measure_Xamx_che,RIGGI20139,PRD_Xmax}. The relationship of the mean value of $X^{max}_{e}$ versus lnA and the relationship of the mean value of $X^{max}_{Cer}$ versus lnA are shown in Fig. \ref{Xmax_A_diffE}. It should be noted that different colors correspond to different energy of primary particles, the lines are the linear fit to the points with the same energy, and the corresponding slopes are indicated next to the lines. As it can be seen in Fig. \ref{Xmax_A_diffE}, $X^{max}_{e}$ and lnA follows a linear relationship, however, the slope is different for different energy. For instance, at $\sim$100 TeV, the slope is $\sim-37$ $g/cm^2$, which is consistent with that of ~\cite{Ne_25Nu}, and at $\sim$1 EeV, the slope is $\sim-24$ $g/cm^2$. This is also true for Cherenkov lights, but the fitted lines are more flat. At higher energy, it shows that $X^{max}_{Cer}$ is less dependent on the lnA (or composition). \indent

\vspace{0pt}

\begin{figure*}[htbp]
\begin{minipage}[t]{1.\linewidth} 
\centerline{\includegraphics[width=0.6\linewidth]{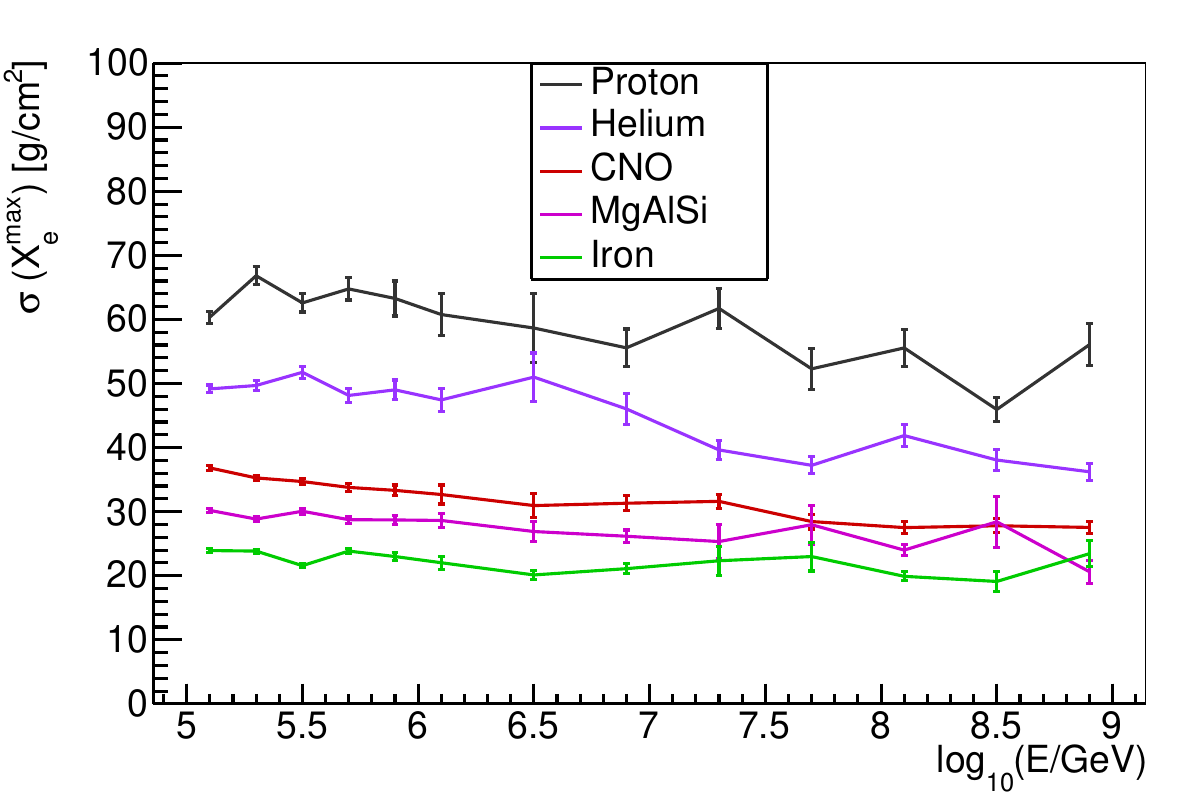}}
\caption{The sigma value of Gaussian function fitted to the $X^{max}_{e}$ distribution versus primary energy. Colored lines indicate different primary particles.}
\label{dev_Xmaxe}
\end{minipage} 
\end{figure*}
\vspace{0pt}

Since shower maximum is a good composition estimator, the shower-to-shower fluctuations of shower maximum is crucial for the composition identification ability. Fig.\ref{dev_Xmaxe} shows the sigma value of the Gaussian function fitted to the $X^{max}_{e}$ distribution. It is clear that, the fluctuations of $X^{max}_{e}$ is smaller for heavier composition. The sigma of $X^{max}_{e}$ distribution is around 50-55$g/cm^2$ for proton, 37-50 $g/cm^2$ for helium, 27-35 $g/cm^2$ for CNO, 25-30 $g/cm^2$ for MgAlSi, and 20-25$g/cm^2$ for iron from $\sim$100 TeV to $\sim$1 EeV. The $X^{max}_{Cer}$ is an estimator of $X^{max}_{e}$. The uncertainty due to the reconstruction from $X^{max}_{Cer}$ to $X^{max}_{e}$ is about 10-15 $g/cm^2$ at high energy. This is demonstrated in Fig. \ref{deltaX}. \indent

\section{Longitudinal development versus lateral distribution}
\label{compa_study}
For ground cosmic ray experiment, there are two kinds of measurements. One is measuring the secondary components through their lateral distribution at a fixed altitude, the properties of which have been studied in detail by ~\cite{EASliu}. Another is measuring the longitudinal developments of secondary components along the trajectory (the properties of longitudinal development have been studied in section \ref{long_dev}.) Energy measurement and composition discrimination ability are two of the most important aspects for cosmic ray measurements. The comparison of those two measurements between variables from longitudinal development and variables from lateral distribution will be discussed below. \indent

\begin{figure*}[htbp]
\begin{minipage}[t]{1.0\linewidth}
\includegraphics[width=0.5\linewidth]{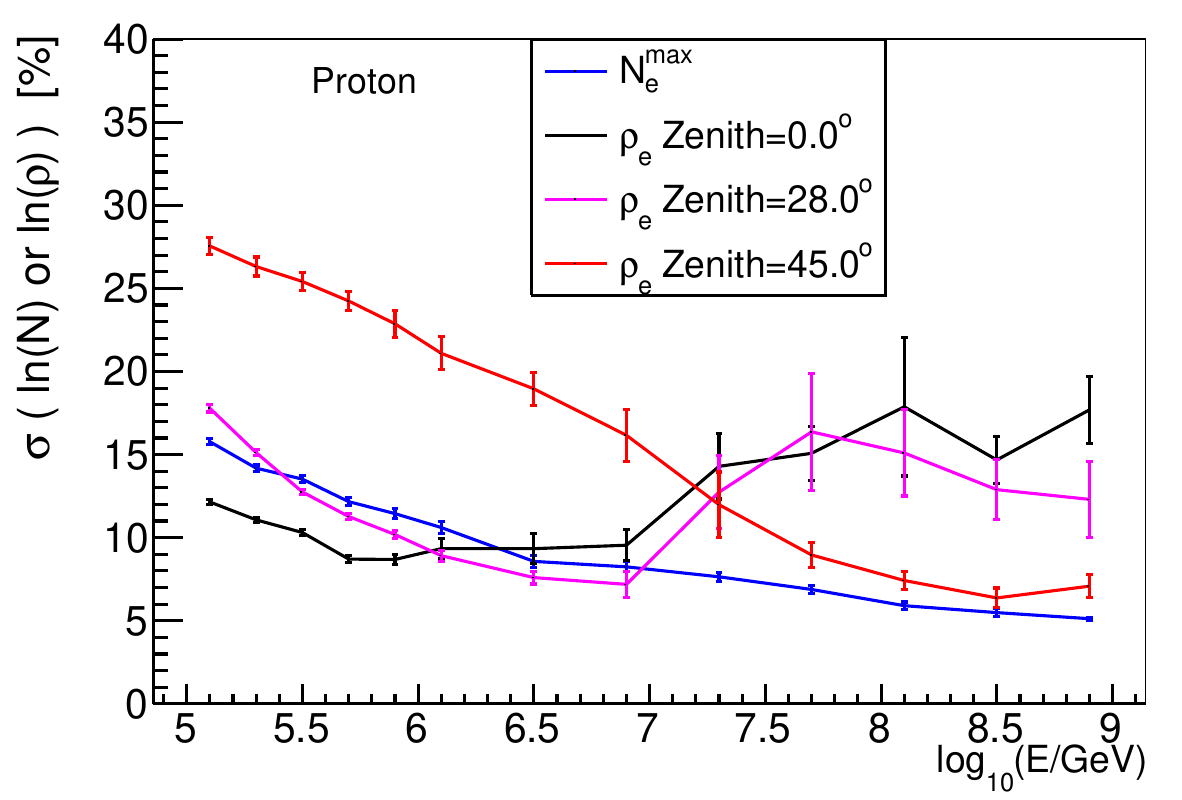}
\hspace{0cm}
\includegraphics[width=0.5\linewidth]{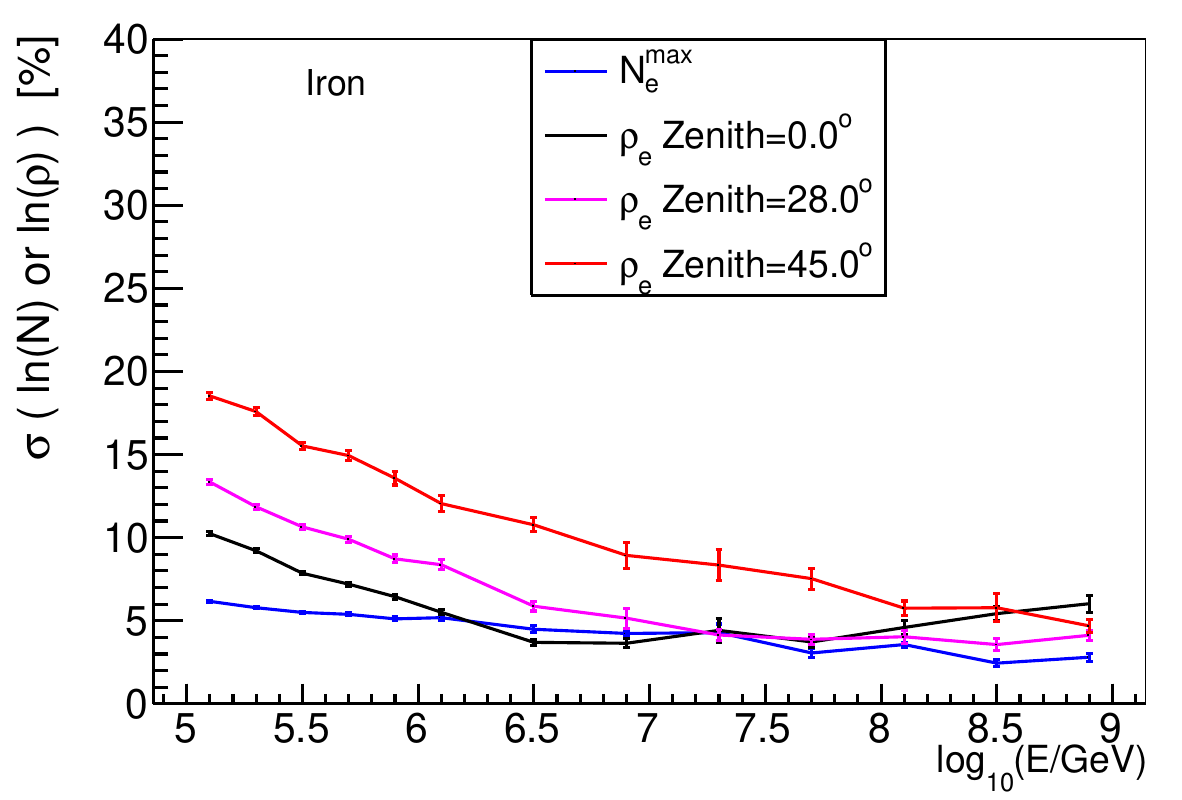}
\caption{The comparison of sigma of the variables ln($N^{max}_{e}$) and ln($\rho_e$) at different zenith angles. The solid blue line indicates $N^{max}_{e}$, other colored lines indicate $\rho_e$ with different zenith angles. The primary particles are proton (left) and iron (right), respectively.}
\label{re_E_diffzen}
\end{minipage}
\end{figure*}
Electron and Cherenkov light from both longitudinal development and lateral distribution are widely used for energy reconstruction. For electron, it was pointed out that the density of electron at 200 m away from the shower axis is less affected by the shower-to-shower fluctuations (see Fig. 11 in reference~\cite{EASliu} for details), and it is a better energy estimator compared to total electron number. The comparison between the sigma value of the Gaussian function fitted to $ln(\rho_{e})$ (electron density) distribution for different zenith angles ($\theta=0^\circ$, $28^\circ$ and $45^\circ$) and the sigma value of the Gaussian function fitted to $ln(N^{max}_{e})$ distribution is shown in Fig. \ref{re_E_diffzen}. Since the energy and zenith angle of primary particles are both fixed, the sigma value represents the shower-to-shower fluctuations, which is related to the reconstructed energy resolution.\indent

The left and right panels of Fig. \ref{re_E_diffzen} correspond to primary proton and iron respectively. As illustrated in this figure, $N^{max}_{e}$ has a smaller fluctuation, and it is quite flat with energy. The shower-to-shower fluctuations of $\rho_e$ has a clear dependence on energy, which is due to the fluctuations of secondary particles having reached minimum at shower maximum. The measurement before and after shower maximum will have larger fluctuations. However, by choosing appropriate zenith angle for the desired energy, the shower-to-shower fluctuations of $\rho_e$ is very close to that of $N^{max}_{e}$. For example, for proton-induced shower below 10 PeV, the shower maximum is smaller than the slant atmospheric depth for all zenith angles, and the $\theta=0^\circ$ is closer to the shower maximum, having a smaller fluctuation of $\rho_e$ than $\theta=45^\circ$. For the proton-induced shower above 50 PeV, the shower maximum is larger than the slant atmospheric depth for $\theta=28^\circ$, and is smaller than that for $\theta=45^\circ$, so $\theta=45^\circ$ has a smaller fluctuation of $\rho_e$ than other zenith angles. For iron below 100 PeV, the situation is different, the shower maximum is always smaller than atmospheric depth for all zenith angles, so that $\theta=0^\circ$ has a smaller fluctuation of $\rho_e$ than other zenith angles.\\\indent

\begin{figure*}[htbp]
\begin{minipage}[t]{1.0\linewidth}
\includegraphics[width=0.45\linewidth]{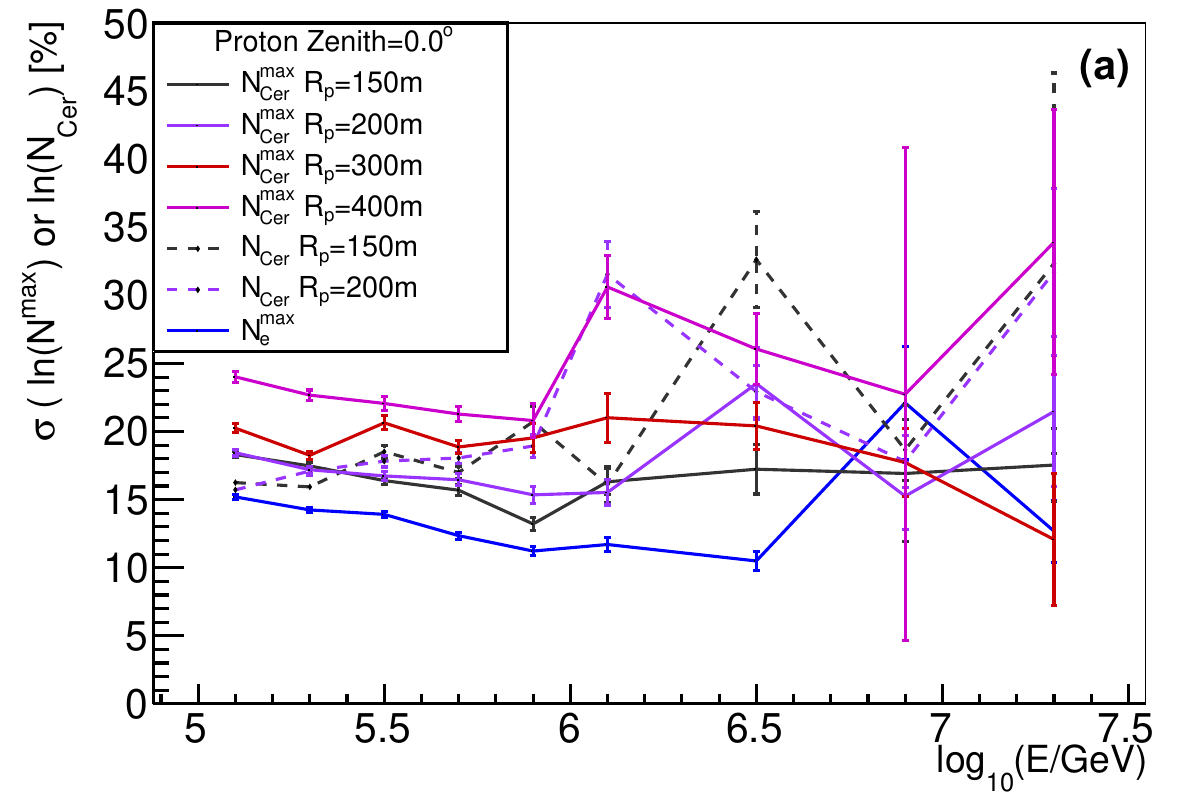}
\hspace{0cm}
\includegraphics[width=0.45\linewidth]{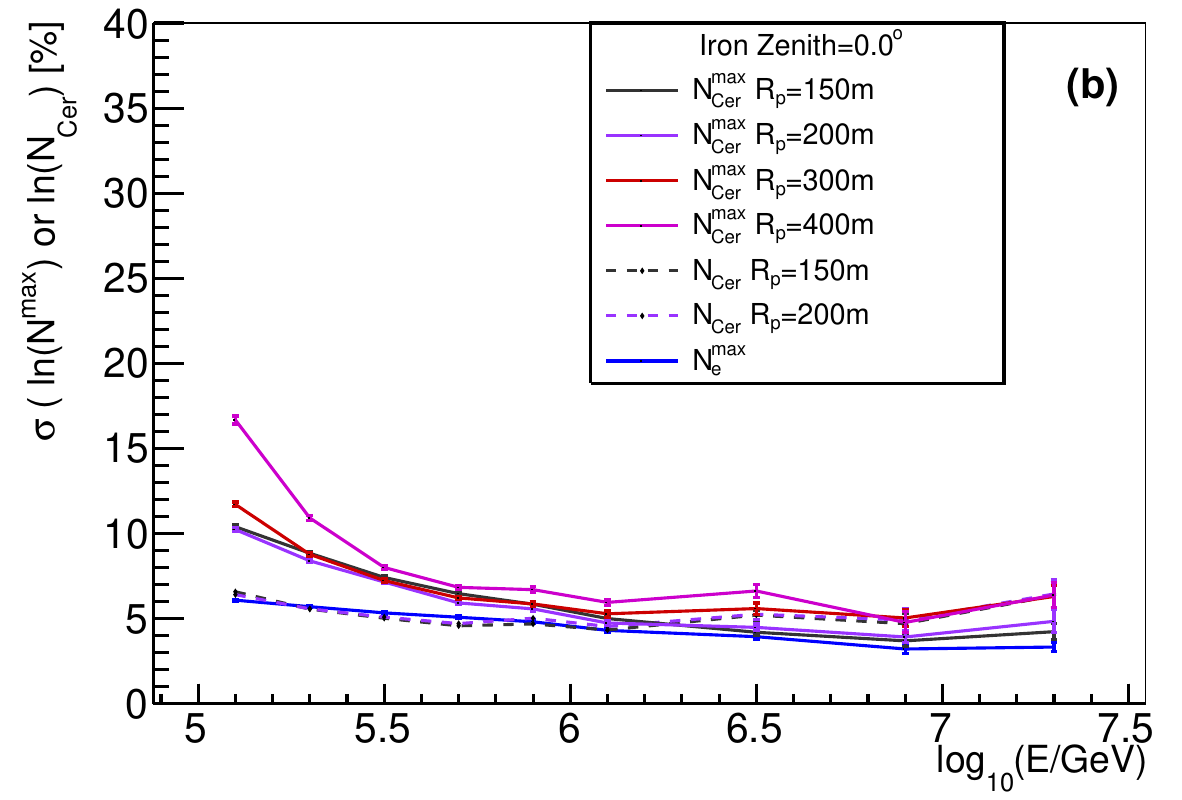} 
\hspace{0cm}
\includegraphics[width=0.45\linewidth]{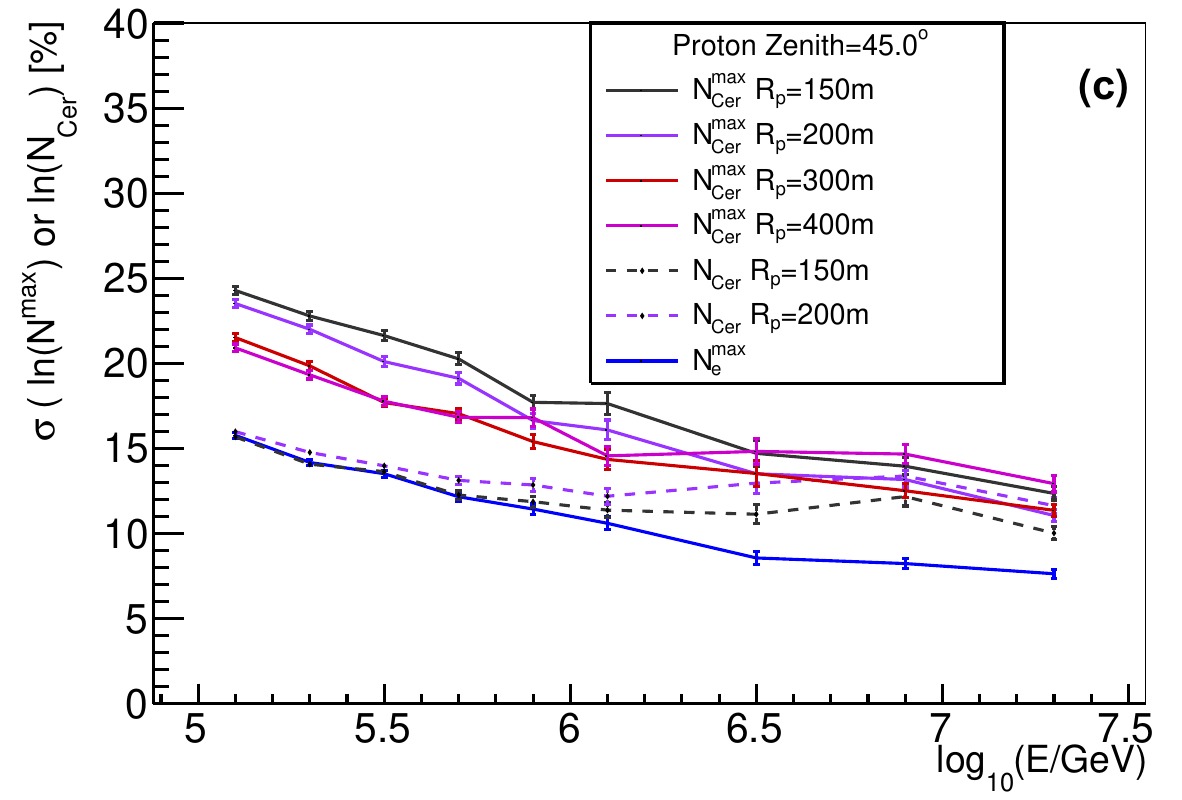}
\hspace{0cm}
\includegraphics[width=0.45\linewidth]{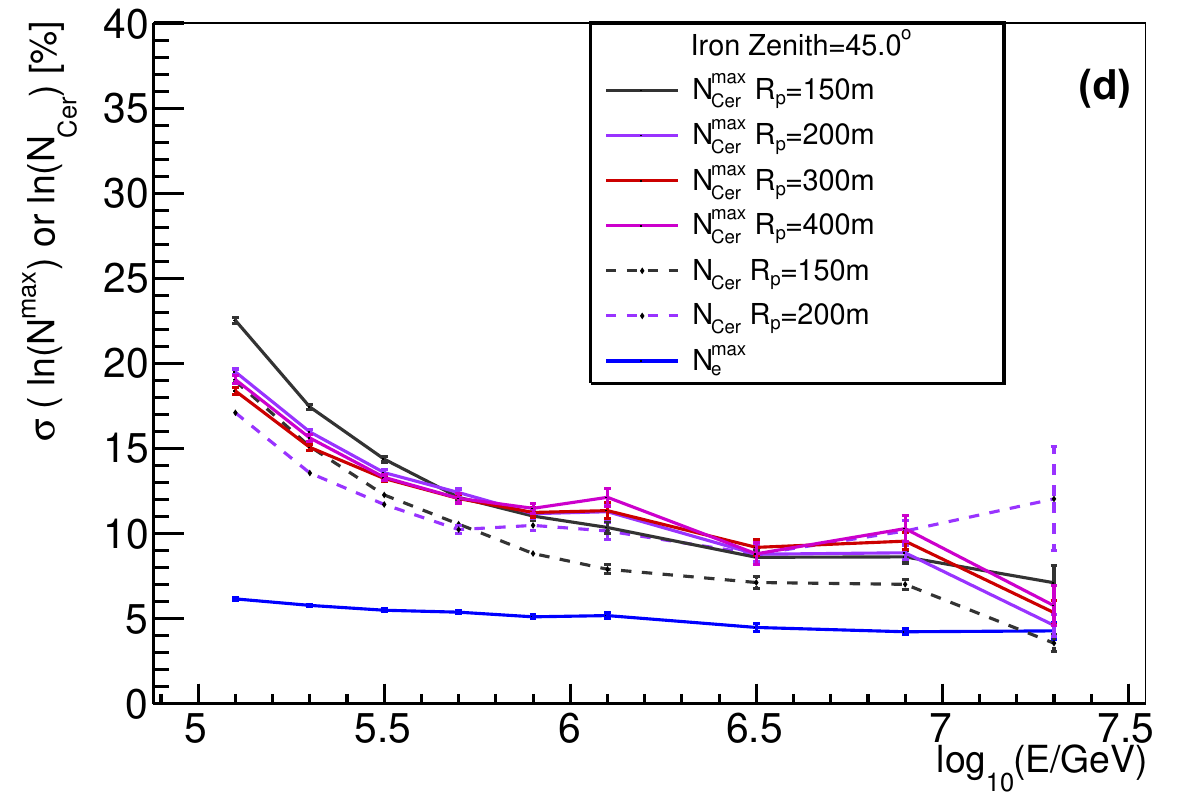}
\caption{The comparison of the shower-to-shower fluctuations between $N^{max}_{e}$, $N^{max}_{Cer}$ from longitudinal development and $N_{Cer}$ on observation level.}
\label{re_E_diffRp}
\end{minipage}
\end{figure*}

Cherenkov light is another widely used method for energy reconstruction. There are two variables for Cherenkov light. The first is the total number of Cherenkov photons at observation level (denoted as $N_{Cer}$ below). Another one is the number of Cherenkov photons per atmospheric depth bin at its shower maximum (or $N^{max}_{Cer}$). The comparison of the shower-to-shower fluctuations between $N^{max}_{e}$, $N^{max}_{Cer}$ and $N_{Cer}$ with different $R_{p}$ values ($R_{p}$=150, 200, 300, 400 m) and different composition for $\theta=0^\circ$ and $\theta=45^\circ$ are shown in Fig. \ref{re_E_diffRp}. For proton with $\theta=0^\circ$, at higher energy, the longitudinal development is more fluctuated, since there are more fraction of events with the shower maximum larger than the vertical atmospheric depth. As clearly seen, $N^{max}_{e}$ from electron longitudinal development has the best resolution in the whole energy range. The Cherenkov light measurement at fixed observation level (or $N_{Cer}$) has better resolution compared to $N^{max}_{Cer}$. This implies that accumulative measurement of Cherenkov light on observation level is a better choice for energy measurement. \\\indent

\begin{figure*}[htbp]
\begin{minipage}[t]{1.0\linewidth}
\includegraphics[width=0.5\linewidth]{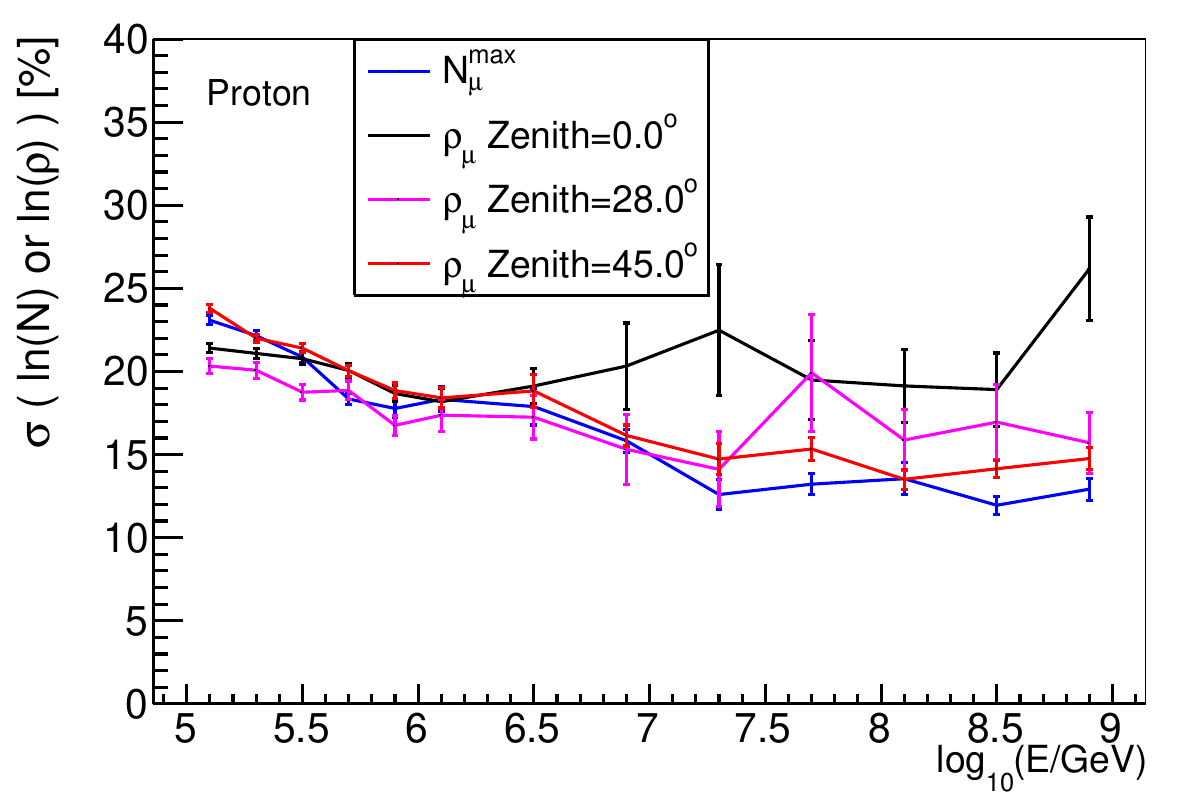}
\hspace{0cm}
\includegraphics[width=0.5\linewidth]{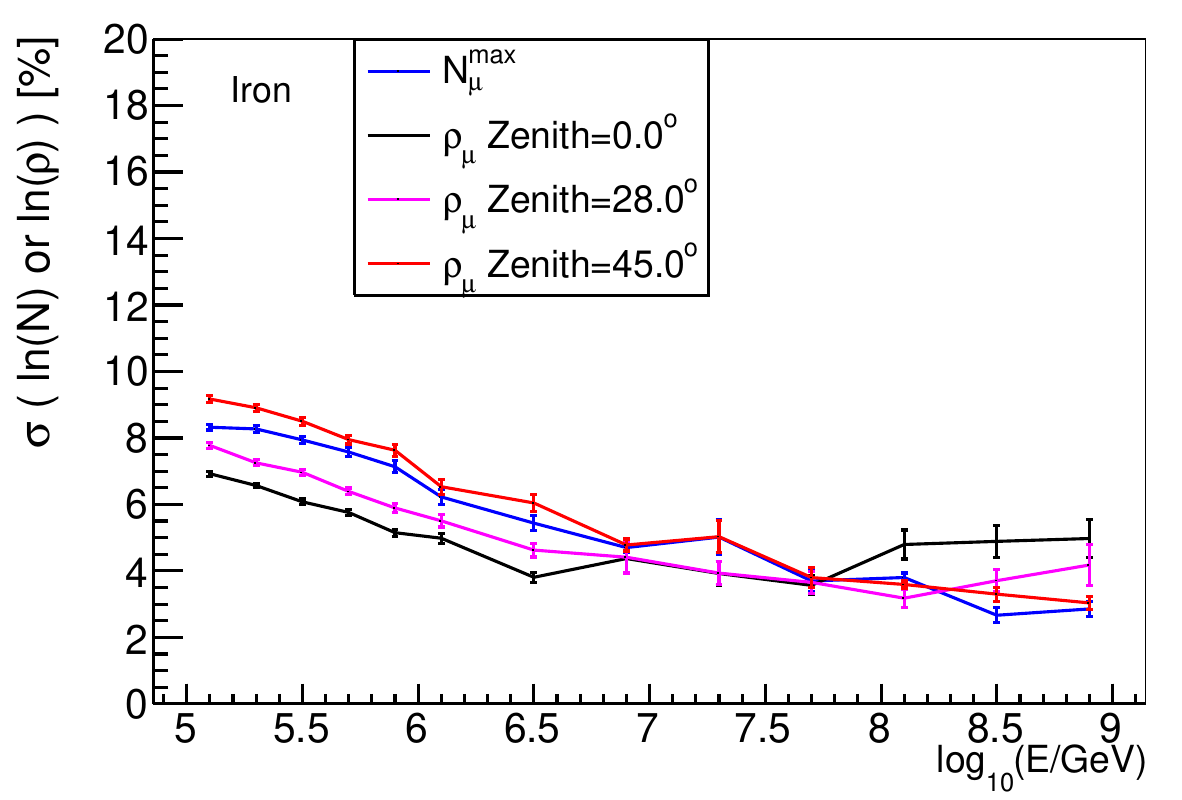}
\caption{The comparison of the shower-to-shower fluctuations of ln($\rho_{\mu}$) and ln($N^{max}_{\mu}$). The solid blue line indicates $N^{max}_{\mu}$, while other colored lines indicate $\rho_{\mu}$ with different zenith angles. The primary particles are protons (left) and irons (right) respectively.}
\label{re_E_diffzen_u}
\end{minipage}
\end{figure*}

For ground cosmic ray experiment, the composition discrimination power is limited compared to space cosmic ray experiment. Therefore, it is important to find powerful variables  for composition discrimination. The variables from lateral distribution for composition discrimination have been studied in detail ~\cite{EASliu}. It was extensively discussed that the density of muon at 250 m away from the shower axis (denoted as $\rho_{\mu}$ below) is the most powerful variable for nuclei discrimination. The shower-to-shower fluctuations of $\rho_{\mu}$ is important for the composition discrimination ability. The comparison of shower-to-shower fluctuations between $\rho_{\mu}$ and $N^{max}_{\mu}$ at different zenith angles for proton and iron is shown in Fig. \ref{re_E_diffzen_u}. \indent

Compared to electron  in Fig. \ref{re_E_diffzen}, the difference in the shower-to-shower fluctuations between the $\rho_{\mu}$ and $N^{max}_{\mu}$ is much smaller in Fig.\ref{re_E_diffzen_u}. The shower-to-shower fluctuations of $\rho_{\mu}$ is closer to that of $N^{max}_{\mu}$. This is due to the fact that muon almost do not interact with atmosphere, and the fluctuations of muon number after shower maximum is similar to the one at shower maximum. Since $N^{max}_{\mu}$ is more difficult to measure compared to $\rho_{\mu}$, the discrimination ability of $\rho_{\mu}$ will be compared to shower maximum from the longitudinal development point of view. \indent

\begin{figure*}[htbp]
\begin{minipage}[t]{1.0\linewidth}
\includegraphics[width=0.34\linewidth]{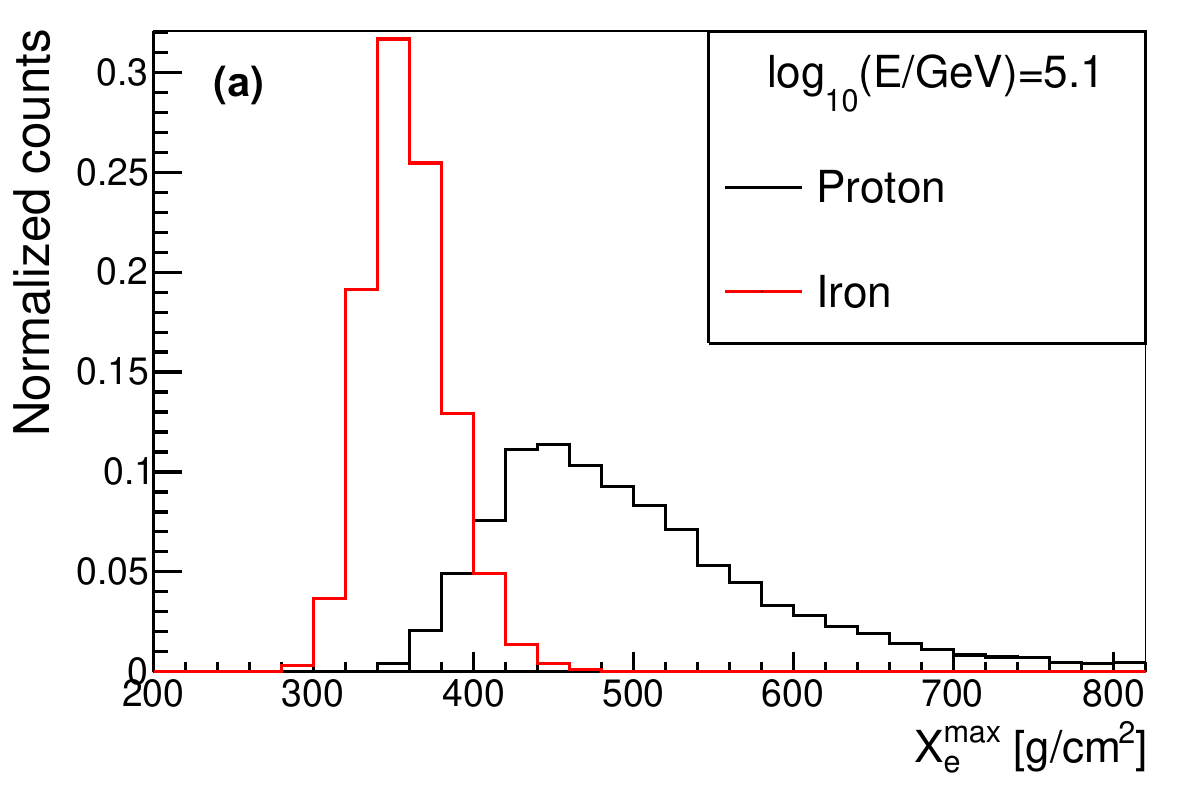}
\hspace{0cm}
\includegraphics[width=0.34\linewidth]{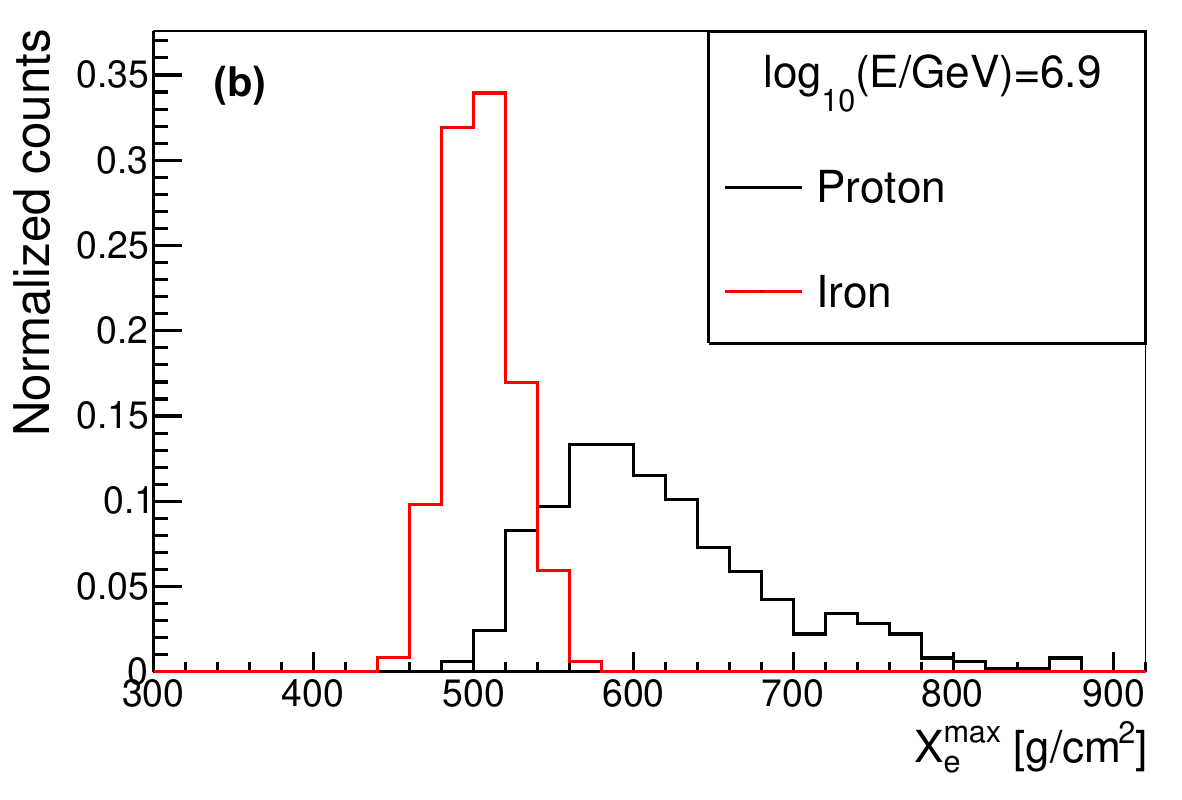}
\hspace{0cm}
\includegraphics[width=0.34\linewidth]{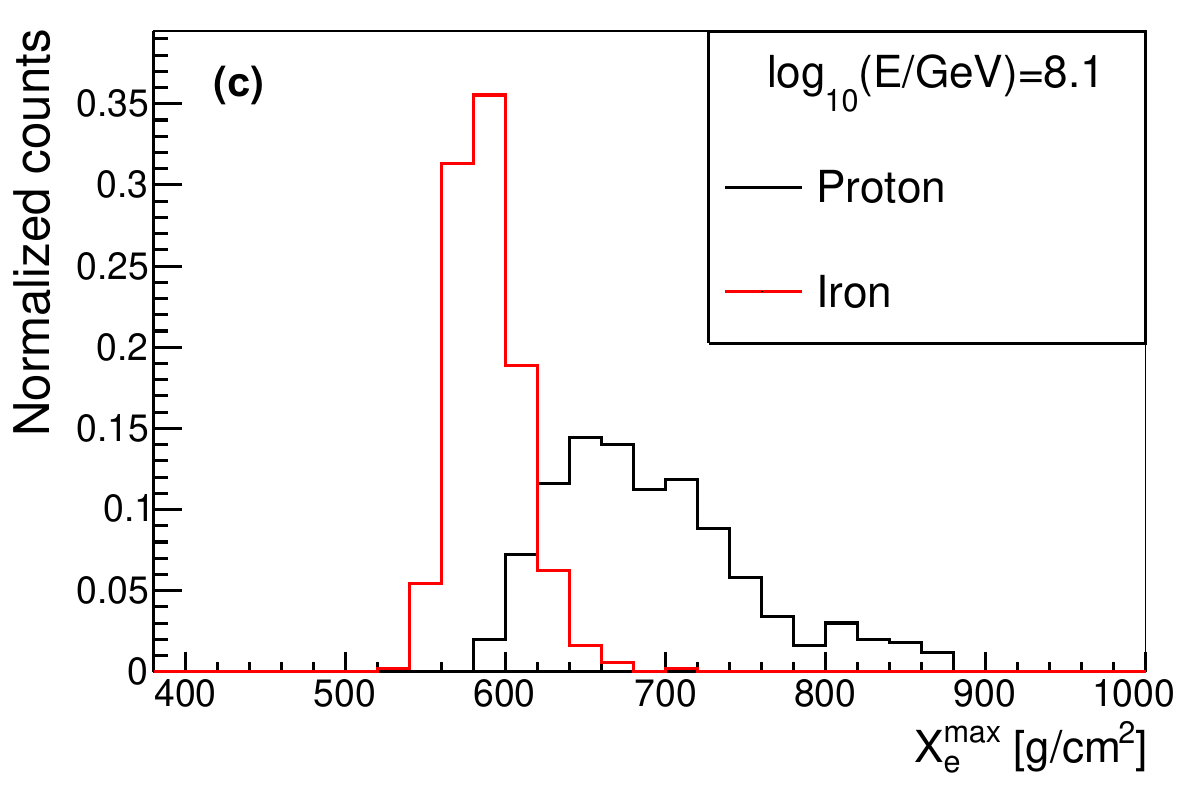} 
\hspace{0cm}
\includegraphics[width=0.34\linewidth]{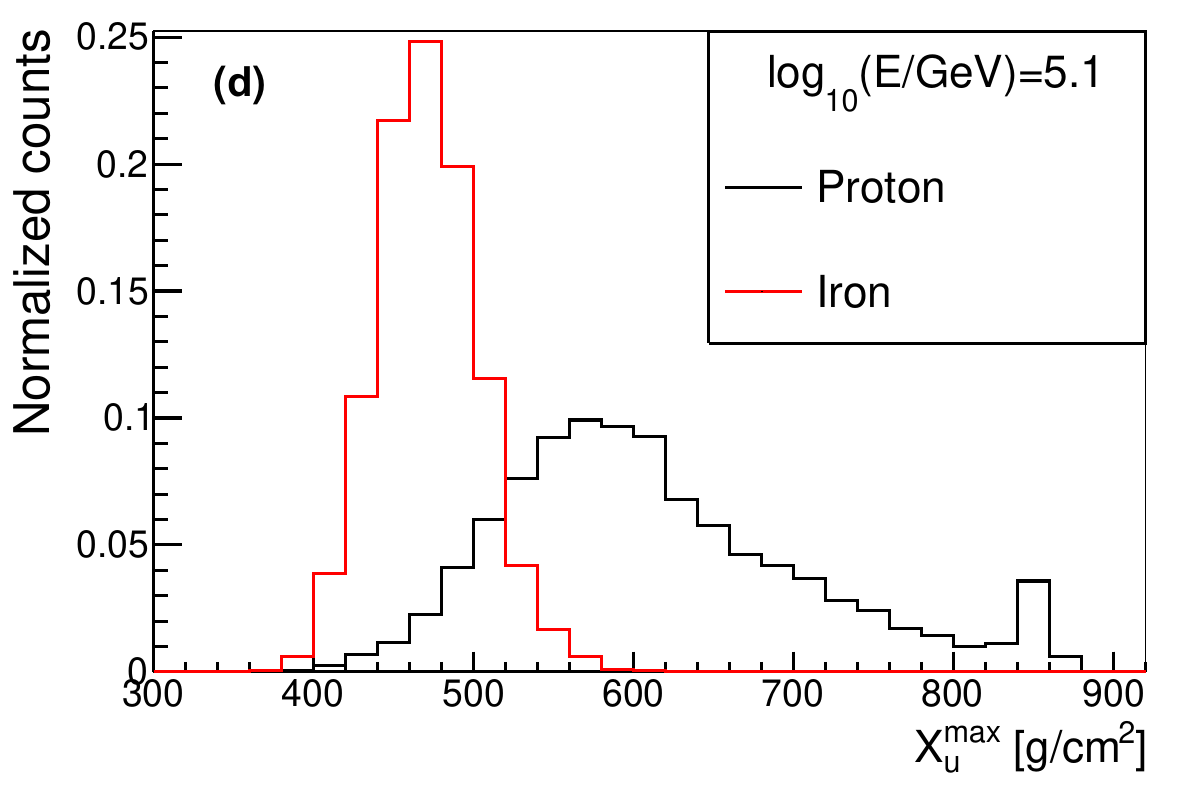} 
\hspace{0cm}
\includegraphics[width=0.34\linewidth]{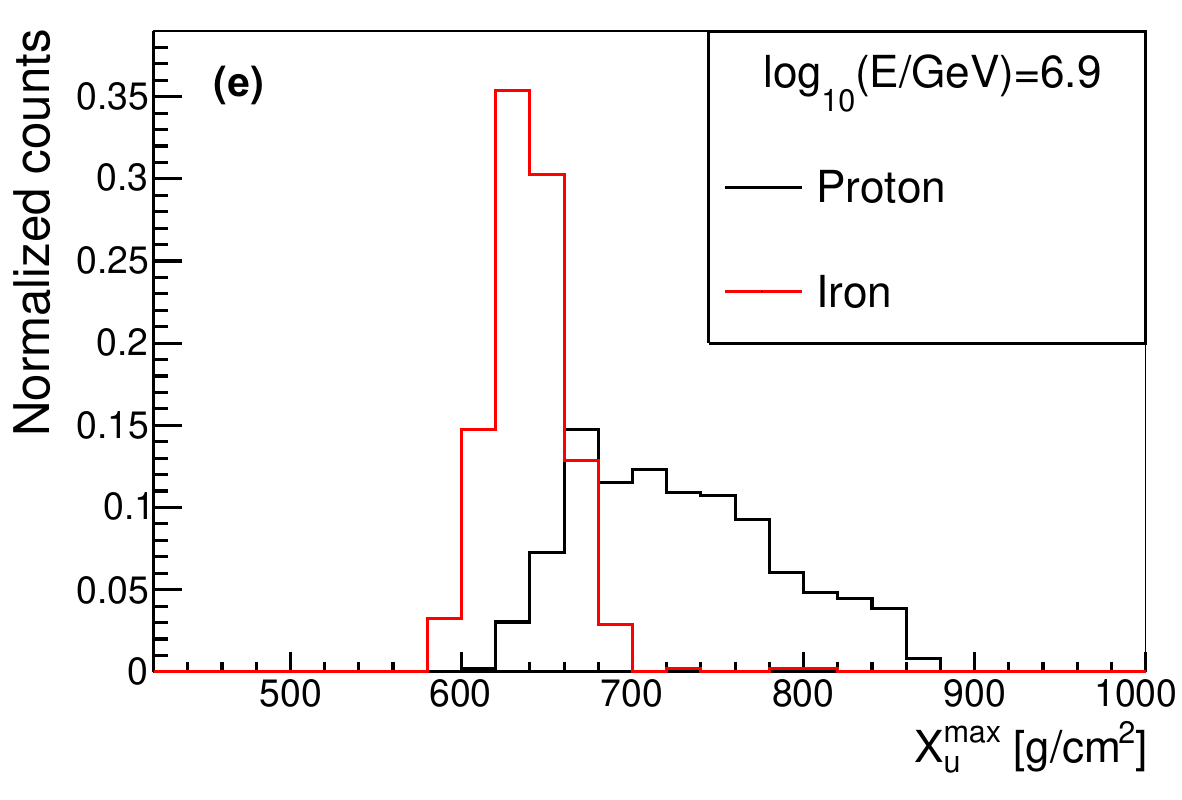}
\hspace{0cm}
\includegraphics[width=0.34\linewidth]{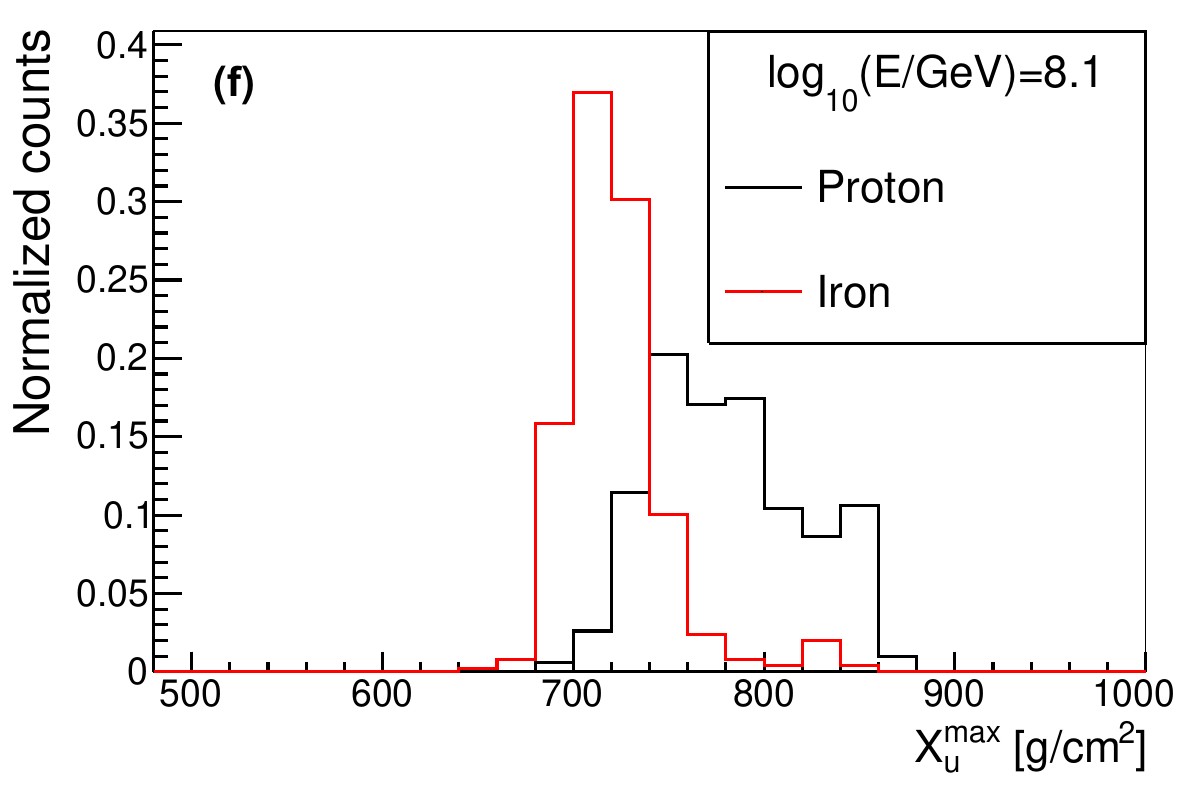}
\hspace{0cm}
\includegraphics[width=0.34\linewidth]{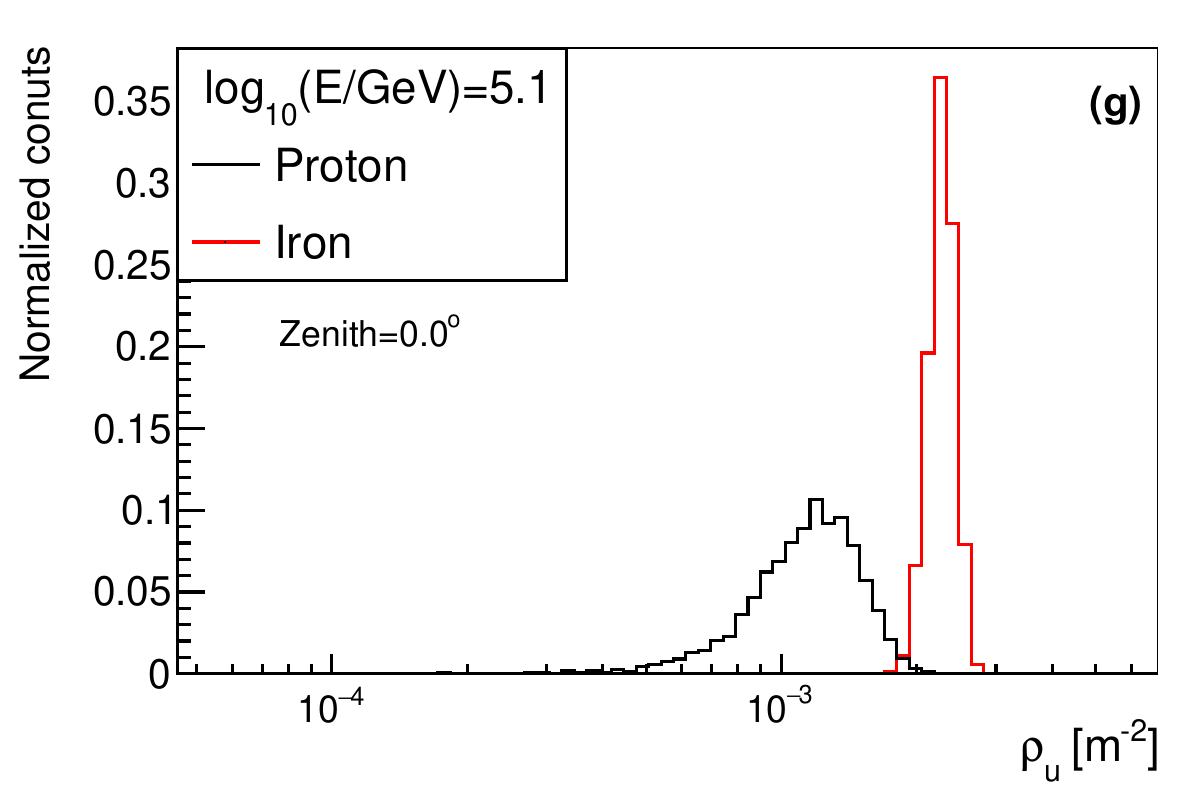} 
\hspace{0cm}
\includegraphics[width=0.34\linewidth]{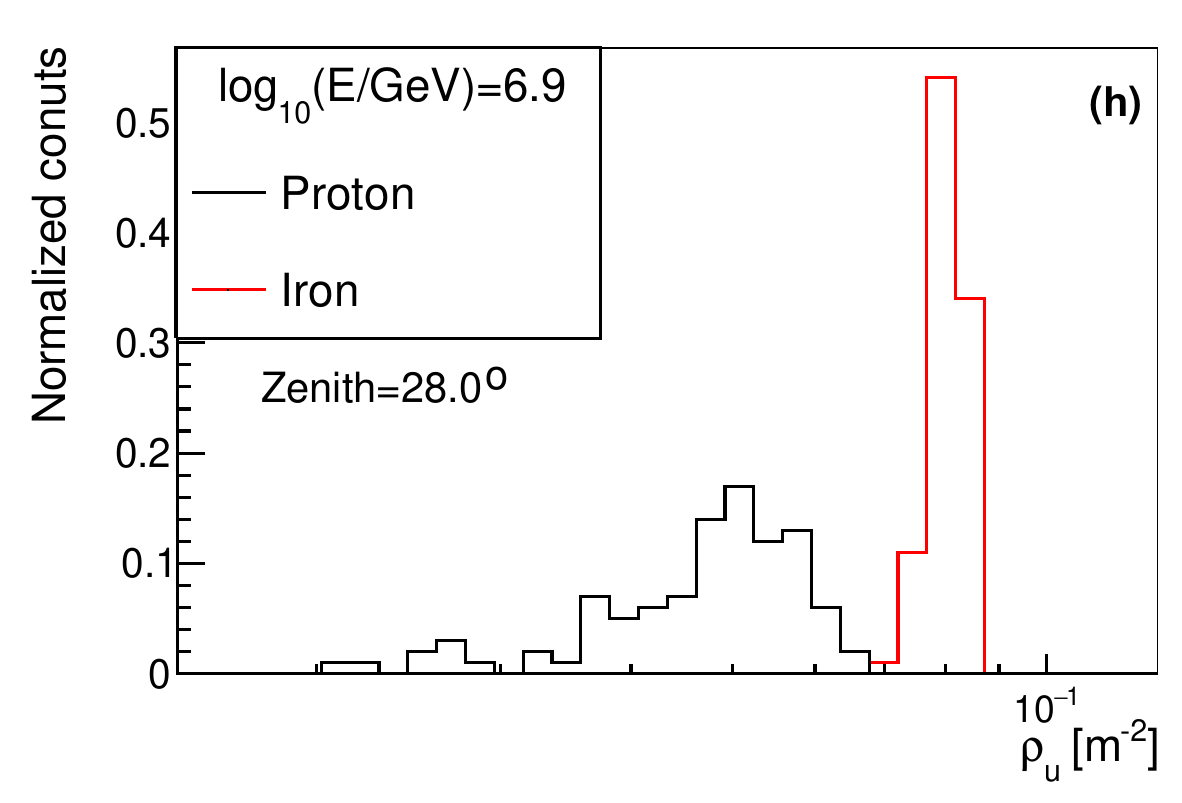}
\hspace{0cm}
\includegraphics[width=0.34\linewidth]{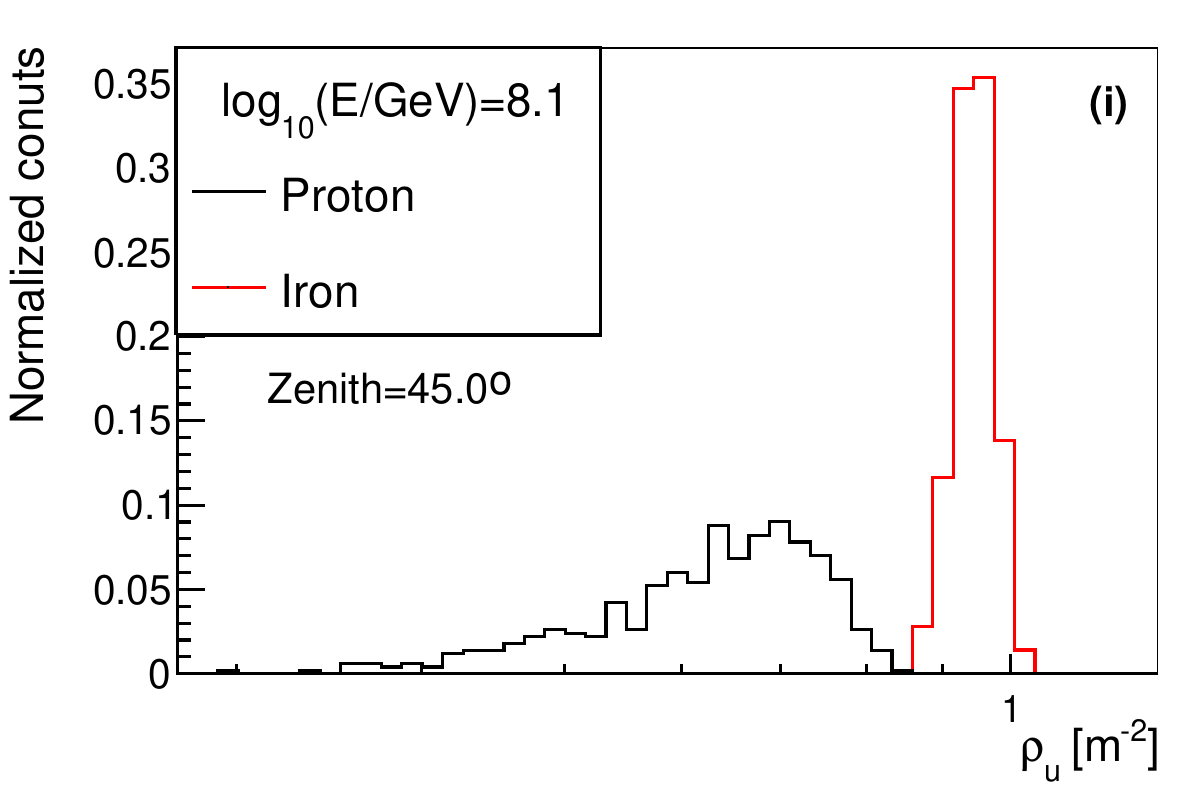}
\caption{ The comparison of distributions between $X^{max}_{e}$ (upper panel), $X^{max}_{\mu}$ (middle panel) and $\rho_{\mu}$ (lower panel) for proton and iron. The energy of primary particles is $log_{10}(E/GeV)$=5.1 (left), $log_{10}(E/GeV)$=6.9 (middle) and $log_{10}(E/GeV)$=8.1 (right), respectively.  The zenith angle of $X^{max}_{e}$ and $X^{max}_{\mu}$ sample is same at $\theta=45^\circ$. The zenith angle of $\rho_{\mu}$ sample is $\theta=0^\circ$ (g), $\theta=28^\circ$ (h) and $\theta=45^\circ$ (i). The black line is the distribution of proton, while the red line is the distribution of iron.}
\label{Identification_eu}
\end{minipage}
\end{figure*}

The variables $X^{max}_{e}$, $X^{max}_{\mu}$ and $X^{max}_{Cer}$ from longitudinal development were widely used as a composition estimators. The distribution of $X^{max}_{e}$, $X^{max}_{\mu}$ from longitudinal development and $\rho_{\mu}$ from lateral distribution for proton and iron at different energies are shown in Fig. \ref{Identification_eu}. As it can be seen, $X^{max}_{e}$ has similar ability with $X^{max}_{\mu}$, but $\rho_{\mu}$ has a much better composition discrimination power than $X^{max}_{e}$ and $X^{max}_{\mu}$. So $\rho_{\mu}$ from lateral distribution is a very powerful composition discrimination variable for cosmic ray physics. \\\indent

\begin{figure*}[htbp]
\begin{minipage}[t]{1.0\linewidth}
\includegraphics[width=0.5\linewidth]{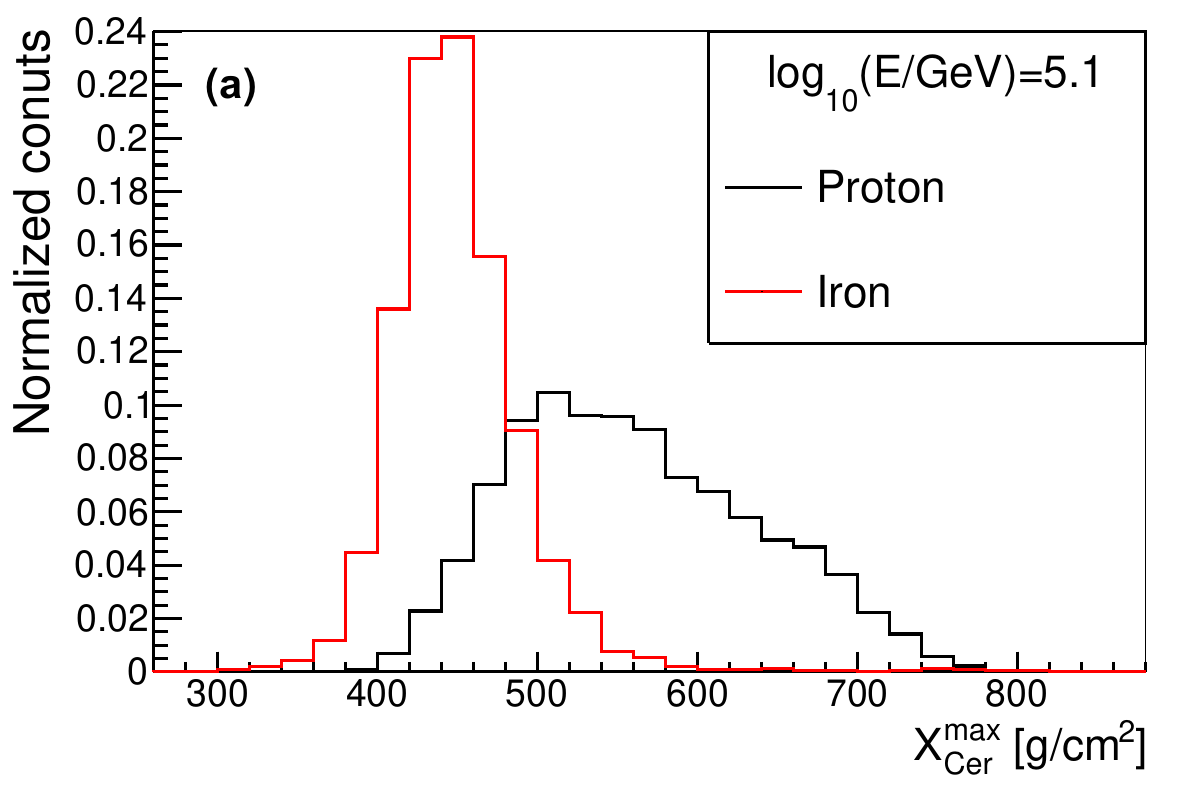}
\hspace{0cm}
\includegraphics[width=0.5\linewidth]{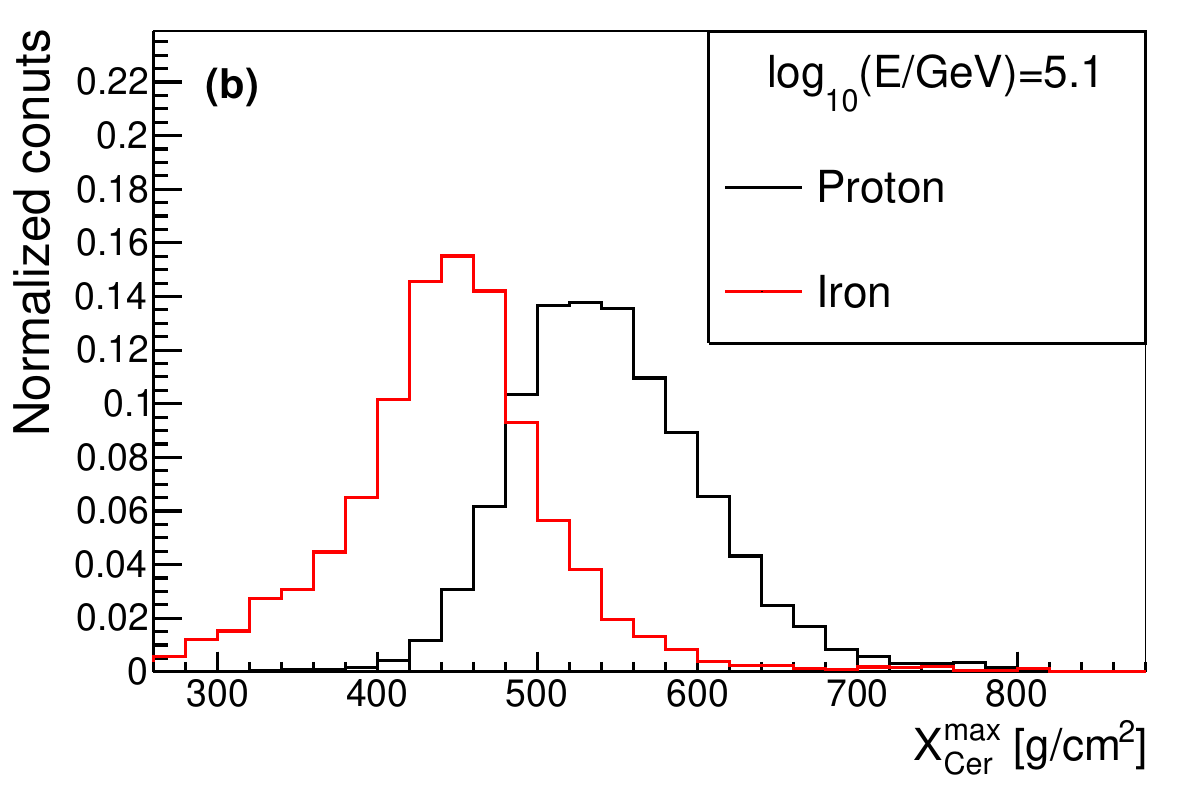}
\hspace{0cm}
\includegraphics[width=0.5\linewidth]{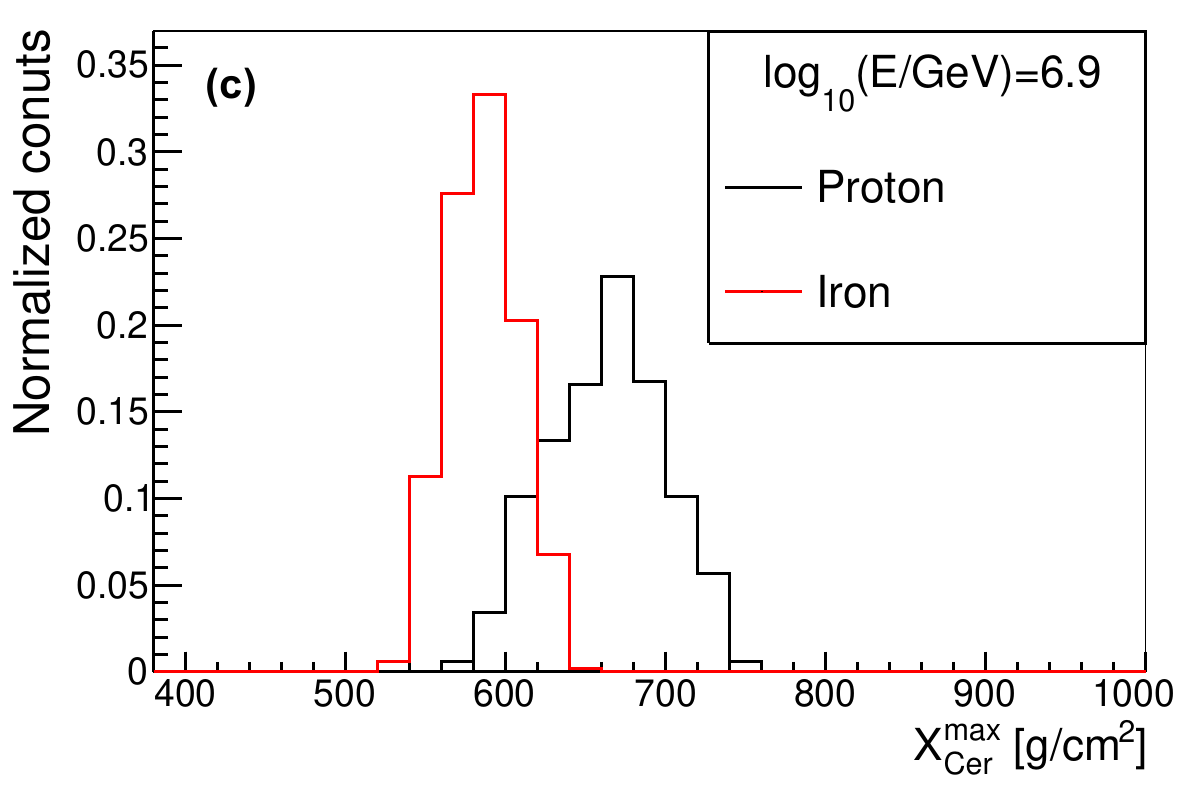} 
\hspace{0cm}
\includegraphics[width=0.5\linewidth]{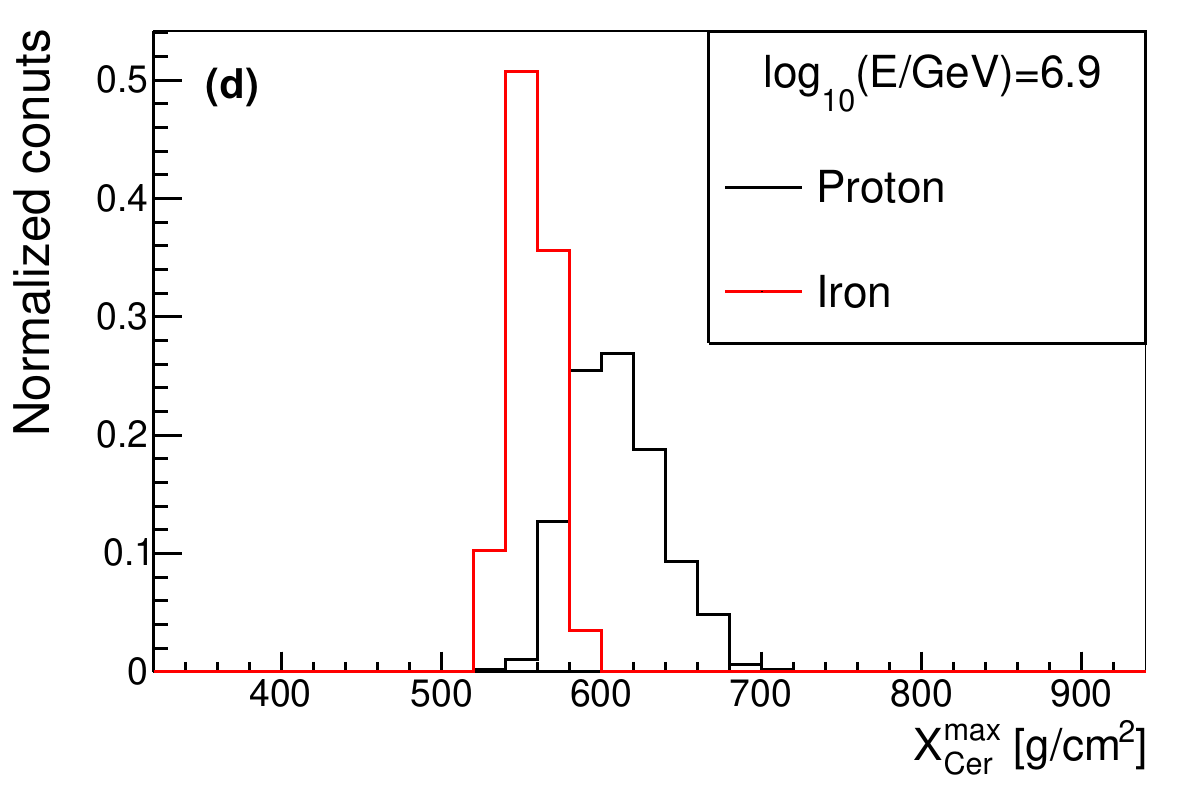} 
\caption{The comparison of distributions of $X^{max}_{Cer}$ for proton and iron. The energies of the shower are $log_{10}(E/GeV)$=5.1 (a,b), $log_{10}(E/GeV)$=6.9 (c, d), and $R_p$=150 m (a, c) and 400m (b, d), respectively. The black line is the distribution of proton, while the red line is the distribution of iron.}
\label{Identification_che}
\end{minipage}
\end{figure*}

The shower maximum measured from Cherenkov light (or $X^{max}_{Cer}$) is an estimator of $X^{max}_{e}$, which can also be used for composition discrimination. The distribution of $X^{max}_{Cer}$ for proton and iron at different $R_{p}$ values and different energies are shown in Fig. \ref{Identification_che}. As seen in the figure, the composition separation power of $X^{max}_{Cer}$ is worse than that of $X^{max}_{e}$ at around 100TeV, while they have similar discrimination power at around 1PeV. \indent

\section{Hadronic interaction model dependence}
\label{diff_hadronic}
Energy spectrum of cosmic ray measured by ground-based experiments depends on hadronic interaction models. For example, the proton energy spectrum measured by KASCADE experiment between $QGSJet-01$ and $Sibyll-2.1$ models are different, with the flux difference as nearly twice~\cite{KASCADEproton}. According to \cite{EASliu}, the differences of lateral distribution of secondary particles between $EPOS-LHC$ and $QGSJet-II-04$ hadronic models have been studied. The effects of hadronic interaction models on longitudinal development will be studied in this section and compared with the effects of hadronic interaction models on lateral distribution.

\begin{figure*}[htbp]
\begin{minipage}[t]{1.0\linewidth}
\includegraphics[width=0.33\linewidth]{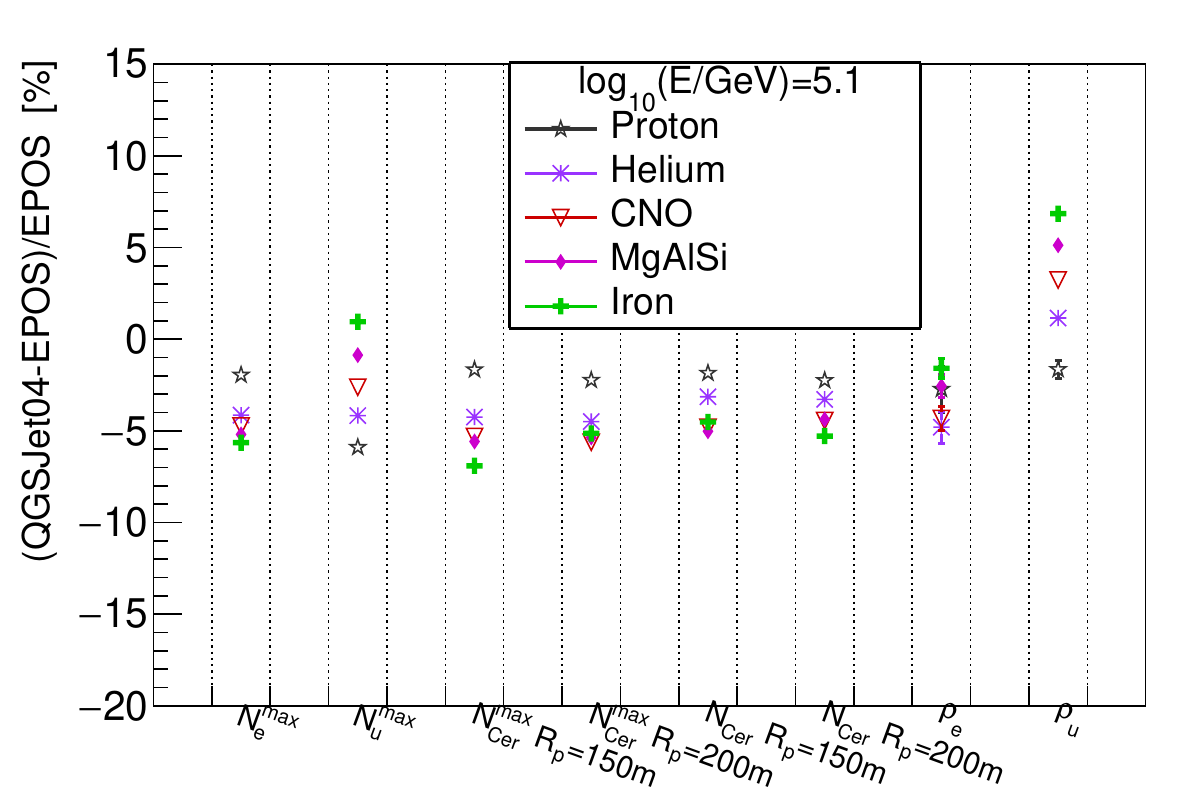}
\hspace{0cm}
\includegraphics[width=0.33\linewidth]{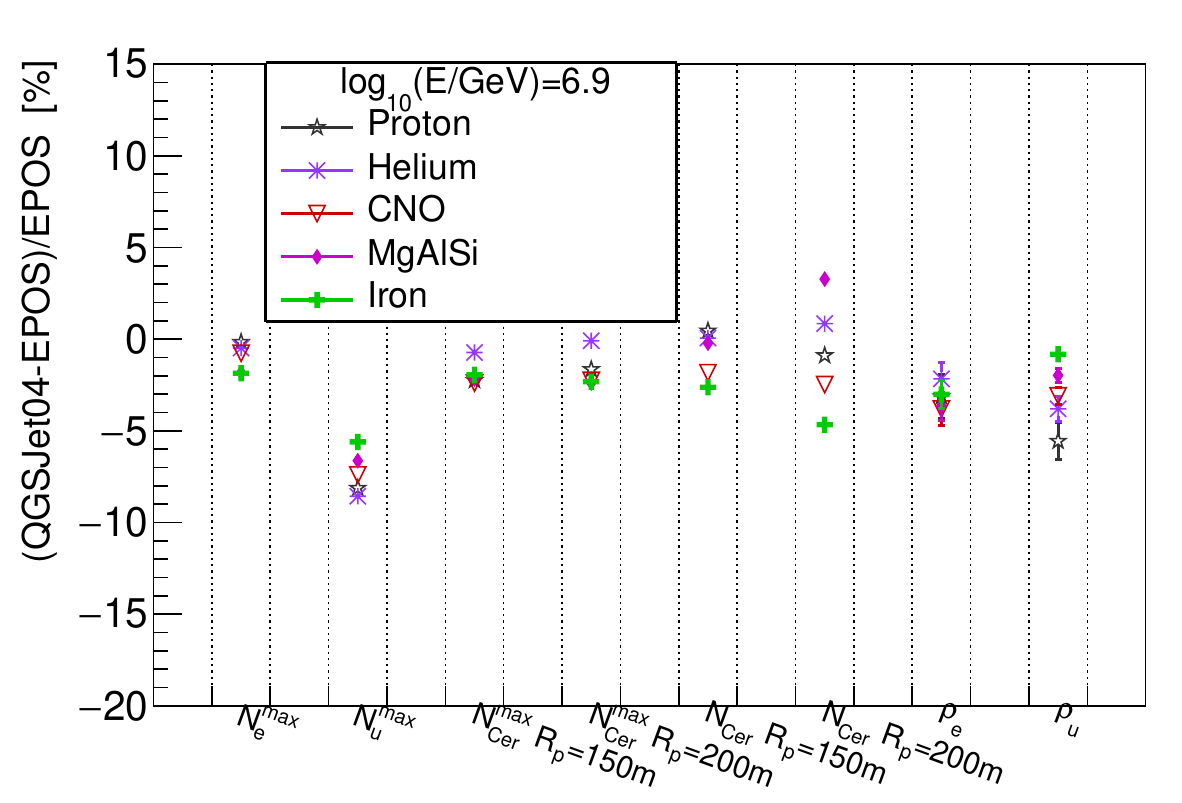}
\hspace{0cm}
\includegraphics[width=0.33\linewidth]{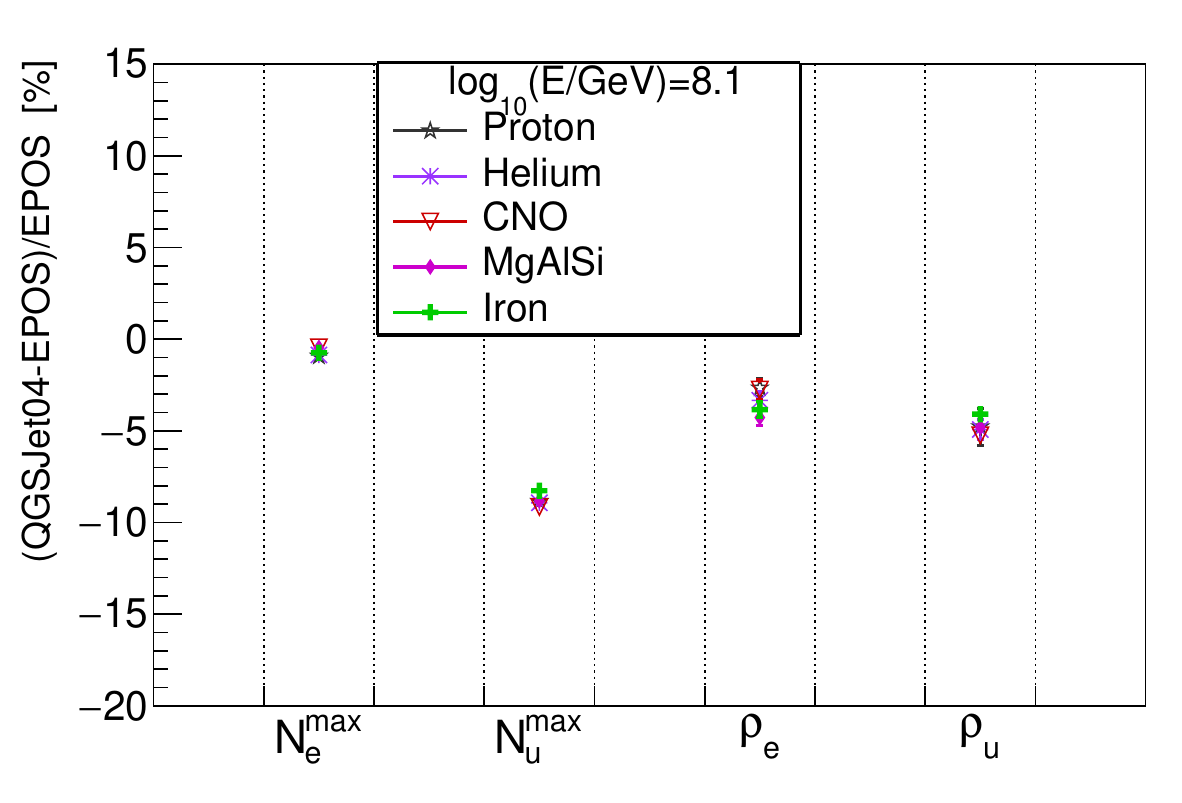} 
\caption{ The differences of $N^{max}_{e}$, $N^{max}_{\mu}$, $N^{max}_{Cer}$, $N_{Cer}$, $\rho_e$ and $\rho_{\mu}$ in percentage between EPOS-LHC model and QGSJet-II-04 model. Different colors and markers indicate different composition of primary particles (proton, helium, cno, mgalsi and iron). The energies are $log_{10}(E/GeV)$=5.1 (left), $log_{10}(E/GeV)$=6.9 (middle) and $log_{10}(E/GeV)$=8.1 (right, no Cherenkov light recorded due to limited computing source and disk size), respectively. }
\label{hmdiffN}
\end{minipage}
\end{figure*}

The difference in percentage (defined as $(QGSJet-EPOS)/EPOS\ \times100\%$) of $N^{max}_{e}$, $N^{max}_{\mu}$, $N^{max}_{Cer}$ , $N_{Cer}$, $\rho_{e}$ and $\rho_{\mu}$ between the two models are shown in Fig.\ref{hmdiffN}. As depicted in figure \ref{hmdiffN}, the differences of all variables except $N^{max}_{\mu}$ is within $\pm6\%$, while for $N^{max}_{\mu}$, the difference is larger for higher energy, around 9\% for $log_{10}(E/GeV)$=8.1. The difference of $\rho_{\mu}$ which is smaller than $N^{max}_{\mu}$ at high energy could be due to the difference of muon number which is mainly in the forward region of the shower (see Fig. 8 of reference \cite{EASliu} for details), and the $\rho_{\mu}$ measured far away from the shower core. It is also noticed that the differences between $EPOS-LHC$ and $QGSJet-II-04$ hadronic models are very similar for different type of nuclei for $log_{10}(E/GeV)$=8.1. \indent

\begin{figure*}[htbp]
\begin{minipage}[t]{1.0\linewidth}
\includegraphics[width=0.33\linewidth]{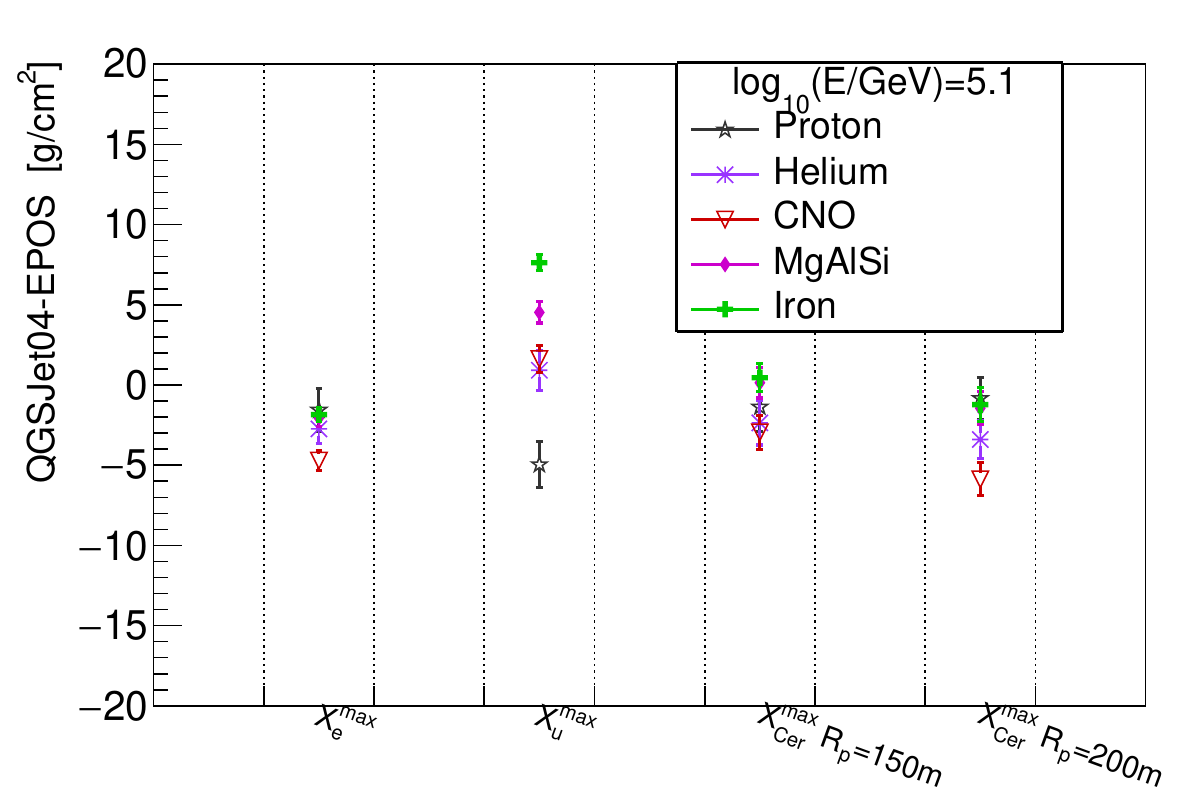}
\hspace{0cm}
\includegraphics[width=0.33\linewidth]{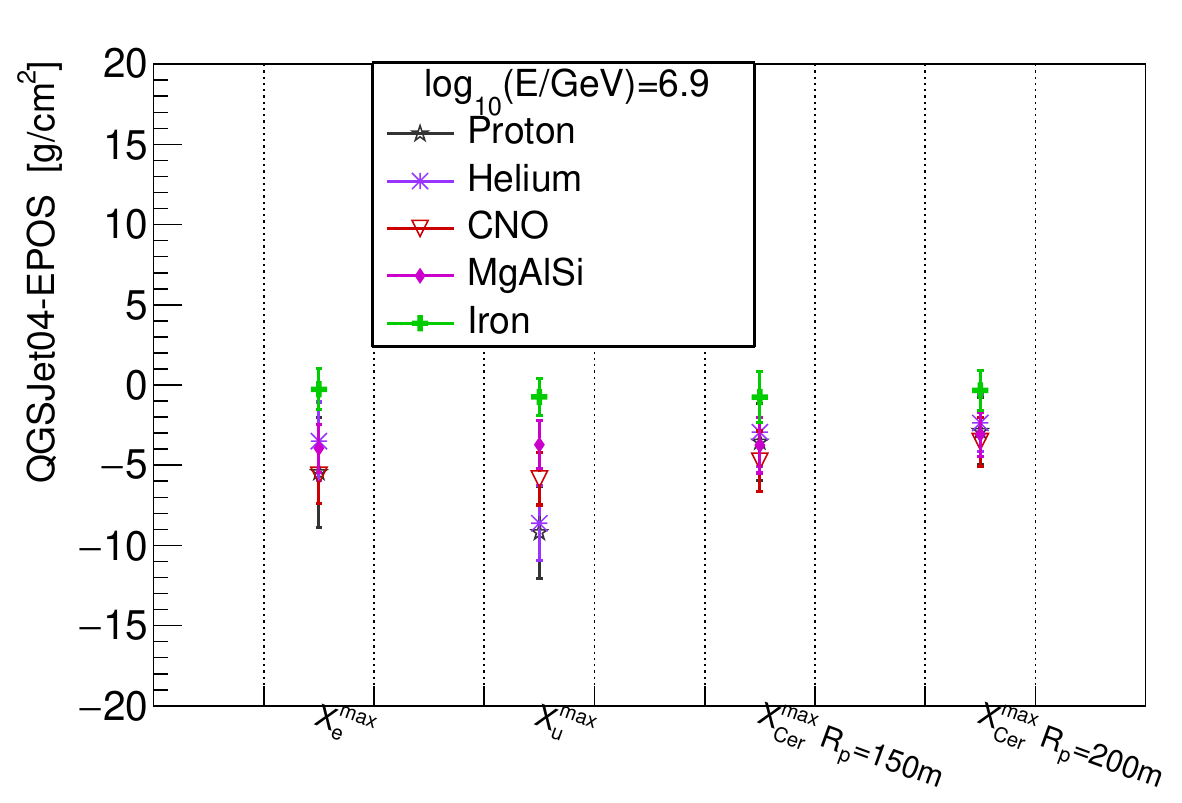}
\hspace{0cm}
\includegraphics[width=0.33\linewidth]{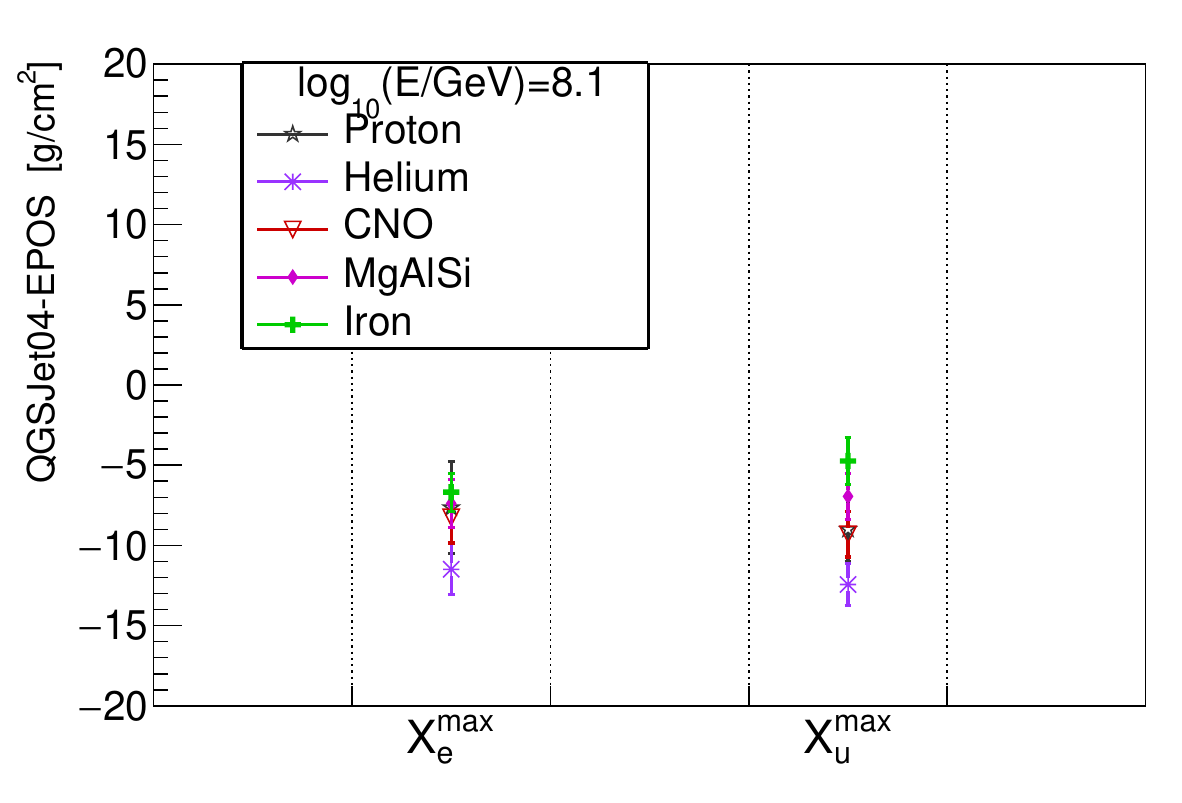} 
\caption{The differences of $X^{max}_{e}$, $X^{max}_{\mu}$ and $X^{max}_{Cer}$ with different $R_{p}$ values between EPOS-LHC model and QGSJet-II-04 model. Different colors and markers indicate different composition of primary particles (proton, helium, cno, mgalsi and iron). The energies of the shower are $log_{10}(E/GeV)$=5.1 (left), $log_{10}(E/GeV)$=6.9 (middle) and $log_{10}(E/GeV)$=8.1 (right, no Cherenkov light recorded due to limited computing source and disk size). }
\label{hmdiffX}
\end{minipage}
\end{figure*}

The location of the shower maximum of an air shower is another measurement that characterizes the properties of longitudinal development. The differences of $X^{max}_{e}$, $X^{max}_{\mu}$ and $X^{max}_{Cer}$ between hadronic models are shown in Fig.\ref{hmdiffX}. As illustrated, when $log_{10}(E/GeV)<=$ 6.9, the differences of $X^{max}_{e}$ and $X^{max}_{\mu}$ is within 6 $g/cm^2$, except $X^{max}_{\mu}$, which is within around 10 $g/cm^2$. \indent

\begin{figure*}[htbp]
\begin{minipage}[t]{1.0\linewidth}
\includegraphics[width=0.5\linewidth]{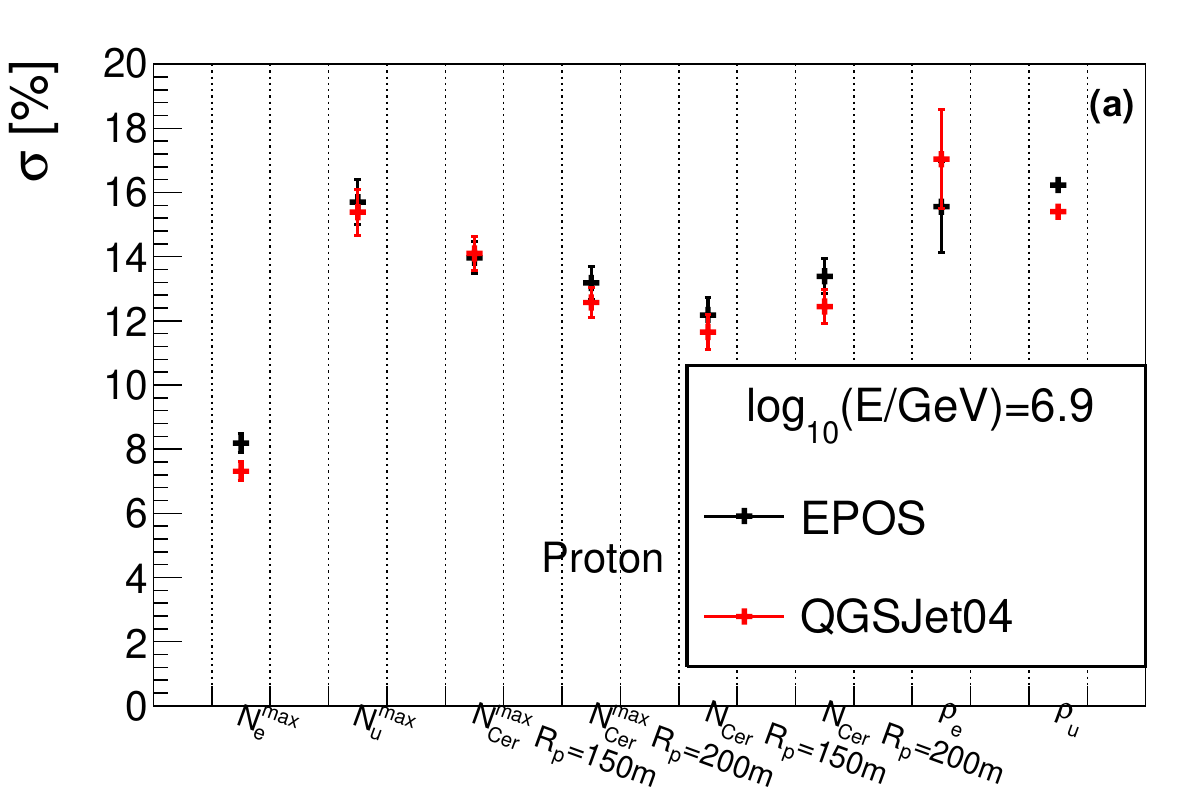}
\hspace{0cm}
\includegraphics[width=0.5\linewidth]{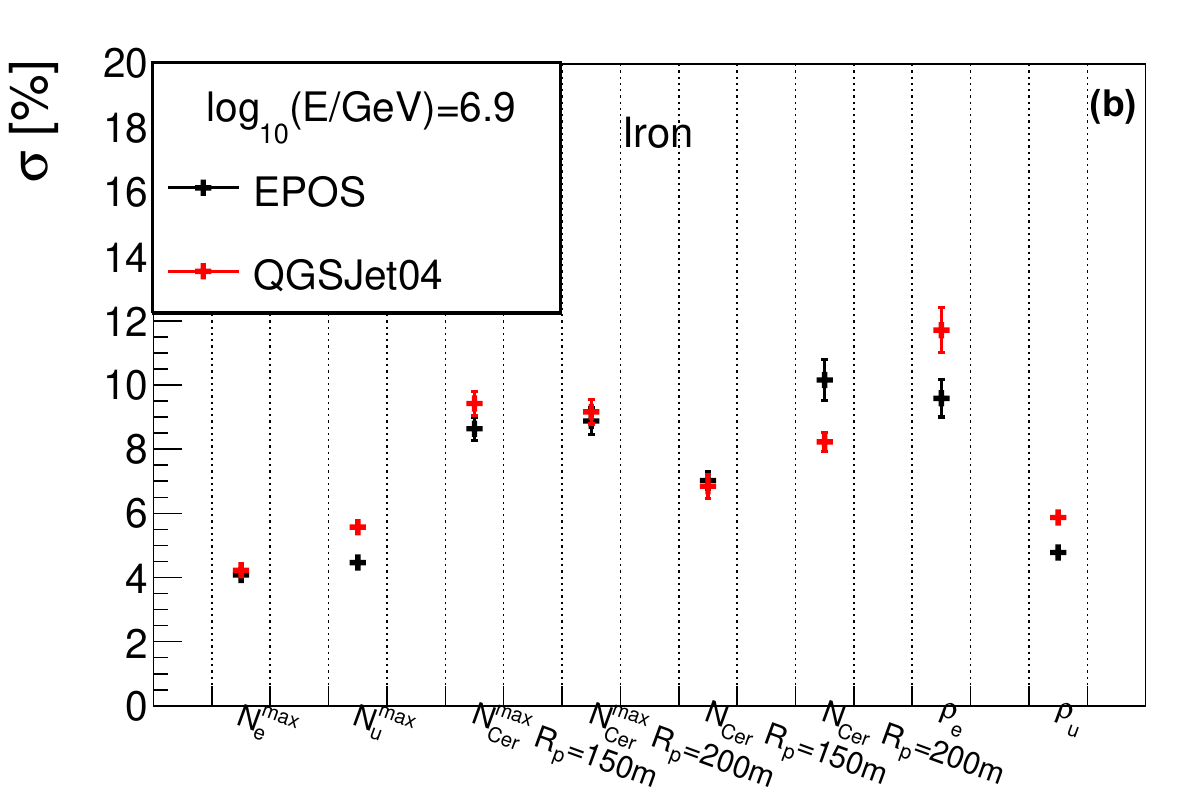}
\hspace{0cm}
\includegraphics[width=0.5\linewidth]{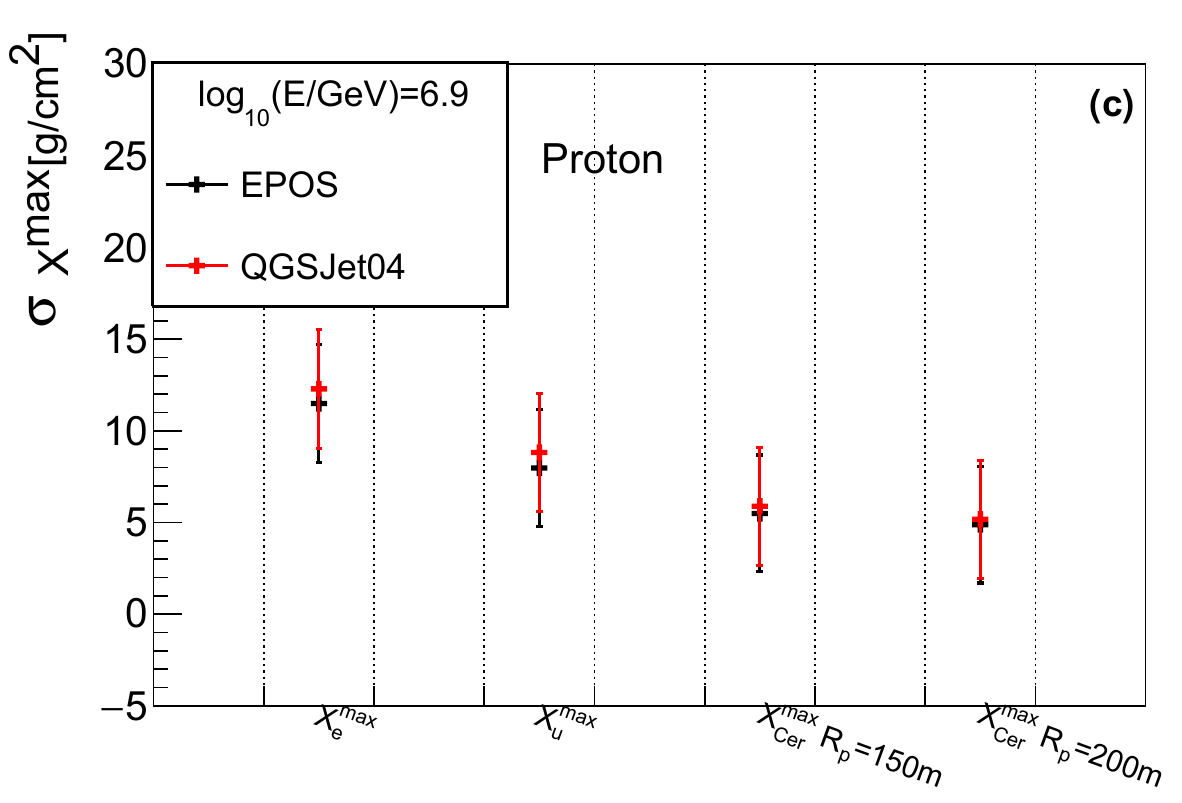} 
\hspace{0cm}
\includegraphics[width=0.5\linewidth]{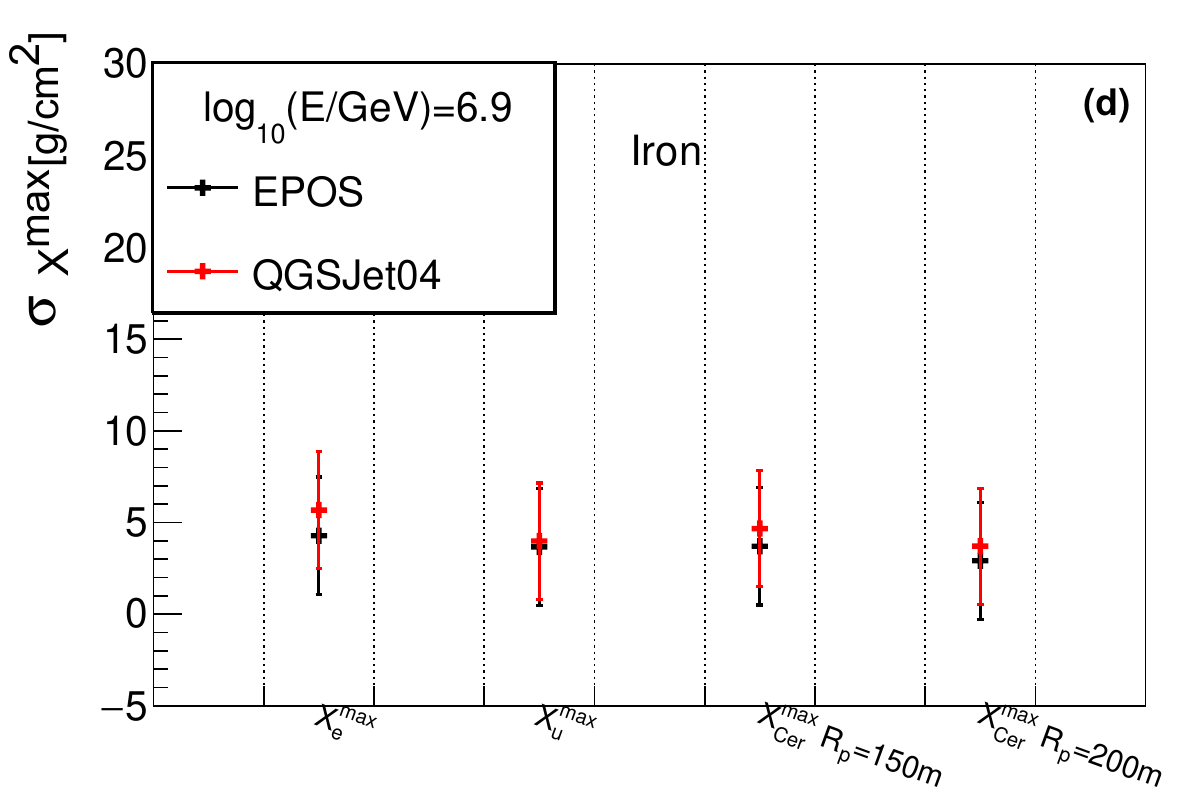} 
\caption{ The upper panel (a,b) show the sigma value of a Gaussian function fitted to the distributions of $N^{max}_{e}$, $N^{max}_{\mu}$, $N^{max}_{Cer}$, $N_{Cer}$, $\rho_e$ and $\rho_{\mu}$ in percentage. The lower panel (c,d) show the sigma value of a Gaussian function fitted to the distributions of of $X^{max}_{e}$, $X^{max}_{\mu}$ and $X^{max}_{Cer}$. The energy of the shower is $log_{10}(E/GeV)$=6.9, and the samples are simulated with $EPOS-LHC$ (black) and $QGSJet-II-04$ (red) hadronic models, respectively.}
\label{sigmadiffH}
\end{minipage}
\end{figure*}

The shower-to-shower fluctuations of shower size will affect the composition separation power and energy resolution. The upper panel of Fig.\ref{sigmadiffH} shows the sigma value of a Gaussian function fitted to the distributions of $N^{max}_{e}$, $N^{max}_{\mu}$, $N^{max}_{Cer}$, $N_{Cer}$, $\rho_e$ and $\rho_{\mu}$ in percentage, while the lower panel shows the sigma value of a Gaussian function fitted to $X^{max}_{e}$, $X^{max}_{\mu}$ and $X^{max}_{Cer}$. The samples are proton and iron nuclei with energy $log_{10}(E/GeV)=$ 6.9 simulated with $EPOS-LHC$ and $QGSJet-II-04$ hadronic models, respectively. As clearly observed, the differences of the sigma value for shower size is within $2.5\%$, while the differences of the sigma value for shower maximum is within $2g/cm^{2}$. Similar conclusions were found for other energies. \indent

In general, for the difference of nuclei between $EPOS-LHC$ and $QGSJet-II-04$ models, the differences of the number of electron and Cherenkov light in longitudinal development are close to the differences of the number from lateral distribution at observation level. The muon density at $R_{p}$=250 m (or $\rho_{\mu}$) has a smaller hadronic interaction model dependence than $N^{max}_{\mu}$. When $log_{10}(E/GeV)<$ 6.9, the differences of $X^{max}_{e}$ and $X^{max}_{Cer}$ are within 6 $g/cm^2$, while the difference of $X^{max}_{\mu}$ is lager, with a maximum of around 10 $g/cm^2$. The shower-to-shower fluctuations both for the shower size and shower maximum are small between $EPOS-LHC$ and $QGSJet-II-04$ models.

\section{Summary}
\label{summary}
In this work, the properties of the longitudinal development of electron, muon and Cherenkov light produced in EAS  in the knee energy region were studied, with a CORSIKA simulated samples for different energy, composition and zenith angles. The shower maximum ($X^{max}_{e}$, $X^{max}_{\mu}$ and $X^{max}_{Cer}$) and the number of secondary particles at shower maximum ($N^{max}_{e}$, $N^{max}_{\mu}$ and $N^{max}_{Cer}$) were derived from the longitudinal development of electron, muon and Cherenkov light by fitting them with a ``Gaussian-in-age'' function. The relationship between $X^{max}_{Cer}$ and $X^{max}_{e}$ is almost composition independent, however, the relationship is dependent on $R_{p}$ and zenith angle, therefore, $X^{max}_{Cer}$ is kind of a complicated composition independent $X^{max}_{e}$ estimator. The uncertainty of $X^{max}_{e}$ reconstruction from $X^{max}_{Cer}$ is about 10-15 $g/cm^{2}$ for all compositions and $R_{p}$ values, when energy is larger than$\sim$1 PeV. From 100 TeV to 1 PeV, the uncertainty is from 10-30 $g/cm^{2}$ for $R_{p}$=150 m, and 10-45 $g/cm^{2}$ for $R_{p}$=400 m. \indent

The performances of energy measurement and composition discrimination ability from $N^{max}_{e}$, $N^{max}_{\mu}$, $N^{max}_{Cer}$, $X^{max}_{e}$, $X^{max}_{\mu}$ and $X^{max}_{Cer}$ from longitudinal development were studied and compared with the variable $\rho_{e}$, $\rho_{\mu}$ from lateral distribution. It was found that $N^{max}_{e}$ has smallest shower-to-shower fluctuations compared to number of Cherenkov light and the density of electron (or $\rho_{e}$) at observation level. The smallest shower-to-shower fluctuations of $N^{max}_{e}$ is around 5\% for nuclei. However, by choosing the appropriate zenith angle at different energy, the fluctuations of $\rho_{e}$ is very close to the fluctuations of $N^{max}_{e}$. Meanwhile, the combination of electron number and muon number at their shower maximum, namely $N^{max}_{e}$+25$N^{max}_{\mu}$, is a very good composition independent energy estimator. $X_{max}$ from longitudinal development can be used for composition discrimination, which is linear with lnA. The shower-to-shower fluctuations of $X^{max}_{e}$ is 50—55 $g/cm^2$ for proton, and 20—25 $g/cm^2$ for iron. For proton/iron separation, the composition separation power of $X^{max}_{Cer}$ is worse than $X^{max}_{e}$ at around 100 TeV, while they have similar discrimination power at around 1PeV. However, the density of muon (or $\rho_{\mu}$) at observation level is much better than $X^{max}_{e}$, $X^{max}_{\mu}$ and $X^{max}_{Cer}$. \\\indent

The differences of the variables both from longitudinal development and lateral distribution between $EPOS-LHC$ and $QGSJet-II-04$ hadronic model were studied. For the differences between $EPOS-LHC$ and $QGSJet-II-04$ models, the differences of the number of electron and Cherenkov light in longitudinal development are close to the differences of the variables from lateral distribution at observation level, while the $N^{max}_{\mu}$ from longitudinal development has a larger model dependencies than $\rho_{\mu}$ from lateral distribution. The differences of $X^{max}_{e}$ were also similar with $X^{max}_{Cer}$, but the difference of $X^{max}_{\mu}$ is larger than both of them. The differences of shower-to-shower fluctuations between $EPOS-LHC$ and $QGSJet-II-04$ models are negligible. \\\indent

Therefore, our results provide important information for the selection of detector type, detector design and data analysis for physical measurements. \\

\section{Acknowledgements}
This work is supported by the Science and Technology Department of Sichuan Province (Grant number 2021YFSY0031), the National Natural Science Foundation of China (Grant number 12205244), the National Key R\&D Program of China (Grant number 2018YFA0404201). The numerical calculations in this paper have been done on Hefei advanced computing center.



\vspace{2.5mm} \centerline{\rule{80mm}{0.1pt}} \vspace{1mm}
\end{document}